\documentclass[preprint,10pt,authoryear]{elsarticle}

\usepackage{amssymb}
\usepackage{amsmath}
\usepackage{tikzfill}
\usepackage{soul} 
\usepackage[margin=1in]{geometry}
\usepackage{mathtools}
\usepackage{amsthm}
\usepackage{amssymb}
\usepackage{tikz}
\usepackage{pgfplots}
\pgfplotsset{compat=1.18} 
\usepackage{caption}
\usepackage{subcaption}
\usepackage{comment}
\usepackage{algorithm}
\usepackage[]{algpseudocode}
\usepgfplotslibrary{fillbetween}
\usepackage{tabularx} 

\newtheorem{theorem}{Theorem}
\newtheorem{lemma}{Lemma}
\theoremstyle{definition}
\newtheorem{definition}{Definition}
\theoremstyle{remark}
\newtheorem{remark}{Remark}

\newcommand{\ee}{\mathbf{e}}
\newcommand{\xx}{\mathbf{x}}
\newcommand{\yy}{\mathbf{y}}

\newcommand{\bb}{\mathbf{b}}
\newcommand{\nn}{\mathbf{n}}
\newcommand{\uu}{\mathbf{u}}
\newcommand{\UU}{\mathbf{U}}
\newcommand{\vv}{\mathbf{v}}
\newcommand{\ww}{\mathbf{w}}
\newcommand{\zz}{\mathbf{z}}

\newcommand{\HH}{\mathbf{H}}

\newcommand{\LL}{\mathcal{L}}
\newcommand{\bzz}{{\boldsymbol{\zeta}}}
\renewcommand{\HH}{\mathcal{H}}
\numberwithin{equation}{section}

\journal{arXiv}
\begin{document}

\begin{frontmatter}



\title{Quasi-Brittle Fracture: The Blended Approach} 


\author[1]{Semsi Coskun}
\author[1]{Davood Damircheli}
\author[1]{Robert Lipton}

\affiliation[1]{organization={Department of Mathematics, Louisiana State University},
            addressline={Lockett Hall}, 
            city={Baton Rouge},
            postcode={70803-4918}, 
            state={LA},
            country={USA}}


\begin{abstract}
A field theory is presented for predicting damage and fracture in quasi-brittle materials. The approach taken here is new and blends a nonlocal constitutive law with a two-point phase field. In this formulation, the material displacement field is uniquely determined by the initial boundary value problem. The theory naturally satisfies energy balance, with positive energy dissipation rate  in accordance with the laws of thermodynamics. Notably, these properties are not imposed but follow directly from the constitutive law and evolution equation when multiplying the equation of motion by the velocity and integrating by parts. In addition to elastic constants, the model requires at most three key material parameters: the strain at the onset of nonlinearity, the ultimate tensile strength, and the fracture toughness. The approach simplifies parameter identification while ensuring  representation of material behavior. The approach seamlessly handles fracture evolution across loading regimes, from quasi-static to dynamic, accommodating both fast crack propagation and quasi-brittle failure under monotonic and cyclic loading. 
Numerical simulations show quantitative and qualitative agreement with experiments, including three-point bending tests on concrete. The model successfully captures the cyclic load-deflection response of crack mouth opening displacement, the structural size-effect related to ultimate load and specimen size, fracture originating from corner singularities in L-shaped domains, and bifurcating fast cracks.
\end{abstract}




 


\end{frontmatter}



\section{Introduction}
\label{sec.intro}

This paper presents a field theory to predict damage and fracture in quasi-brittle materials. The approach taken here is new and blends a nonlocal constitutive law with a two-point phase field, handling fracture evolution across quasi-static to dynamic loading regimes seamlessly.
In this formulation, we establish an initial boundary value problem for which the material displacement field and the damage field are shown to be uniquely determined. The damage field, which serves as the phase field for the problem, is exclusively a function of the material strain with values between zero and one. The energy balance between the elastic potential energy and the damage energy follows directly from the equations of motion, with positive energy dissipation rate  consistent with  thermodynamics. Moreover, the failure energy can be directly and explicitly computed for a flat crack. Here, the failure energy factors into a product of two terms: one providing an explicit formula for the energy release rate, and the other giving the crack length.  What is novel is that these properties are not imposed, but follow directly from the constitutive law and evolution equations on multiplying the equation of motion by the velocity and integrating by parts. No assumptions beyond the evolution equation are required or used. Numerical simulations are developed using discrete approximations of the proposed model. In addition to elastic constants, the model requires at most three key material parameters: the strain at the onset of nonlinearity, the ultimate tensile strength, and the fracture toughness. The goal is to achieve an accurate representation of the quasi-brittle material behavior using the least number of parameters. 

The space-time nonlocality of the field theory presented here is motivated by the nonlocal computational models of free fracture given by peridynamics, \cite{silling2000reformulation}, \cite{silling2007peridynamic},  phase field methods, \cite{Bourdin-Francfort-EtAl-2000}, \cite{Aranson}, \cite{Karma}, \cite{miehe2010thermodynamically}, cohesive crack and element modeling, \cite{HillerborgPetersson1976}, \cite{XuNeedeleman1994}, \cite{OrtizPandolfi}, and pseudo-elastic damage accumulation models \cite{Carpinteri}.
In the blended approach, we take the peridynamic viewpoint and the force between pairs of points is related to the strain through a nonlinear constitutive law depending on the strain. Strains are calculated as difference quotients of displacements between pairs of points. Damage localization emerges solely from dynamics through the two-point phase field, which assesses force degradation between pairs of points; see Figure \ref{Failure envelope} and \eqref{twopointphase}. Unlike traditional phase field methods requiring an additional diffusive equation, our phase field and strain fields are directly coupled through nonlocal weighted averaging of pairwise interactions resulting in a single evolution equation in terms of the displacement field. In the blended approach, the critical energy release rate $\mathcal{G}_c$ follows from the evolution equations and is the energy per unit area needed to eliminate all the tensile force interaction between points lying on either side of a straight crack. Direct mathematical analysis shows that this result is independent of the length scale of nonlocality, see \eqref{crofton}. However, there remains an elastic resistance to interpenetration of cracks lips, as shown in Section \ref{sec.Straight Cracks Compression}. 

The nonlocal approach chosen here is based on peridynamics for the following reasons:  It is now known  that for suitable peridynamic models that any sequence of solutions associated with vanishing nonlocality have subsequences converging to cluster points that are solutions of the dynamic Navier equation away from cracks (see \cite{lipton2014dynamic} and \cite{lipton2016cohesive}). Additionally, the associated peridynamic energy $\Gamma$-converges to the Griffith fracture energy as shown by \cite{lipton2014dynamic} and \cite{lipton2016cohesive}. Moreover, the elastic interaction between the nonlocal crack and surrounding peridynamic field converges to the classic zero traction condition on the crack lips, this is demonstrated for straight cracks in \cite{liptonjha2021}. The crack tip driving force is found to match the product of the energy release rate and crack tip velocity in the limit of vanishing nonlocality for suitable peridynamic models. (Assuming uniform regularity of solutions away from the crack tip.) 
This approach bypasses Mott's hypothesis, and instead the result emerges directly from peridynamics, see \cite{jhalipton2020}. In turn, this delivers the explicit formula that relates the crack velocity and the energy flowing through the crack tip, see \cite{freund2}. In summary, peridynamics replaces difference quotients with differences and the Navier operator with a nonlocal integral operator. The aforementioned analysis shows that the peridynamic solutions converge in the limit of vanishing nonlocality to the solution of the Navier equation away from the crack and the associated limit fields interact elastically with the crack according to the classical theory, see \cite{freund1990dynamic}.

The proposed blended model handles quasi-brittle fracture across loading conditions, from quasi-static to dynamic fracture evolution, accommodating both monotonic and cyclic loading cases. We are motivated by previous quasi-brittle models developed in peridynamics by \cite{hobbs},  \cite{Gal}, \cite{NIAZI}, in the phase field models of \cite{Verhoosel}, \cite{Wu} and cohesive zone models \cite{OrtizPandolfi}, \cite{FalkNeedlemanRice2001}, \cite{bio} and the  pseudo-elastic damage accumulation models \cite{Carpinteri}. The literature is vast and these citations are by no means exhaustive. 

In this model, an explicit length scale  $L$  that is characteristic of the evolution is found. The ratio of fracture toughness to material strength together with the failure envelope determine $L$, see Section \ref{criticalenergyrelease}. 
Notably, the deterministic size effect \cite{BazantPlanas1998}, \cite{Carpinteri},  emerges directly from our initial boundary value problem and is observed in numerical simulations, consistent with size-effects reported by \cite{hobbs} using an exponential peridynamic potential. We work with a bilinear constitutive relation provided in \eqref{bilinear} and Figure \ref{Failure envelope} to see for the simplest case that the blended approach recovers quasi-brittle phenomena including the size-effects and accurate load-displacement curves.

The distinguishing features of this approach in relation to existing approaches are now outlined. To the best of our knowledge, for the first time in free fracture modeling, we present a mathematically well-posed, {\em strength}-based evolution that delivers a displacement-damage pair from an initial boundary-value problem described exclusively by momentum balance. The phase field associated with material damage is not the solution of a separate diffusion equation but instead is recovered  from  the solution of the momentum balance equation. The energy balance between the elastic potential energy and the damage energy follows directly from the equations of motion, with a positive energy dissipation rate  consistent with  thermodynamics. The tensile strength is free to degrade with time, while the the compressive strength of the material does not. (Although one can include material degradation in compression as well.) Classic phase field models for quasi-brittle fracture are not able to fully model strength \cite{miehe2010thermodynamically}, \cite{Wu}. This is because  classical variational phase-field models cannot account for material strength as a material property independent of fracture toughness. The new observation made in this paper is that material degradation models can be described by a two-point history-dependent phase field description of the bond force. Moreover, for this model, the elasticity, strength, and fracture toughness can be prescribed independently of each other. We are quantitatively able to capture the crack mouth opening versus load for cyclic three-point bending experiments using this model. To the best of our knowledge, no phase field model has attempted this. For this model, only essential material properties are used and no less.  Here the bond degradation is calibrated only by the specimen strength, critical energy release rate, and the material stiffness.

Lastly, we point out the recently introduced strength-based phase field model of \cite{KumarFrancfortPalmes}  based on the stationary conditions of a min-max scheme employing the Ambrosio-Tortarelli functional. This physically insightful and novel scheme is adapted to elastic-plastic materials in the quasi-static-quasi-brittle regime and is extended to evolve the phase field subject to an anisotropic strength surface in \cite{KumarFrancfortPalmes2}.  For comparison, the method developed here is dynamic and evolves fracture in materials for which the strength constraints are primarily known for uniaxial tensile stress and compression. This includes geological and structural materials such as rock and concrete. Most importantly, our strength domain is history-dependent and evolves in time as the material degrades.  Our motivation lies in the earlier approaches of \cite{Carpinteri}, \cite{HillerborgPetersson1976}, \cite{XuNeedeleman1994}, \cite{OrtizPandolfi} that naturally accommodate history-dependent strength.

It should be noted that the method used to calibrate the aforementioned peridynamic models and the Blended model to known values of elastic moduli, fracture toughness, and strength do not correspond to those commonly used in peridynamics. In our approach, the calibration of fracture toughness and elasticity coefficients follow directly from  $\Gamma$-convergence and geometric measure theory, and the material strength is calibrated as an independent parameter as in \cite{Carpinteri}, \cite{HillerborgPetersson1976}, \cite{XuNeedeleman1994}, \cite{OrtizPandolfi}. Moreover, motivated by the aforementioned investigations, we treat the blended model as a well-posed approximation theory for fracture evolution.

In this model, the solution of the initial boundary value problem is the displacement field $\uu(t,\xx)$. The damage and fracture sets are deduced directly from the solution of the initial value problem. This is done using the difference quotient of the solution and the history-dependent constitutive law. This is explained as follows: The difference quotient is considered only for two points $\yy$ and $\xx$ points closer than some fixed length $\epsilon>0$ called the horizon. The force interaction between the two points depends on the strain through a linear elastic constitutive law and is called a bond. In our model, when the force exceeds the history-dependent tensile strength of the bond as prescribed by a memory function \eqref{scaledmax} and the failure envelope \eqref{failure envelope}, the moduli of elasticity begins to soften in tension. When the elastic modulus is softened to zero, the bond is said to be failed (in tension but not in compression). In this model, zones of damage and fracture can then be deduced from the a posteriori examination of the elastic moduli of each bond. Most importantly, the elastic moduli associated with two end points of a bond between  $\yy$ and $\xx$ is simply the product of the two-point phase field and the elastic moduli of an undamaged bond. Because of this, the evolution of the two-point phase field is directly coupled to the evolution of two-point strain and does not evolve through a diffusion equation as in conventional  phase field fracture formulations. 

We describe the failure set $\Gamma^\epsilon(t)$ associated with a flat crack given by a two-dimensional piece of surface $R_t$, where $t$ denotes the time dependence.  A crack is defined as an internal boundary where the points $\yy$ above the surface are no longer influenced by the forces due to the points $\xx$ below the surface and vice versa. This is the case of failed bond alignment, i.e., all bonds connecting points $\yy$  above  $R_t$ to points $\xx$ below are broken. A calculation of the failure energy of $\Gamma^\epsilon(t)$ required to break this set of bonds is given in Section \ref{criticalenergyrelease}. It shows that the failure energy is the product of the critical energy release rate multiplied by the two-dimensional surface measure of $R_t$. Displacement jumps can only occur across $R_t$, and nonlocal traction forces are zero on either side of $R_t$. Since a bond broken in tension still elastically resists compression, there is an elastic force resisting the interpenetration of crack lips, see Section \ref{sec.Straight Cracks Compression}.

The geometry of the failure set is controlled by how it grows dynamically. Growth is governed by the rate of work done against boundary forces and the dynamic interaction between elastic displacement and bond failure. Although the interaction is captured implicitly through the evolution equations, one can now apply the rate form of energy balance to explicitly deliver the time rate of the damage energy and characterize the location of the region undergoing damage. This region is called the process zone $PZ^\epsilon(t)$ and from the constitutive law, corresponds to the regions of highest strain, see Figure \ref{Failure envelope}. In this way, we see that cracks naturally grow dynamically from their tips in this model, as demonstrated in the simulations given in Section \ref{sec.numerics}.

The solutions of the initial boundary value problem are obtained through numerical simulation and presented in Section \ref{sec.numerics}. We begin with mode-I fracture simulations on a high-strength concrete beam under monotonic and cyclic loading. We then simulate mixed-mode fracture tests on concrete beams under cyclic loading. We follow with a mixed-mode fracture simulation for an L-shaped panel. We then examine fast fracture, exhibiting crack branching based on the intensity of the loading and compare the numerical solution  to its phase field counterpart presented in \cite{borden2012phase}. The section concludes by demonstrating the structural size effect for quasi-brittle fracture.

In conclusion, we deliver a well-posed boundary value problem for general degradation envelopes beyond bilinear, trilinear, and exponential, even envelopes that allow strain hardening before failure. In this paper, the material is assumed to be damaged under tensile stress; however, the approach proposed here extends without modification to include compressive damage. Future work will incorporate elastic anisotropy using interactions mediated by state-based peridynamic models, see \cite{silling2007peridynamic}.

The paper is organized as follows: The initial boundary value problem for the fracture evolution of quasi-brittle material is formulated in the next section. Section \ref{wellposedness} establishes existence and uniqueness of the evolution problem using standard fixed-point methods for integral equations posed in Banach space.  Next, we examine load controlled evolutions in Section \ref{energybalanceprocesszone}. We begin by multiplying the evolution equation by the velocity and integrating by parts. Since the strain in every bond is history-dependent, the analysis proceeds using the history-dependent two-point phase field. Because of this, we put the bonds into different groups: undamaged, damaged but not damaging, and damaging. Then for each group, we examine the damage power $\dot{\mathcal{D}}(t)$ and the elastic power $\dot{\mathcal{E}}(t)$ to arrive at explicit formulas and discover that the evolution is consistent with thermodynamics. Next, we integrate the stress power in each bond with respect to time and then integrate in space with respect to the endpoints of the bonds $y$ and $x$. This gives us the forms of the energies $\mathcal{E}(t)$ and $\mathcal{D}(t)$ and the energy balance. In Section \ref{energybalanceprocesszone2}, we outline the same steps for displacement-controlled evolutions and summarize our results for general evolutions.
In Section \ref{sec.Straight Cracks}, it is shown that the model calibration follows directly from the two-point interaction potential and energy balance and given values of elastic moduli, stress at which there is an onset of nonlinearity,  strength, and energy release rate. Section \ref{nondimensionalization} provides the dimension-free groups associated with the dynamics. The results of numerical experiments are presented in Section \ref{sec.numerics}.


\section{A nonlocal phase field formulation}
\label{sec:nonlocal:model}

The body containing the damaging material $\Omega$ is a bounded domain in two or three dimensions.  The horizon  $\epsilon$ is chosen  smaller than the characteristic length $ L$.  Nonlocal interactions between a point $\xx$ and its neighbors $\yy$ are confined to the sphere (disk) of radius $\epsilon$ denoted by $\HH_\epsilon(\xx)=\{\yy:\,|\yy-\xx|<\epsilon\}$. The length scale $\epsilon$ is taken to be significantly smaller than $L$ and small enough to resolve the process zone.  
 Here $V^\epsilon_d=\omega_d\epsilon^d$ is the $d$-dimensional volume of the ball $\HH_\epsilon(\xx)$ centered at $\xx$ where $\omega_d$ is the volume of the unit ball in $d$ dimensions.  The elastic displacement $\uu(t,\xx)$ is defined for $0\leq t\leq T$ and $\xx$ in $\Omega$. We write $\uu(t)=\uu(t,\cdot)$ and introduce the 
two-point strain $S(\yy, \xx, \uu(t))$ between the point $\xx$ and any point $\yy \in \HH_\epsilon(\xx)$  resolved along the direction $\ee$ given by
\begin{align}\label{strain}
    S(\yy, \xx, \uu(t)) = \frac{(\uu(t,\yy)-\uu(t,\xx))}{|\yy-\xx|}\cdot\ee, & \hbox{  where  } \ee=\frac{\yy-\xx}{|\yy-\xx|},
\end{align}	
and introduce the scaled strain
\begin{align}\label{def of r}
r:=r(\yy,\xx,\uu(t))=\sqrt{\frac{|\yy-\xx|}{L}}S(\yy,\xx,\uu(t)).
\end{align}
The strain satisfies the symmetry $S(\yy, \xx, \uu(t)) = S(\xx, \yy, \uu(t)) $.
The scaled nonlocal kernel is introduced and is given by
\begin{equation}\label{scale}
\rho^\epsilon(\yy,\xx)=\frac{{\chi_\Omega(\yy)}J^\epsilon(|\yy-\xx|)}{ \epsilon V^\epsilon_d},
\end{equation}
where {$\chi_\Omega$} is the characteristic function of $\Omega$, $J^\epsilon(|\yy-\xx|)$ is the influence function, a positive function on the ball of radius $\epsilon$ centered at $\xx$ and is radially decreasing taking the value $M$ at the center of the ball  and $0$  for $|\yy-\xx|\geq\epsilon$. The radially symmetric influence function is written $J^\epsilon(|\yy-\xx|)=J(|\yy-\xx|/\epsilon)$. The nonlocal kernel is scaled by $\epsilon^{-1}(V_d^{\epsilon})^{-1}$ enabling the model to be characterized by, 1) the  linear elastic shear $G$ and Lam\'e $\lambda$ moduli  in regions away from the damage,  2) the critical energy release rate ${G}_c$ and, 3) the  strength of the material, this is done in Section \ref{sec.Straight Cracks}.

\subsection{Constitutive laws with memory, Initial Boundary Value problem}
\label{sec.Constiutivelaws BVP}

\begin{figure}
    \centering
    \begin{subfigure}{.45\linewidth}
        \begin{tikzpicture}[xscale=0.75,yscale=0.75]
		    \draw [<->,thick] (0,5) -- (0,0) -- (3.0,0);
			\draw [-,thick] (0,0) -- (-3.5,0);
		
			\draw [-,thick] (0,0) to [out=0,in=-180] (2.5,2.15);
			\draw  [-,thick] (0,0) parabola (-2,4);
			
			\draw (2.5,-0.2) -- (2.5, 0.0);
			\node [above] at (2.5,2.0) {$g(r^F)$};
			\node [below] at (2.5,-0.1) {${r}^F$};

			\draw (1.25,-0.2) -- (1.25, 0.0);
			\node [below] at (1.4,-0.1) {${r}^C$};
			
			\draw (0.8,-0.2) -- (0.8, 0.0);
			\node [below] at (0.8,-0.1) {${r}^L$};

			\node [right] at (3,0) {$r$};
			
			\node [right] at (0,4.20) {$g(r)$};
			
		  \end{tikzpicture}
		  \caption{}
		  \label{ConvexConcavea}
    \end{subfigure}
    \hskip2em
    \begin{subfigure}{.45\linewidth}
        \begin{tikzpicture}[xscale=0.8,yscale=0.8]
		    \draw [<-,thick] (0,3) -- (0,-3);
			\draw [->,thick] (-5,0) -- (3.5,0);
            \draw [-,ultra thick,red] (0,0) -- (3.5,0);
			;
            \draw [dashed,thick] (0,0) to [out=60,in=140] (1.5,1.5); 
            \draw [dashed,thick] (1.5,1.5) to [out=-45,in=180] (3,0.0);
						
			\draw [-,ultra thick] (-1.735,-3.22214) -- (0,0); 
			
			\draw [-,ultra thick,blue] (0,0) -- (1.332,1.653); 
			\draw [-,ultra thick] (0,0) -- (0.7,1.3); 
			\draw [-,ultra thick,blue] (0,0) -- (1.8,1.119); 
			\draw (2.8,-0.2) -- (2.8, 0.0);
			
			\draw (1.2,-0.2) -- (1.2, 0.0);
			\draw (0.75,-0.2) -- (0.75, 0.0);
			\node [below] at (1.4,-0.2) {${r}^C$};
			\node [below] at (0.75,-0.2) {${r}^L$};
			\node [below] at (2.8,-0.2) {${r}^F$};
			
			\node [right] at (3.5,0) {${r}$};
			\node [right] at (0,2.2) {$g'(r)$};
		  \end{tikzpicture}
		   \caption{}
		   \label{ConvexConcaveb}
    \end{subfigure}
    \caption{  (a) The profile $g(r)$ used in characterizing the failure envelope. (b) Bond stiffness on the failure envelope of unloading laws $g'(r)$ given by dashed curve. The bond stiffness becomes nonlinear at $r^L$, exhibits strain hardening between $r^L$ and $r^C$ and goes smoothly to zero at $r^F$. Unloading laws are linear elastic with a softer stiffness controlled by the phase field and shown in blue. The bond offers zero tensile stiffness for bonds broken in tension (red), see Definition \ref{def: Second}. However the bond continues to elastically resist negative strains. The bond stiffness given in \eqref{constpos}, \eqref{constneg}.}
    \label{ConvexConcave}
\end{figure}
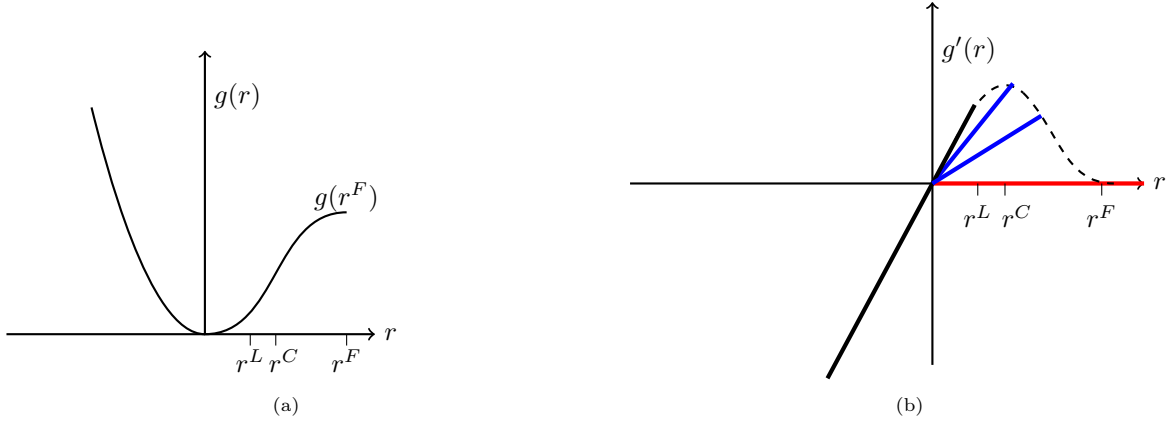

We introduce the envelope associated with the nonlinear (convex-concave) potential function $g(r)$ given by $\partial_r g(r)$, see Figure \ref{ConvexConcaveb}.   The force between two points is a function of the scaled strain. In the blended model, the force is characterized by a  stiffness between two points. Initially, the stiffness remains a constant value with tensile strain but becomes nonlinear when the tensile strain exceeds $S^L$, exhibits strain hardening between $S^L$ and $S^C$ and goes smoothly to zero at $S^F$. On the other hand, the force between two points is given by a constant stiffness for compressive strain. 

\begin{definition}[Bond]
\label{First}
The force associated with the interval between two points $\yy$ and $\xx$ is referred to as a bond.
\end{definition}

The bond stiffness at the strains $S^L$, $S^C$, and $S^F$ depend on the scaled bond length $\sqrt{|\yy-\xx|/L}$ and are given by the parameters $r^L$, $r^C$, and $r^F$ determined by the material and the method of calibration is described in Section \ref{sec.Straight Cracks}. The associated strains are given by:
\begin{align}\label{strains}
S^L=&\frac{r^L}{\sqrt{|\yy-\xx|/L}}\nonumber\\
S^C=&\frac{r^C}{\sqrt{|\yy-\xx|/L}}\nonumber\\
S^F=&\frac{r^F}{\sqrt{|\yy-\xx|/L}}.
\end{align}
The square root dependence on relative bond length is introduced in
the regularized bond model \cite{lipton2014dynamic}, \cite{lipton2016cohesive} and provides the envelope for fast fracture in \cite{lipton2024energy}.

\begin{figure}
    \centering
			

			

        \begin{tikzpicture}[xscale=1.0,yscale=1.0]
		    \draw [<-,thick] (0,3) -- (0,-3);
			\draw [->,ultra thick,red] (2.8,0) -- (3.5,0);
            \draw [-,ultra thick,red] (0,0) -- (2.8,0);
              \draw [-,thick] (-5,0) -- (0,0);
            \usetikzlibrary{decorations.markings}
             \begin{scope}[very thick,decoration={
    markings,
    mark=at position 0.5 with {\arrow{<}}}
    ] 
            \draw [-,ultra thick,red,postaction={decorate}] (0,0) -- (2.8,0);
            \end{scope}
            
            \draw [dashed,ultra thick] (0,0) to [out=60,in=140] (1.5,1.5); 
             \begin{scope}[very thick,decoration={
    markings,
    mark=at position 5pt with {\arrow{>}}, mark=at position 35pt with {\arrow{>}}}
    ] 
            \draw [dashed, ultra thick, postaction={decorate}] (1.5,1.5) to [out=-45,in=180] (3,0.0);
            \end{scope}
             \begin{scope}[very thick,decoration={
    markings,
    mark=at position 30pt with {\arrow{<}}, mark=at position 45pt with {\arrow{>}}}
    ] 
    	      \draw [-,ultra thick, postaction={decorate}] (-1.735,-3.22214) -- (0,0); 
			\draw [-,ultra thick,blue,postaction={decorate}] (0,0) -- (1.332,1.653); 
			\draw [-,ultra thick,blue,postaction={decorate}] (0,0) -- (1.8,1.119); 
            \end{scope}
               \begin{scope}[very thick,decoration={
    markings,
    mark=at position 40pt with {\arrow{>}}, mark=at position 20pt with {\arrow{<}}}
    ] 
            \draw [-,ultra thick,postaction={decorate}] (0,0) -- (0.7,1.3); 
            \end{scope}
			\draw (2.8,-0.2) -- (2.8, 0.0);
			\draw (1.2,-0.2) -- (1.2, 0.0);
			\draw (0.75,-0.2) -- (0.75, 0.0);
			\node [below] at (1.4,-0.2) {${r}^C$};
			\node [below] at (0.75,-0.2) {${r}^L$};
			\node [below] at (2.8,-0.2) {${r}^F$};
			\node [right] at (3.5,0) {${r}$};
			\node [right] at (0,2.75) {force per unit volume$^2$};
		  \end{tikzpicture}
    \caption{ The constitutive law for a bond. The failure envelope is the black dashed curve. The linear unloading is blue. The slope of the linear unloading curve decreases when the bond strain at the present time is equal to its maximum over all earlier times. This is indicated by the arrows on the dashed curve. The force - strain constitutive law for a bond unable resist tension is  red.
    All bonds are linear elastic under compressive strain.}
    \label{Failure envelope}
\end{figure}
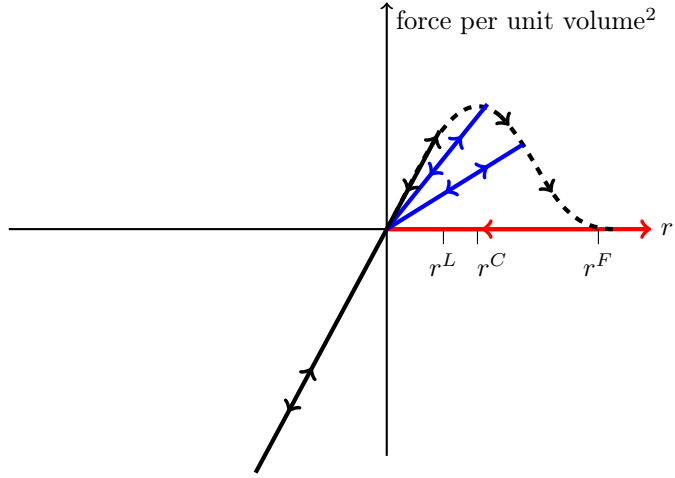

We introduce the memory function given by the maximum strain up to time $t$ given by
\begin{align}
\label{scaledmax}
S^*(t,\yy,\xx,\uu)=\max_{0\leq \tau\leq t}\{ S(\yy, \xx, \uu(\tau))\}.
\end{align}
and set $$r^\ast:=r^\ast(t,\yy,\xx,\uu)=\sqrt{\frac{|\yy-\xx|}{L}}S^*(\yy,\xx,\uu(t)).$$
The failure envelope (see Figure \ref{Failure envelope}) is defined by
\begin{equation}\label{failure envelope}
\rho^\epsilon(\yy,\xx)\frac{\partial_{r^*} g(r^*)}{r^*}S^*(t,\yy,\xx,\uu)=\rho^\epsilon(\yy,\xx)\frac{\partial_{r^*} g(r^*)}{\sqrt{{|\yy-\xx|}/L}}.
\end{equation}
On the failure envelope, the present strain at time $t$ is the maximum strain and  $r^*=r$ 
and the strain-dependent bond stiffness is proportional to the slope of the linear unloading curve and is $\frac{\partial_{r^*} g(r^*)}{r^*}$ and 
\begin{equation}\label{degradezero}
\frac{\partial_r g(r)}{r}=\frac{\partial_{r^*} g(r^*)}{r^*}.
\end{equation}

For positive strain the slope is constant $\partial_r g(r)/r=\overline{\mu}$ in a neighborhood of $r=0$ and represents the undamaged bond it then decreases as the strain increases and is written
$$\frac{\partial_{r^*} g(r^*)}{r^*}=\overline{\mu}\gamma(\uu)(\yy,\xx,t),$$
where  the two-point phase field satisfies $0\leq \gamma(\uu)(\yy,\xx,t)\leq 1$  and is given by
\begin{align}\label{twopointphase}
\gamma(\uu)(\yy,\xx,t)=\frac{(\overline{\mu})^{-1}\partial_{r^*} g(r^*)}{r^*}.
\end{align}
 For positive strains the constitutive relationship is defined by  
 \begin{align}\label{constituitive}
 \hbox{bond force per unit volume$^2$}=\rho^\epsilon(\yy,\xx)\mu(\gamma(\uu)(\yy,\xx,t))S(\yy,\xx,\uu(t))\boldsymbol{e}
 \end{align}
 with
\begin{align}\label{constpos}
\mu(\gamma(\uu)(\yy,\xx,t)):=\overline{\mu}\gamma(\uu)(\yy,\xx,t),
\end{align}
for negative strains bonds do not soften i.e.,
\begin{align}\label{constneg}
\mu(\gamma(\uu)(\yy,\xx,t))=\overline{\mu},
\end{align}
and
\begin{align}\label{constituitivep2}
 \hbox{bond force per unit volume$^2$}=\rho^\epsilon(\yy,\xx)\overline{\mu} S(\yy,\xx,\uu(t))\boldsymbol{e}.
 \end{align}
 \begin{definition}[Broken Bond]
 \label{def: Second}
In this treatment, we say a bond between two points $\yy$ and $\xx$ ``fails in tension'' or ``broken in tension''  if and only if its stiffness is $\mu(\gamma(\uu)(\yy,\xx,t)=0$ for positive strains and $\overline{\mu}$ for negative strains.
\end{definition}
\noindent This notion of broken bond still allows the ``bond'' to resist compressive force, see Figure \ref{ConvexConcaveb}. 

The stiffness $\overline{\mu}$ is calibrated to the undamaged material stiffness in Section \ref{sec.Straight Cracks}.
\begin{figure}
    \centering
 
        \begin{tikzpicture}[xscale=0.8,yscale=0.8]
		    \draw [<-,thick] (0,3) -- (0,-3);
			\draw [->,thick] (-5,0) -- (3.5,0);
            \draw [-,thick] (-5,1.5) -- (1.5,1.5) to [out=-45,in=180] (3,0.0);

            \draw [-,thick] (-5,1.5) -- (1.5,1.5); 
			\draw (0.0,0.2) -- (0.0, -0.2);
			\node [below] at (-0.2,-0.2) {${0}$};
			\draw (2.8,-0.2) -- (2.8, 0.0);
			\draw (1.5,-0.2) -- (1.5, 0.0);
			\draw (2.0,-0.2) -- (2.0, 0.0);
			\node [below] at (1.5,-0.2) {${r}^L$};
			\node [below] at (2.0,-0.2) {${r}^C$};
			\node [below] at (2.8,-0.2) {${r}^F$};
			\node [right] at (3.5,0) {${r^\ast}$};
            \node [right] at (0,2.2) {$\gamma(\uu)(\yy,\xx,t))$};
		  \end{tikzpicture}
    \caption{ $\gamma(\uu)(\yy,\xx,t))$. It is one for $r^\ast\leq r^L$ then decays to zero for $r^L<r^\ast(t,\yy,\xx,\uu)<r^F$.}
    \label{ConvexConcaveC}
\end{figure}
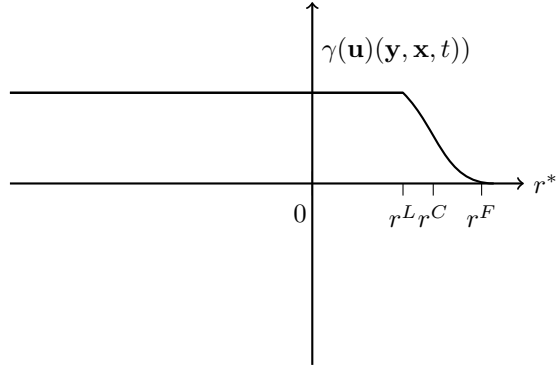
As an example, consider the bilinear model for positive strains. For this case $r^L=r^C$ and 
\begin{equation}\label{bilinear}
\partial_{r^*} g(r^*)=\left\{\begin{array}{lc}\overline{\mu}r^*,& r^*< r^C,\\
\overline{\mu}r^C\frac{r^F-r^*}{r^F-r^C},&r^C\leq r^*\leq r^F,\\
0,&r^F<r^*.\end{array}\right.
\end{equation}
Then the two-point phase field is
\begin{equation}\label{bilinearphasefield}
\gamma(\uu)(\yy,\xx,t)=\left\{\begin{array}{lc} 1,&r^*< r^C,\\
(\frac{r^F}{r^*}-1)\frac{r^C}{(r^F-r^C)}, & r^C\leq r^*\leq r^F,\\
0, &r^F<r^*. \end{array}\right.
\end{equation}
see Figure \ref{ConvexConcaveC}.  As before, the undamaged bond force per unit volume$^2$ is $\rho^\epsilon(\yy,\xx)\overline{\mu}S(\yy,\xx,\uu(t))\ee$ and no bonds damage for negative strains.

The set of bonds corresponding to negative two-point strain is denoted $N^\epsilon(t)$ and is the collection of pairs $(\xx,\yy)$ in $\Omega\times(\Omega\cap \HH_\epsilon(\xx))$ such that the bond between them sustains negative strain is
\begin{align}\label{minus}
N^\epsilon(t)=\{(\xx,\yy)\hbox{ in }\Omega\times(\Omega\cap \HH_\epsilon(\xx)); \, S(\yy,\xx,\uu(t))\leq 0\}
\end{align}
and has with indicator function
\begin{equation}\label{negatave}
\chi_N:=\chi_N(S(\yy,\xx,\uu(t)))=\left\{\begin{array}{ll}1,&\hbox{if  $(\xx,\yy)$ in $N^\epsilon(t)$}\\
0,& \hbox{otherwise}.\end{array}\right.
\end{equation}
Similarly, denote the collection of pairs $(\xx,\yy)$ in $\Omega\times(\Omega\cap \HH_\epsilon(\xx))$ such that the bond between them sustains non-negative strain by
\begin{align}\label{plus}
P^\epsilon(t)=\{(\xx,\yy)\hbox{ in }\Omega\times(\Omega\cap \HH_\epsilon(\xx)); \, S(\yy,\xx,\uu(t))\geq 0\}.
\end{align}
This set has indicator function
\begin{equation}\label{positive}
\chi_P:=\chi_P(S(\yy,\xx,\uu(t)))=\left\{\begin{array}{ll}1,&\hbox{if  $(\xx,\yy)$ in $P^\epsilon(t)$}\\
0,& \hbox{otherwise}.\end{array}\right.
\end{equation}
Collecting definitions, we have $\mu(\gamma(\uu)(\yy,\xx,t))$ defined for all strains and given by
\begin{align}
    \label{const0}
    \mu(\gamma(\uu)(\yy,\xx,t)):=\chi_N\overline{\mu}+\chi_P\mu(\gamma(\uu)(\yy,\xx,t))
\end{align}
In summary, the constitutive law relating force to strain and subsequent bond failure in tension is described by

\begin{equation}
	\boldsymbol{f}^\epsilon(t,\yy,\xx,\uu)=\rho^\epsilon(\yy,\xx)\mu(\gamma(\uu)(\yy,\xx,t))S(\yy,\xx,\uu(t))\boldsymbol{e} ,
	\label{Eqn.const}
\end{equation}
see Figure \ref{Failure envelope}.
The nonlocal force density $\LL^\epsilon[\uu](t,\xx)$ defined for all points $\xx$ in $\Omega$  is given by
\begin{align}
      \label{eq: force}
 \LL^\epsilon [\uu](t,\xx) = -\int_{\Omega} \boldsymbol{f}^\epsilon(t,\yy,\xx,\uu) \,d\yy.
\end{align}
The material is assumed to be homogeneous and the balance of linear momentum for each point $\xx$ in the body $\Omega$  is given by
\begin{align}\label{eq: linearmomentumbal}
\rho\ddot{\uu}(t,\xx)+ \LL^\epsilon [\uu](t,\xx) =\bb(t,\xx),
\end{align}
where  $\bb(t,\xx)$ is a prescribed body force density and $\rho$ is 
is the mass density of the material.  The linear momentum balance is supplemented with the initial conditions on the displacement and velocity given by
\begin{align}
\uu(0,\xx)=\uu_0(\xx),\nonumber\\
\dot{\uu}(0,\xx)=\vv_0(\xx),\label{initialconditions}
\end{align}
and we look for a solution $\uu(t,\xx)$ on a time interval $0<t<T$.
This completes the problem formulation for the load-controlled fracture evolution where the body force $\bb(t,\xx)$ for $0\leq t\leq T$ is prescribed.

In summary, the failure envelope for the Blended model corresponds to $r=r^\ast$ and is given by 
\begin{align}\label{failureenv}
 \hbox{ bond force per unit volume$^2$}=\rho^\epsilon(\yy,\xx)\frac{\partial_r g(r(\yy,\xx,\uu(t)))}{\sqrt{\frac{{|\yy-\xx|}}{L}}}\boldsymbol{e}.
 \end{align}
Collecting observations we have
\begin{equation}\label{bilinear2}
\begin{array}{lc}\partial_t\gamma(\uu)(\yy,\xx,t)=0,& r< r^*,\\
\partial_t\gamma(\uu)(\yy,\xx,t)\leq 0,&r=r^*.\end{array}.
\end{equation}
The degradation of the phase field ensures that the damage energy is non-decreasing; therefore, the process is thermodynamically consistent, see \eqref{damagepower}. It is also consistent with the time dependence of a generic damage state proposed in \cite{sillinglehoucq2010}.

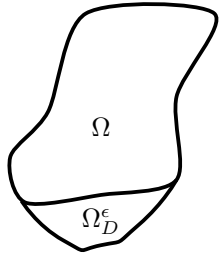
\begin{figure}
    \centering
    
      \begin{tikzpicture}[>=latex]
 \draw[line width=.5mm] plot [smooth cycle] coordinates {(0,0) (1,0.1) (2,0.3) (2,1.4) (2.5,2.5) (0.8,2.5) (0.3,1.2) (-0.2,0.6) } node at (1,1) {$\Omega$};
 \draw[line width=.5mm] plot [smooth,tension=1.2] coordinates {(-0.05,0.02) (0.5,-0.5) (1.0,-0.565) (1.5,-0.295) (2,0.3)} node at (1.0,-0.25) {$\Omega_D^\epsilon$};
\end{tikzpicture}
    \caption{ Domain $\Omega$ with prescribed Dirichlet data on $\Omega_D^\epsilon$. The union is denoted by $\Omega^*$}
    \label{Dirichlet}
\end{figure}

A displacement load can be prescribed on the boundary of the specimen $\Omega$. For the nonlocal model, the displacement $\UU(t,\xx)$ is prescribed on a layer $\Omega_D^\epsilon$ surrounding part or all of the specimen see, e.g., \cite{DuTaoTian}. We write $\Omega^*=\Omega\cup\Omega_D^\epsilon$ where the layer is of maximum thickness greater than or equal to $\epsilon$, see Figure \ref{Dirichlet} . With this in mind, we write
\begin{equation}\label{scale2}
\rho^{\ast,\epsilon}(\yy,\xx)=\frac{{\chi_{\Omega^*}(\yy)}J^\epsilon(|\yy-\xx|)}{ \epsilon V^\epsilon_d},
\end{equation}
and
\begin{equation}
	\boldsymbol{f}^\epsilon(t,\yy,\xx,\uu)=\rho^{\ast,\epsilon}(\yy,\xx)\mu(\gamma(\uu)(\yy,\xx,t))S(\yy,\xx,\uu(t))\boldsymbol{e} ,
	\label{Eqn.const2}
\end{equation}
with the nonlocal force density $\LL^\epsilon[\uu](t,\xx)$ defined for all points $\xx$ in $\Omega$  given by
\begin{align}
      \label{eq: force2}
 \LL^\epsilon [\uu](t,\xx) = -\int_{\Omega^*} \boldsymbol{f}^\epsilon(t,\yy,\xx,\uu) \,d\yy.
\end{align}
For this case, the initial boundary value problem for fracture evolution is to
find a displacement $\uu(t,\xx)$ on $\Omega^*$ that satisfies $\uu(t,\xx)=\UU(t,\xx)$ in $\Omega^\epsilon_D$ and for $\xx$ in
$\Omega$  satisfies the evolution equation 
\begin{align}\label{eq: linearmomentumbal2}
\rho\ddot{\uu}(t,\xx)+ \LL^\epsilon [\uu](t,\xx) =\bb(t,\xx),
\end{align}
for $0<t<T$, with the initial conditions on the displacement and velocity given by
\begin{align}
\uu(0,\xx)=\uu_0(\xx),\nonumber\\
\dot{\uu}(0,\xx)=\vv_0(\xx).\label{initialconditions2}
\end{align}
To illustrate ideas, we fix all bonds crossing the interface between $\Omega$ and $\Omega^\epsilon_D$ to be undamaged and have stiffness $\overline{\mu}$ as well as all bonds between points in $\Omega^\epsilon_D$.

\subsection{Elastic Resistance to Compression Across a Crack}
\label{sec.Straight Cracks Compression}

In this model, cracks are created by bonds that offer zero stiffness under tension. On the other hand, bonds under compressive strains continue to resist elastically. When a bond ceases to resist tension the constitutive relation is illustrated by the red branch of the force-strain relation in Figure \ref{Failure envelope}. Let $\nn$ denote the unit normal vector perpendicular to the crack. Let $\yy-\xx$ denote all line segments that cross a planar crack such that $(\yy-\xx)\cdot\nn>0$ and $\epsilon>|\yy-\xx|$. For deformation fields such that $(\uu(t,\yy)-\uu(t,\xx))\Vert\nn$ and $(\uu(t,\yy)-\uu(t,\xx))\cdot\nn<0$, basic geometric considerations show that  the strain is compressive, ie., $S(\yy,\xx,\uu(t))<0$ for  all bonds $|\yy-\xx|<\epsilon$. The constitutive relation \eqref{constneg} shows that the bond force opposes the compressive strain for all bonds crossing the crack.

\subsection{Damage evolution}
\label{damageevol}
We observe that nonlocal dynamics for $\uu(t)$ over $\Omega$ in $\mathbb{R}^d$ presents as dynamics over $$\{(\yy,\xx)\in\Omega\times\Omega\,;\,|\yy-\xx|<\epsilon\}\subset\mathbb{R}^d\times\mathbb{R}^d$$ for the two-point strain $S(\yy,\xx,\uu(t))$.  In the next section, we show that $\partial_t\gamma(\uu)(\yy,\xx,t)$ is well defined almost everywhere on $\Omega\times\Omega$, i.e., here $\uu$ can have jumps in the $\xx$ variable for $t$ fixed.
The strain dynamics generates sets where force is related to strain and an evolution emerges for the phase field.
The set of undamaged bonds at time $t$ is given by
\begin{align}\label{undamage}
UZ^\epsilon(t) =\{(\yy,\xx)\in\,\Omega\cup\Omega\,; \, |\yy-\xx|<\epsilon,\, \hbox{and}\, \, \gamma(\uu)(\yy,\xx,t))=1\hbox{ and }\partial_t\gamma(\uu)(\yy,\xx,t)=0\},
\end{align}
with indicator function
\begin{equation}\label{undamagechar}
\chi_{UZ^\epsilon(t)}(\yy,\xx)=\left\{\begin{array}{ll}1,&\hbox{if  $(\yy,\xx)$ in $UZ^\epsilon(t)$}\\
0,& \hbox{otherwise}.\end{array}\right.
\end{equation}
The set of pairs $(\yy,\xx)$ corresponding to damaged bonds but not yet irrevocably broken at time $t$ and are not damaging at time $t$ are given by the set
\begin{align}\label{damaged}
DZ^\epsilon(t) &=\{(\yy,\xx)\in\,\Omega\cup\Omega \,; \, |\yy-\xx|<\epsilon,\, \hbox{and}\, \, 1>\gamma(\uu)(\yy,\xx,t))>0\hbox{ and }\partial_t\gamma(\uu)(\yy,\xx,t)=0,\}
\end{align}
with indicator function
\begin{equation}\label{damagedchar}
\chi_{DZ^\epsilon(t)}(\yy,\xx)=\left\{\begin{array}{ll}1,&\hbox{if  $(\yy,\xx)$ in $DZ^\epsilon(t)$}\\
0,& \hbox{otherwise}.\end{array}\right.
\end{equation}
The set of bonds undergoing damage at time $t$ are given by the set
\begin{align}\label{damaging}
PZ^\epsilon(t) &=\{(\yy,\xx)\in\,\Omega\cup\Omega\,; \, |\yy-\xx|<\epsilon,\, \hbox{and}\, \, \partial_t\gamma(\uu)(\yy,\xx,t)<0.\}
\end{align}
with indicator function
\begin{equation}\label{processchar}
\chi_{PZ^\epsilon(t)}(\yy,\xx)=\left\{\begin{array}{ll}1,&\hbox{if  $(\yy,\xx)$ in $PZ^\epsilon(t)$}\\
0,& \hbox{otherwise}.\end{array}\right.
\end{equation}
The failure set $\Gamma^\epsilon(t)$ is the collection of pairs $(\yy,\xx)$ in $\Omega\times\Omega$ such that $|\yy-\xx|<\epsilon$ and the bond between them was broken at some time $\tau$, $0<\tau\leq t$. 
The failure set is written 
\begin{align}\label{damageset}
\Gamma^\epsilon(t)=\{(\yy,\xx)\hbox{ in }\Omega\times\Omega\, ; \, |\yy-\xx|<\epsilon,\, \hbox{and}\, \, \gamma(\uu)(\yy,\xx,t)=0\}
\end{align}
with indicator function
\begin{equation}\label{busted}
\chi_{\Gamma^\epsilon(t)}(\yy,\xx)=\left\{\begin{array}{ll}1,&\hbox{if  $(\yy,\xx)$ in $\Gamma^\epsilon(t)$}\\
0,& \hbox{otherwise}.\end{array}\right.
\end{equation}

Next, we consider the jump set of the displacement $\uu(t)$ along the direction $\ee$ at time $t$ and introduce the relation between jump sets of $\uu(t)$ and $\Gamma^\epsilon(t)$. 
\begin{definition}[Jump set]
\label{jump}
    \begin{align*}
        J_{\ee}^+(\uu(t))=\{\xx\in \Omega\,;\,\lim_{\yy\rightarrow\xx}(\uu(t,\yy)-\uu(t,\xx))\cdot \ee>0, \hbox{ for }(\yy-\xx)\cdot\ee>0,\hbox{ and }(\yy-\xx)\|\ee\}
    \end{align*}
\end{definition}
It now follows that:
\begin{lemma}[Relationship between jump discontinuities and bonds broken in tension]
\label{damageandjump}
If $\lim_{\yy\rightarrow\xx}S(\yy, \xx, \uu(t))$ is positive and the displacement suffers a jump discontinuity at $\xx$, then there exists an  interval $(0,\eta)$  with $\epsilon\geq\eta>0$, for which the bonds $(\xx+s\ee,\xx)$ with $0< s<\eta$ belong to $\Gamma^\epsilon(t)$, i.e., if $\xx\in J_{\ee}^+(\uu(t))$ then $\xx$ is the end point of the family of all broken bonds parallel to $\ee$ of length less than $\eta$.
\end{lemma}
\noindent The lemma follows immediately from the definitions of $J^+_\ee(\uu(t))$ and $\Gamma^\epsilon(t)$. \\

The set of undamaged bonds at time $t$ is given by
\begin{align*}
UZ^\epsilon(t) =\{(\xx,\yy)\in\,\Omega\cup\Omega\cap \HH_\epsilon(\xx)\hbox{ such that }\gamma(\uu)(\yy,\xx,t))=1\},
\end{align*}
and the set of pairs $(\yy,\xx)$ corresponding to damaged and unloaded but not yet irrevocably broken at time $t$ is given by the set
\begin{align}\label{damages}
DZ^\epsilon(t) &=\{(\xx,\yy)\in\,\Omega\cup\Omega\cap \HH_\epsilon(\xx)\hbox{ such that }1>\gamma(\uu)(\yy,\xx,t))\geq 0, \hbox{ with } \partial_t\gamma(\uu)(\yy,\xx,t))=0.\}
\end{align}
The set of pairs $(\yy,\xx)$ currently in the process of damaging at time $t$ is the process zone and given by the set
\begin{align*}
PZ^\epsilon(t) &=\{(\xx,\yy)\in\,\Omega\cup\Omega\cap \HH_\epsilon(\xx)\hbox{ with }-\partial_t\gamma(\uu)(\yy,\xx,t))>0.\}
\end{align*}
We show in Section \ref{wellposedness} that this nonlinear nonlocal initial boundary problem can be solved and has a unique displacement-damage set pair.


\section{Existence and uniqueness of displacement and damage evolution.}
\label{wellposedness}

In this section, we state and prove the existence of solution for the force-controlled fracture evolution with homogeneous nonlocal boundary traction. Here, the nonlocal homogeneous boundary traction is enforced by the kernel \eqref{scale}. We then state the existence of solution for the general initial boundary value problem noting its proof follows methods identical to the force controlled problem. Body forces are easily chosen for which one can find a unique solution $\uu(t,\xx)$ of the initial value problem in \eqref{eq: linearmomentumbal} and \eqref{initialconditions}. We introduce the subspace $\mathcal{U}$ of rigid body motions defined by \eqref{rigidmotion}. 
\begin{equation}\label{rigidmotion}
\mathcal{U}=\{\ww:\,\ww=\mathbb{Q}\xx+\mathbf{c};\, \mathbb{Q} \in \mathbb{R}^{d\times d},\,\,\mathbb{Q}^T=-\mathbb{Q};\,\,\mathbf{c}\in\mathbb{R}^d\}.
\end{equation}
The calculation shows $S(\yy,\xx,\ww)=0$ for $\ww\in\mathcal{U}$, which is the null space of the strain operator. Solvability in the presence of homogeneous traction requires $\int_\Omega\ww\cdot\bb(t,\xx)\,d\xx=0$ for all $\ww\in\mathcal{U}$ and $0\leq t \leq T$. We denote the subspace of $L^\infty(\Omega;\mathbb{R}^d)$ orthogonal to $\mathcal{U}$ as $\dot L^\infty(\Omega;\mathbb{R}^d)$ and to expedite presentation, we denote $\dot L^\infty(\Omega;\mathbb{R}^d)$ by $X$. The norm on $C^2([0,T];X)$ is given by \eqref{norms}. 
\begin{align}\label{norms}
\Vert \uu(t,\xx)\Vert_{C^2([0,T]; X)}=\sup_{0\leq t\leq T}\{ \sum_{i=0}^2||\partial_t^i\uu(t,\xx)||_{X}\}.
\end{align}
To simplify the notation, we denote the space $C([0,T];X)$ as $CX$ with the corresponding norm $\Vert\cdot \Vert_{CX}$. Similarly, we write $C^2([0,T];X)$ as $C^2X$. The existence and uniqueness of the solution is stated below.

\begin{theorem}[{\bf Existence and Uniqueness of Solution of Force-Controlled Fracture Evolution}]
\label{exist}
 The initial value problem given by \eqref{eq: linearmomentumbal} and \eqref{initialconditions} with initial data in $X$ and $\bb(t,\xx)$ belonging to $CX$
has a unique solution $\uu(t,\xx)$ in $C^2X$ with two strong derivatives in time.
 \end{theorem}

The initial value problem delivers a unique displacement-damage set pair: ${\uu^\epsilon(t), DZ^\epsilon(t)}$. 
The key points are that the operator $\LL^\epsilon\uu$ is a map from $CX$ into itself, and it is Lipschitz continuous in $\uu$ with respect to the $CX$ norm. The theorem then follows from the Banach fixed-point theorem. To establish these properties, we summarize the differentiability and Lipschitz continuity of the damage factor.



\begin{lemma}
    For $\uu\in CX$, the mapping $(\yy,\xx)\mapsto \gamma(\uu)(\yy,\xx,t):\Omega\times \Omega\rightarrow\mathbb{R}$ is measurable for every $t\in[0,T]$ and the mapping $t\mapsto \gamma(\uu)(\yy,\xx,t):[0,T]\rightarrow\mathbb{R}$ is continuous for almost all $(\yy,\xx)$. Moreover, for almost all $(\yy,\xx)\in (\Omega\times \Omega\cap H_\epsilon(\xx))$ and all $t\in [0,T]$, the map $\uu\mapsto \gamma(\uu)(\yy,\xx,t):CX\rightarrow\mathbb{R}$ is Lipschitz continuous with:
\begin{align}\label{properties 4}
| \gamma(\uu)(\yy,\xx,t)-\gamma(\ww)(\yy,\xx,t)|\leq \Vert \uu-\ww\Vert_{CX} \frac{C}{{|\yy-\xx|}},
\end{align}
where $C$ is a constant independent of vector fields $\uu$ and $\ww$ in $CX$.
\end{lemma}

\noindent The claims in the lemma are straightforward. Equation \eqref{properties 4} follow as in \cite{lipton2024energy}, along with the fact that 
 $$|S^*(\yy,\xx,\uu,t) - S^*(\yy,\xx,\ww,t)| \leq \max_{0\leq\tau \leq t}|S(\yy,\xx,\uu(\tau)-\ww(\tau))|.$$


\begin{lemma}[{\bf Lipschitz continuity of \( \mathcal{L}^\epsilon [\cdot]\)}]
\label{linftcnttheta} 
Given two functions $\uu$ and $\ww$  in $CX$ and a constant $C$ independent of $\uu$ and $\ww$ such that
\begin{align}\label{lipshitzg}
|\mathcal{L}^\epsilon[\mathbf{u}](t, \mathbf{x}) - \mathcal{L}^\epsilon[\mathbf{w}](t, \mathbf{x})| \leq C \|\mathbf{u} - \mathbf{w}\|_{CX} \int_\Omega \frac{\rho^\epsilon(\mathbf{y}, \mathbf{x})}{|\mathbf{y} - \mathbf{x}|} \, d\mathbf{y},
\end{align}
 \end{lemma}
 \noindent{\em Proof.}\,\,
 To establish the Lipschitz continuity of \( \mathcal{L}^\epsilon \), we bound \( |\mathcal{L}^\epsilon[\mathbf{u}](t, \mathbf{x}) - \mathcal{L}^\epsilon[\mathbf{w}](t, \mathbf{x})| \) by decomposing the integrand and applying intermediate estimates. Starting from
\begin{align*}
|\mathcal{L}^\epsilon[\mathbf{u}](t, \mathbf{x}) - \mathcal{L}^\epsilon[\mathbf{w}](t, \mathbf{x})| = \left| \int_\Omega \left( \boldsymbol{f}^\epsilon(t, \mathbf{y}, \mathbf{x}, \mathbf{u}) - \boldsymbol{f}^\epsilon(t, \mathbf{y}, \mathbf{x}, \mathbf{w}) \right) d\mathbf{y} \right|,
\end{align*}
we split \( \boldsymbol{f}^\epsilon \) using \eqref{Eqn.const}
\begin{align}\label{const1}
\boldsymbol{f}_P^\epsilon(t,\yy,\xx,\uu)&= {\rho^\epsilon(\yy,\xx)}\chi_P(S(\yy,\xx,\uu(t)))\mu(\gamma(\uu)(\yy,\xx,t))S(\yy,\xx,\uu(t))\boldsymbol{e}\nonumber\\
\boldsymbol{f}_N^\epsilon(t,\yy,\xx,\uu)&={\rho^\epsilon(\yy,\xx)}\chi_N(S(\yy,\xx,\uu(t)))\overline{\mu}S(\yy,\xx,\uu(t))\boldsymbol{e},
\end{align}
and
\begin{align*}
\boldsymbol{f}^\epsilon(t, \mathbf{y}, \mathbf{x}, \mathbf{u}) = \boldsymbol{f}^\epsilon_P(t, \mathbf{y}, \mathbf{x}, \mathbf{u}) + \boldsymbol{f}^\epsilon_N(t, \mathbf{y}, \mathbf{x}, \mathbf{u}).
\end{align*}
Thus, we write:
\begin{align*}
|\mathcal{L}^\epsilon[\mathbf{u}](t, \mathbf{x}) - \mathcal{L}^\epsilon[\mathbf{w}](t, \mathbf{x})| \leq |I| + |II| +|III|,
\end{align*}
where
\begin{align*}
I = \int_\Omega \left( \boldsymbol{f}^\epsilon_P(t, \mathbf{y}, \mathbf{x}, \mathbf{u}) - \boldsymbol{f}^\epsilon_P(t, \mathbf{y}, \mathbf{x}, \mathbf{w}) \right) d\mathbf{y},
\end{align*}
and
\begin{align*}
II = \int_\Omega \left( \boldsymbol{f}^\epsilon_N(t, \mathbf{y}, \mathbf{x}, \mathbf{u}) - \boldsymbol{f}^\epsilon_N(t, \mathbf{y}, \mathbf{x}, \mathbf{w}) \right) d\mathbf{y},
\end{align*}
and
\begin{align*}
III = \int_\Omega \left( \boldsymbol{f}^\epsilon_P(t, \mathbf{y}, \mathbf{x}, \mathbf{u}) - \boldsymbol{f}^\epsilon_N(t, \mathbf{y}, \mathbf{x}, \mathbf{w}) \right) d\mathbf{y},
\end{align*}
and
\begin{align*}
IV = \int_\Omega \left( \boldsymbol{f}^\epsilon_N(t, \mathbf{y}, \mathbf{x}, \mathbf{u}) - \boldsymbol{f}^\epsilon_P(t, \mathbf{y}, \mathbf{x}, \mathbf{w}) \right) d\mathbf{y}.
\end{align*}
We can then decompose \( I \) into two integrals, \( I_1 \) and \( I_2 \), as follows:
\begin{align*}
I_1 = \int_\Omega \rho^\epsilon(\mathbf{y}, \mathbf{x}) \mu(\gamma(\mathbf{u})(\mathbf{y}, \mathbf{x}, t)) \left( S(\mathbf{y}, \mathbf{x}, \mathbf{u}(t)) - S(\mathbf{y}, \mathbf{x}, \mathbf{w}(t)) \right) \mathbf{e} \, d\mathbf{y},
\end{align*}
\begin{align*}
I_2 = \int_\Omega  \rho^\epsilon(\mathbf{y}, \mathbf{x})  (\mu(\gamma(\mathbf{u})(\mathbf{y}, \mathbf{x}, t)) -\mu(\gamma(\mathbf{w})(\mathbf{y}, \mathbf{x}, t)))  S(\mathbf{y}, \mathbf{x}, \mathbf{w}(t))  \mathbf{e} \, d\mathbf{y}.
\end{align*}
Using the Lipschitz continuity of \( S \), we find $C_1$ independent of $\uu$ and $\ww$ in $CX$ such that

\begin{align*}
I_1 \leq C_1 \|\mathbf{u} - \mathbf{w}\|_{CX} \int_\Omega \frac{\rho^\epsilon(\mathbf{y}, \mathbf{x})}{|\mathbf{y} - \mathbf{x}|} \, d\mathbf{y}.
\end{align*}
For \( I_2 \), using the Lipschitz continuity of \( \mu \), we get a constant $C_2$ independent of $\uu$ and $\ww$ in $CX$ such that

\begin{align*}
I_2 \leq C_2 \overline{\mu} \|\mathbf{u} - \mathbf{w}\|_{CX} \int_\Omega \frac{\rho^\epsilon(\mathbf{y}, \mathbf{x})}{|\mathbf{y} - \mathbf{x}|} \, d\mathbf{y}.
\end{align*}
Similarly, we write
\begin{align*}
II = \int_\Omega \frac{ \rho^\epsilon(\mathbf{y}, \mathbf{x})}{{|\mathbf{y} - \mathbf{x}|}}  \bar{\mu}\left( S(\mathbf{y}, \mathbf{x}, \mathbf{u}(t)) -  S(\mathbf{y}, \mathbf{x}, \mathbf{w}(t)) \right) \mathbf{e} \, d\mathbf{y}.
\end{align*}
Using the Lipschitz continuity of \( S \), we obtain
\begin{align*}
II \leq C \|\mathbf{u} - \mathbf{w}\|_{CX} \int_\Omega \frac{\rho^\epsilon(\mathbf{y}, \mathbf{x})}{|\mathbf{y} - \mathbf{x}|} \, d\mathbf{y}.
\end{align*}
For the case $III$, one can see that based on the nature of the model,
\begin{align*}
    |\boldsymbol{f}^\epsilon_P(t, \mathbf{y}, \mathbf{x}, \mathbf{u}) - \boldsymbol{f}^\epsilon_N(t, \mathbf{y}, \mathbf{x}, \mathbf{w})|&\leq C_1\Bar{\mu} |S(u)-S(w)|\int_\Omega \frac{\rho^\epsilon(\mathbf{y}, \mathbf{x})}{|\mathbf{y} - \mathbf{x}|} \, d\mathbf{y},\\
    &\leq C_2\Bar{\mu}  \|\mathbf{u} - \mathbf{w}\|_{CX}\int_\Omega \frac{\rho^\epsilon(\mathbf{y}, \mathbf{x})}{|\mathbf{y} - \mathbf{x}|} \, d\mathbf{y},
\end{align*}
and case $IV$ is handled similary.
Combining the bounds for  \( I \) through \( IV\), 
we conclude
\begin{align}\label{Lipschitz}
|\mathcal{L}^\epsilon[\mathbf{u}](t, \mathbf{x}) - \mathcal{L}^\epsilon[\mathbf{w}](t, \mathbf{x})| \leq C \|\mathbf{u} - \mathbf{w}\|_{CX} \int_\Omega \frac{\rho^\epsilon(\mathbf{y}, \mathbf{x})}{|\mathbf{y} - \mathbf{x}|} \, d\mathbf{y}.
\end{align}
Note that the last integral in \eqref{Lipschitz} is bounded in all dimensions and the Lipschitz continuity of \( \mathcal{L}^\epsilon \) with respect to \( \mathbf{u} \) and \( \mathbf{w} \) is established.

To see that the operator $\LL^\epsilon\uu$ is a map from $CX$ into itself,
first set $\ww=0$ in \eqref{Lipschitz} to get $\Vert\mathcal{L}^\epsilon[\mathbf{u}]\Vert_{CX}\leq L\Vert \uu\Vert_{CX}$. 
The time continuity of $\mathcal{L}^\epsilon[\mathbf{u}](t)$ can be seen from Lipschitz continuity with respect to the $\Vert\cdot\Vert_X$ norm given by
\begin{align}\label{Lipschitz2}
|\mathcal{L}^\epsilon[\mathbf{u}](t, \mathbf{x}) - \mathcal{L}^\epsilon[\mathbf{w}](t, \mathbf{x})| \leq C \|\mathbf{u} - \mathbf{w}\|_{X},
\end{align}
then choosing $\ww=\uu(t')$ $t\not=t'$ and noting $\uu$ is in $CX$.


\noindent{\em Proof.} (Proof of Theorem \ref{exist})\,\,
We can now show that the solution \( \uu \) is the unique fixed-point of the mapping \( \uu(t) = (I \uu)(t) \), where \( I \) is a mapping from \( CX \) to itself, defined by
\begin{equation}\label{Fixedpointmap}
\begin{aligned}
(I\uu)(t) = \uu_0 + t\vv_0 + \rho^{-1} \int_0^t (t - \tau) \, \mathcal{L}^\epsilon[\uu](\tau) + \bb(\tau) \, d\tau.
\end{aligned}
\end{equation}
This formulation is equivalent to finding the unique solution to the initial value problem given by equations \eqref{eq: linearmomentumbal} and \eqref{initialconditions}. By incorporating the factor \( \rho^{-1} \) into \( \mathcal{L}^\epsilon[\uu] + \bb \) and denoting the Lipchitz constant for $\rho^{-1}\mathcal{L}^\epsilon$ by $L$, we proceed to show that \( I \) is a contraction mapping. To demonstrate that \( I \) is a contraction, we introduce an equivalent norm:
\begin{equation}\label{Equivnorm}
\begin{aligned}
|||\uu|||_{CX} = \max_{t \in [0, T_0]} \left\{ e^{-2LTt} \Vert \uu(t) \Vert_{X} \right\}.
\end{aligned}
\end{equation}
For \( t \in [0, T] \), we have
\begin{equation}\label{Contrction}
\begin{aligned}
\Vert (I\uu)(t) - (I\ww)(t) \Vert_{X} &\leq \int_0^t (t - \tau) \Vert \mathcal{L}[\uu](\tau) - \mathcal{L}[\ww](\tau) \Vert_{X} \, d\tau \\
&\leq LT \int_0^t \Vert \uu - \ww \Vert_{C([0, \tau]; X)} \, d\tau \\
&\leq LT \int_0^t \max_{s \in [0, \tau]} \left\{ \Vert \uu(s) - \ww(s) \Vert_{X} e^{-2LTs} \right\} e^{2LT\tau} \, d\tau \\
&\leq \frac{e^{2LTt} - 1}{2} |||\uu - \ww|||_{CX},
\end{aligned}
\end{equation}
which leads to
\begin{equation}\label{ContrctionD}
\begin{aligned}
|||(I\uu)(t) - (I\ww)(t)|||_{CX} \leq \frac{1}{2} |||\uu - \ww|||_{CX}.
\end{aligned}
\end{equation}
Thus, \( I \) is a contraction. By the Banach Fixed-Point Theorem, there exists a unique fixed-point \( \uu(t) \) in \( CX \), and from \eqref{Fixedpointmap}, it follows that \( \uu(t) \) also belongs to \( C^2X \).

We conclude the section by stating a similar existence theorem for the initial boundary value problem for fracture evolution. To do this, we introduce the Banach spaces $X=L^\infty(\Omega,\mathbb{R}^d)$, $X^*=L^\infty(\Omega^*,\mathbb{R}^d)$, and the Banach space $C^1([0,T];X^*)$. The imposed boundary displacement $\UU(t,\xx)$ belongs to $C^1([0,T];X^*)$ and is zero for $\xx$ in $\Omega$ and is a prescribed nonzero function $\UU(t,\xx)$ on $\Omega_D^\epsilon$. Here we choose $\UU(t,\xx)$ not to be a rigid rotation or translation. So, the solvability constraint on the force $\bb(t,\xx)$ is no longer needed, as rigid rotations are ruled out.

\begin{theorem}[{\bf Existence and Uniqueness of Solution of the Displacement and Force Controlled Fracture Evolution}]
\label{exist2}
 The initial boundary value problem  given by \eqref{eq: linearmomentumbal2} and \eqref{initialconditions2} with initial data in $X^*$, $\bb(t,\xx)$ belonging to $CX$, imposed boundary displacement $\UU(t,\xx)$ in $C^1X^*$,
has a unique solution $\uu(t,\xx)$ belonging to $C^1X^*$ and $C^2X$ with two strong derivatives in time.
 \end{theorem}

\section{Energy balance for force-controlled fracture evolutions}
\label{energybalanceprocesszone}

In this section, we establish the power and energy balance for force-controlled fracture evolutions. Multiplying the linear momentum equation given in \eqref{eq: linearmomentumbal} with $\dot{\uu}(\xx, t)$ and integrating over the domain results in
\begin{equation}
	\int_{\Omega} \rho \ddot{\uu}(\xx, t) \cdot \dot{\uu}(\xx, t) \,d\xx + 
	\int_{\Omega} \LL^\epsilon [\uu](t,\xx) \cdot \dot{\uu}(\xx, t) \,d\xx = 
	\int_{\Omega} \bb(t,\xx) \cdot \dot{\uu}(\xx, t) \,d\xx.
 \label{eq: linearmomentumbalInt}
\end{equation}

Integrating by parts in the second term of \eqref{eq: linearmomentumbalInt} gives a product of force and strain rate,  and we obtain
\begin{align}\label{eq:prebalance}
&\partial_t\left(\int_\Omega\, \rho\,\frac{|\dot{\uu}(t)|^2}{2}\,d\xx \right)\nonumber\\
&+\frac{1}{2}\int_\Omega\int_\Omega\,|\yy-\xx|\rho^\epsilon(\yy,\xx)\mu(\gamma(\uu)(\yy,\xx,t)) S(\yy,\xx,\uu(t)) S(\yy,\xx,\dot\uu(t))\,d\yy\,d\xx\nonumber\\
& =\int_\Omega\,\bb(t)\cdot\dot\uu(t)\,d\xx.
\end{align}

To obtain power balance, we partition the domain  as
$$\{(\yy,\xx)\in\Omega\times\Omega\,;\,|\yy-\xx|<\epsilon\}=UZ^\epsilon(t)\cup DZ^\epsilon(t)\cup PZ^\epsilon(t). $$
Now define
\begin{align}\label{const3}
F'_{s}(S(\yy,\xx,\uu(t)))&:=\frac{1}{2}\rho^\epsilon(\yy,\xx)|\yy-\xx|\chi_P(\yy,\xx,t)\,\mu(s)\,S(\yy,\xx,\uu(t))\nonumber\\
&+\frac{1}{2}\rho^\epsilon(\yy,\xx)|\yy-\xx|\chi_N(\yy,\xx,t)\,\overline{\mu}\,S(\yy,\xx,\uu(t)),
\end{align}
where $s$ is viewed as a parameter and
\begin{align*}
\int_0^t\,F'_{s}(S(\yy,\xx,\uu(\tau)))S(\yy,\xx,\dot\uu(\tau))\,d\tau=F_{s}(S(\yy,\xx,\uu(t)))-F_{s}(S(\yy,\xx,\uu(0))).
\end{align*}

The kinetic energy is
\begin{align}
\label{kenitic}
    \mathcal{K}(t)=\int_\Omega\frac{\rho|\dot\uu(t)|^2}{2}\,d\xx,
\end{align}
and equation \eqref{eq:prebalance} can be written in the form
\begin{align}\label{eq:abbbalance}
&\dot{\mathcal{K}}(t)+\int_\Omega\int_\Omega\,F'_{\gamma(\uu)(\yy,\xx,t)}(S(\yy,\xx,\uu(t)))S(\yy,\xx,\dot\uu(t))\,d\yy\,d\xx\nonumber\\
& =\int_\Omega\,\bb(t)\cdot\dot\uu(t)\,d\xx.
\end{align}
Observe that for all $(\yy,\xx)$ bonds in $UZ^\epsilon(t)$  have $\gamma(\uu)(\yy,\xx,t)=1$ so  $\mu(\gamma(\uu)(\yy,\xx,t))=\overline{\mu}$ and
\begin{align}
\label{identiy1}
\overline{\mu} S(\yy,\xx,\uu(t)) S(\yy,\xx,\dot\uu(t))=\partial_t\left(\overline{\mu} \frac{S^2(\yy,\xx,\uu(t))}{2}\right) 
\end{align}
for positive strain and negative strain. So, for bonds in $UZ^\epsilon(t)$, the stress power is the elastic power given by
\begin{align}\label{stresspoewrundamaged}
 F'_1(S(\yy,\xx,\uu(t))S(\yy,\xx,\dot\uu(t)).
\end{align}
Observe for all $(\yy,\xx)$ bonds in $DZ^\epsilon(t)$  have $\partial_t\gamma(\uu)(\yy,\xx,t)=0$ so $\partial_t\mu(\gamma(\uu)(\yy,\xx,t))=0$ and
\begin{align}
\label{identiy2}
\mu(\gamma(\uu)(\yy,\xx,t)) S(\yy,\xx,\uu(t)) S(\yy,\xx,\dot\uu(t))=\partial_t\left(\mu(\gamma(\uu)(\yy,\xx,t)) \frac{S^2(\yy,\xx,\uu(t))}{2}\right) 
\end{align}
for positive strain and  negative strain.
So for bonds in $DZ^\epsilon(t)$ the stress power is the elastic power given by
\begin{align}\label{stresspowerdamaged}
F'_{\gamma(\uu)(\yy,\xx,t)}(S(\yy,\xx,\uu(t))S(\yy,\xx,\dot\uu(t)).
\end{align}
With this in mind we find that the total elastic power $\dot{\mathcal{E}}_1^\epsilon(t)$ in $UZ^\epsilon\cup DZ^\epsilon$  is
\begin{align}\label{elasticpower}
\dot{\mathcal{E}}_1^\epsilon(t)&=\int_{UZ^\epsilon(t)\cup DZ^\epsilon(t)}\, F'_{\gamma(\uu)(\yy,\xx,t)}(S(\yy,\xx,\uu(t))(S(\yy,\xx,\dot\uu(t))\,d\yy\,d\xx.
\end{align}
When  $(\yy,\xx)$ belongs to $PZ^\epsilon(t)$, the stress power of the bond connecting $\yy$ to $\xx$ is a combination of elastic power and damage power. Here $S^*(\yy,\xx,\uu,t)=S(\yy,\xx,\uu(t))$ and
\begin{align}
\label{identiy3}
\mu(\gamma(\uu)(\yy,\xx,t)) S^\ast(\yy,\xx,\uu(t)) S^\ast(\yy,\xx,\dot\uu(t))&=\partial_t\left(\mu(\gamma(\uu)(\yy,\xx,t)) \frac{S^2(\yy,\xx,\uu(t))}{2}\right)\nonumber\\
&-\partial_t\mu(\gamma(\uu)(\yy,\xx,t))\frac{S^2(\yy,\xx,\uu(t))}{2}.
\end{align}
Hence, the total stress power $\dot{\mathcal{S}}^\epsilon(t)$ in $PZ^\epsilon(t)$ is a combination of elastic and damage power given by
\begin{align}\label{totaldamagepower}
\dot{\mathcal{S}}^\epsilon(t)=\dot{\mathcal{E}}_2^\epsilon(t)+\dot{\mathcal{D}}^\epsilon(t),
\end{align}
where the elastic power is
\begin{align}\label{elasticpzpower2}
\dot{\mathcal{E}}_2^\epsilon(t)=\frac{1}{2}\int_{PZ^\epsilon(t)}\,|\yy-\xx|\rho^\epsilon(\yy,\xx)\partial_t\left(\mu(\gamma(\uu)(\yy,\xx,t)) \frac{S^2(\yy,\xx,\uu(t))}{2}\right) \,d\yy\,d\xx.
\end{align}
and the damage power (energy dissipation rate) is
\begin{align}\label{damagepower}
\dot{\mathcal{D}}^\epsilon(t)=-\frac{1}{2}\int_{PZ^\epsilon(t)}\,|\yy-\xx|\rho^\epsilon(\yy,\xx)\partial_t\mu(\gamma(\uu)(\yy,\xx,t))\frac{S^2(\yy,\xx,\uu(t))}{2} \,d\yy\,d\xx,
\end{align}
where $\dot{\mathcal{D}}^\epsilon>0$ since $\partial_t\gamma(\uu)(\yy,\xx,t)<0$. Importantly, note that for bonds in $PZ^\epsilon(t)$ that $\mu(\gamma(\uu)(\yy,\xx,t))$ is decreasing with $t$.
On setting  $\dot{\mathcal{E}}={\mathcal{E}}_1^\epsilon+{\mathcal{E}}_2^\epsilon$
and collecting results, we have the power balance

\noindent{\bf Power Balance for load controlled fracture evolution}
\begin{align}\label{eq:rateenergybalance}
\dot{\mathcal{K}}(t) + \dot{\mathcal{E}}^\epsilon(t) + \dot{\mathcal{D}}^\epsilon(t)=\int_\Omega\,\bb(t)\cdot\dot\uu(t)\,d\xx.
\end{align}

It is clear that the damage energy only changes when  $\partial_t\gamma(\uu)(\yy,\xx,t)<0$ and is determined by the evolving displacement field, through the change in elastic energy, kinetic energy and work done against the load.  Rearranging terms these observations are summarized in the following Lemma.
\begin{lemma}[{\bf Growth of the Damage Energy and Process Zone}]
\label{damagezonegrowth}
We have the power balance:
\begin{align}\label{eq:processzonepowerbalance}
\dot{\mathcal{D}}^\epsilon(t) =\int_\Omega\,\bb(t)\cdot\dot\uu(t)\,d\xx-\dot{\mathcal{K}}(t) - \dot{\mathcal{E}}^\epsilon(t)  \geq 0.
\end{align}
\end{lemma}
It is observed that conditions for which damage nucleates and propagates follows directly from \eqref{eq:processzonepowerbalance} and is given below.
\begin{remark}[{\bf  Condition for Damage}]
\label{damagecondition}
{\em If the rate of energy put into the system exceeds the material's capacity to generate kinetic and elastic energy through  displacement and velocity, then damage occurs.}
\end{remark}
Note here that \eqref{eq:processzonepowerbalance} is  consistent with thermodynamics and this formulation follows directly  from the equations of motion \eqref{eq: linearmomentumbal}.
In summary, power balance  shows that $\dot{\mathcal{D}}^\epsilon(t)>0$ only on  $PZ^\epsilon(t)$ and zero elsewhere. If a crack exists, the strain is greatest in the neighborhood of its tips and the location of the process zone is at the crack tips.

To obtain an  explicit formula for the elastic energy and damage energy, one exchanges time and space integrals in $\int_0^t\dot{\mathcal{E}}(\tau)+\dot{\mathcal{D}}(\tau)\,d\tau$. 
To further expedite the time integration, we partition the domain of spatial integration  as
$$\{(\yy,\xx)\in\Omega\times\Omega\,;\,|\yy-\xx|<\epsilon\}=UZ^\epsilon(t)\cup DZ^\epsilon(t)\cup PZ^\epsilon(t). $$

For bonds in $UZ^\epsilon(t)$ the stress work is given by
\begin{align}\label{stressworkundamaged}
w_{UD}(t,\yy,\xx)&=\int_0^t\chi_{UZ^\epsilon(t)}(\yy,\xx)\,F'_1(S(\yy,\xx,\uu(\tau))S(\yy,\xx,\dot\uu(\tau))\,d\tau\nonumber\\
&=\chi_{UZ^\epsilon(t)}(\yy,\xx)\,\left(F_1(S(\yy,\xx,\uu(t)))-F_1(S(\yy,\xx,\uu(0)))\right).
\end{align}
Next, for  bonds in $DZ^\epsilon(t)$ we define $t^L_{\yy,\xx}$  to be the instant when $S(\yy,\xx,\uu(t^L_{\yy,\xx}))=S^L$. For bonds in $DZ^\epsilon(t)$ we also define  $t^U_{\yy,\xx}\leq t$ to be the most recent time of bond unloading $S(\yy,\xx,\dot\uu(t^U_{\yy,\xx} ))\leq 0$  with $S(\yy,\xx,\uu(t^U_{\yy,\xx})) \geq S^C$.
 For bonds in $DZ^\epsilon(t)$,  the stress work is given by
\begin{align}\label{stressworkdamaged}
w_{DZ}(t,\yy,\xx)&=\int_0^{t^L_{\yy,\xx}}\chi_{DZ^\epsilon(t)}(\yy,\xx)F'_1(S(\yy,\xx,\uu(\tau))S(\yy,\xx,\dot\uu(\tau))d\tau\nonumber\\
&+\frac{1}{2}\int_{t^L_{\yy,\xx}}^{t^U_{\yy,\xx}}\chi_{DZ^\epsilon(t)}(\yy,\xx)|\yy-\xx| \rho^\epsilon(\yy,\xx)\mu(\gamma(\uu)(\yy,\xx,\tau)) S^\ast(\yy,\xx,\uu(\tau)) S^\ast(\yy,\xx,\dot\uu(\tau) )d\tau\nonumber\\
&+\int_{t^U_{\yy,\xx}}^t\chi_{DZ^\epsilon(t)}(\yy,\xx) F'_{\gamma(\uu)(\yy,\xx,t^U_{\yy,\xx})}(S(\yy,\xx,\uu(\tau))S(\yy,\xx,\dot\uu(\tau)) d\tau,
\end{align}
The middle term in \eqref{stressworkdamaged} is given by the non-negative quantity $w_{IDZ}(t,\yy,\xx)$ defined as

\begin{align}\label{product2}
w_{IDZ}(t,\yy,\xx)&=\frac{1}{2}\int_{t^L_{\yy,\xx}}^{t^U_{\yy,\xx}}|\yy-\xx| \rho^\epsilon(\yy,\xx)\mu(\gamma(\uu)(\yy,\xx,\tau)) S^\ast(\yy,\xx,\uu(\tau)) S^\ast(\yy,\xx,\dot\uu(\tau) )d\tau\nonumber\\
&=\frac{1}{2}\int_{t^L_{\yy,\xx}}^{t^U_{\yy,\xx}} |\yy-\xx| \rho^\epsilon(\yy,\xx)\partial_{\tau}\left(\mu(\gamma(\uu)(\yy,\xx,\tau)) \frac{S^2(\yy,\xx,\uu(\tau))}{2}\right)d{\tau}\nonumber\\
&+\frac{1}{2}d_{DZ}(t,\yy,\xx).
\end{align}
where
\begin{align}\label{product4}
d_{DZ}(t,\yy,\xx)=-\int_{t^L_{\yy,\xx}}^{t^U_{\yy,\xx}} |\yy-\xx| \rho^\epsilon(\yy,\xx)\partial_{\tau}\mu(\gamma(\uu)(\yy,\xx,\tau))\frac{S^2(\yy,\xx,\uu(\tau))}{2}d\tau.
\end{align}
Integrating the first term on the right hand side of \eqref{product2} gives
\begin{align}\label{product10}
&\int_{t^L_{\yy,\xx}}^{t^U_{\yy,\xx}} |\yy-\xx| \rho^\epsilon(\yy,\xx)\partial_{\tau}\left(\mu(\gamma(\uu)(\yy,\xx,\tau) \frac{S^2(\yy,\xx,\uu(\tau))}{2}\right)d{\tau}=\nonumber\\
&=|\yy-\xx| \rho^\epsilon(\yy,\xx)(\mu(\gamma(\uu)(\yy,\xx,t^U_{\yy,\xx}))) \frac{S^2(\yy,\xx,\uu(t^U_{\yy,\xx}))}{2}\nonumber\\
&-|\yy-\xx| \rho^\epsilon(\yy,\xx)(\mu(\gamma(\uu)(\yy,\xx,t^L_{\yy,\xx})) \frac{S^2(\yy,\xx,\uu(t^L_{\yy,\xx}))}{2},
\end{align}
so, integrating the first and third terms of \eqref{stressworkdamaged} and applying \eqref{product2}  gives
\begin{align}\label{stressworkdamagedintegrated}
w_{DZ}(t,\yy,\xx)&= \chi_{DZ^\epsilon(t)}(\yy,\xx)\frac{1}{2} d_{DZ}(t,\yy,\xx)\nonumber\\
&+\chi_{DZ^\epsilon(t)}(\yy,\xx) \left( F_{\gamma(\uu)(\yy,\xx,t^U_{\yy,\xx})}(S(\yy,\xx,\uu(t))) - F_1(S(\yy,\xx,\uu(0)))\right).
\end{align}

 For bonds in $PZ^\epsilon(t)$,  the total elastic and inelastic stress work is given by
\begin{align}\label{stressworkundamaging}
w_{PZ}(t,\yy,\xx)&=\int_0^{t^L_{\yy,\xx}}\chi_{PZ^\epsilon(t)}(\yy,\xx)F'_1(S(\yy,\xx,\uu(\tau))S(\yy,\xx,\dot\uu(\tau))d\tau\nonumber\\
&+\frac{1}{2}\int_{t^L_{\yy,\xx}}^{t}\chi_{PZ^\epsilon(t)}(\yy,\xx)|\yy-\xx| \rho^\epsilon(\yy,\xx)\mu(\gamma(\uu)(\yy,\xx,\tau)) S^\ast(\yy,\xx,\uu(\tau)) S^\ast(\yy,\xx,\dot\uu(\tau) )d\tau.
\end{align}
The last term of \eqref{stressworkundamaging}  is given by the non-negative quantity $w_{IPZ}(t,\yy,\xx)$ defined as
\begin{align}\label{product3}
w_{IPZ}(t,\yy,\xx)&=\frac{1}{2}\int_{t^L_{\yy,\xx}}^t |\yy-\xx| \rho^\epsilon(\yy,\xx)\mu(\gamma(\uu)(\yy,\xx,\tau)) S^\ast(\yy,\xx,\uu(\tau)) S^\ast(\yy,\xx,\dot\uu(\tau) )d\tau\nonumber\\
&=\frac{1}{2}\int_{t^L_{\yy,\xx}}^t \partial_{\tau}\left(|\yy-\xx| \rho^\epsilon(\yy,\xx)\mu(\gamma(\uu)(\yy,\xx,\tau)) \frac{S^2(\yy,\xx,\uu(\tau))}{2}\right)d \tau\nonumber\\
&+\frac{1}{2}d_{PZ},(t,\yy,\xx)
\end{align}
where
\begin{align}\label{product5}
d_{PZ}(t,\yy,\xx)= -\int_{t^L_{\yy,\xx}}^t|\yy-\xx| \rho^\epsilon(\yy,\xx) \partial_{\tau}\mu(\gamma(\uu)(\yy,\xx,\tau))\frac{S^2(\yy,\xx,\uu(\tau))}{2} d\tau.
\end{align}
Applying the fundamental theorem of calculus, the first term of \eqref{stressworkundamaging} and to  \eqref{product3}  gives
\begin{align}\label{stressworkundamagingintegrated}
w_{PZ}(t,\yy,\xx)&=\chi_{PZ^\epsilon(t)}(\yy,\xx) \left( F_{\gamma(\uu)(\yy,\xx,t)}(S(\yy,\xx,\uu(t))- F_1(S(\yy,\xx,\uu(0))\right)\nonumber\\
&+\chi_{PZ^\epsilon(t)}(\yy,\xx)\frac{1}{2}d_{PZ}(t,\yy,\xx).
\end{align}
Having integrated over time for $\yy$ and $\xx$ fixed, we integrate \eqref{stressworkundamaged} \eqref{stressworkdamagedintegrated} and \eqref{stressworkundamagingintegrated} over $\yy$ and $\xx$ variables to obtain
the elastic energy at the present time $t$:
\begin{align}\label{elastict}
\mathcal{E}^\epsilon(t)&=\int_{\Omega\times\Omega} (\chi_{UZ^\epsilon(t)}(\yy,\xx)+\chi_{PZ^\epsilon(t)}(\yy,\xx))F_{\gamma(\uu)(\yy,\xx,t)}(S(\yy,\xx,\uu(t))) \,d\yy d\xx\nonumber\\
&+\int_{\Omega\times\Omega} \chi_{DZ^\epsilon(t)}(\yy,\xx)F_{\gamma(\uu)(\yy,\xx,t^U_{\yy,\xx})}(S(\yy,\xx,\uu(t))) \,d\yy d\xx,
\end{align}
where the initial elastic energy is
\begin{align}\label{elasticti}
{\mathcal{E}}^\epsilon(0)&=\int_{\Omega\times\Omega}\chi_{UZ^\epsilon(t)}(\yy,\xx)\,F_1(S(\yy,\xx,\uu(0)))\, d\yy d\xx\nonumber\\
&+\int_{\Omega\times\Omega}\chi_{DZ^\epsilon(t)}(\yy,\xx)\,F_1(S(\yy,\xx,\uu(0)))\,d\yy\,d\xx\nonumber\\
&+\int_{\Omega\times\Omega}\chi_{PZ^\epsilon(t)}(\yy,\xx)\,F_1(S(\yy,\xx,\uu(0)))\,d\yy\,d\xx\nonumber\\
&=\int_{\Omega\times\Omega}\,F_1(S(\yy,\xx,\uu(0)))\,d\yy\,d\xx.
\end{align}

The non-negativity of the elastic energy ${\mathcal{E}}^\epsilon(t)$ follows from the non-negativity of $F_{\gamma(\uu)(\yy,\xx,t)}(S(\yy,\xx,\uu(t)))$, i.e.,
\begin{align}\label{pos}
{\mathcal{E}}^\epsilon(t)\geq 0.
\end{align}

Recalling that $\mathcal{D}(0)=0$ and collecting results shows the damage energy expended from $0$ to $t$ is given by the non-negative quantity
\begin{align}\label{damageenergy}
{\mathcal{D}}^\epsilon(t)&=\frac{1}{2}\int_{\Omega\times\Omega}(\chi_{DZ^\epsilon(t)}(\yy,\xx)d_{DZ}(t,\yy,\xx)+\chi_{PZ^\epsilon(t)}(\yy,\xx)d_{PZ}(t,\yy,\xx))\,d\yy\,d\xx
\end{align}

The kinetic energy at time $t$ is
\begin{align}\label{eq: balancek}
\mathcal{K}(t)=&\frac{\rho}{2}\int_\Omega\, |\dot{\uu}(t)|^2\,d\xx.
\end{align}
The energy balance then follows and is given by

\noindent{\bf Energy balance for load-controlled fracture evolution}
\begin{align}\label{eq:processzoneenergybalance}
{\mathcal{D}}^\epsilon(t) =\int_0^t\int_\Omega\,\bb(\tau)\cdot\dot\uu(\tau)\,d\xx\, d\tau-({\mathcal{K}}(t) + {\mathcal{E}}^\epsilon(t)  - {\mathcal{K}}(0) - {\mathcal{E}}^\epsilon(0) ),
\end{align}
where
 $\mathcal{E}^\epsilon(0)$, $\mathcal{E}^\epsilon(t)$, $\mathcal{D^\epsilon}(t)$,  $\mathcal{K}(t)$,   are given by \eqref{elasticti},  \eqref{elastict}, \eqref{damageenergy},   \eqref{eq: balancek} and  $\mathcal{D}^\epsilon(0)=0$, 

Next, we demonstrate that the evolution that delivers the displacement-damage pair has bounded elastic, potential, and damage energy given by
 \vspace*{1\baselineskip}

\begin{theorem}[{\bf Energy bound}]
\label{energybound}
\begin{equation}\label{boundedenergy}
\max_{0<t<T}\left\{ \mathcal{K}(t)+\mathcal{E}^\epsilon(t)+\mathcal{D}^\epsilon(t)\right\}< C.
\end{equation}
where the constant $ C$ only depends on the initial conditions and the load history.
\end{theorem}

 \noindent{\em Proof.}\,\,
To find this bound,  write
\begin{align}\label{W}
W(t)=\mathcal{K}(t)+ \mathcal{E}^\epsilon(t)+ \mathcal{D}^\epsilon(t)+\eta
\end{align}
Here, all terms are non-negative and $\eta>0$. Note that the rate form of energy balance gives
\begin{align}\label{eq:ineqlity}
\partial_t W(t)=\int_\Omega\,\bb(t)\cdot\dot\uu(t)\,d\xx\leq \Vert\bb(t)\Vert_{L^2(\Omega,\mathbb{R}^d)}\Vert\dot\uu(t)\Vert_{L^2(\Omega,\mathbb{R}^d)}\leq\sqrt{\frac{2}{\rho}}\sqrt{W(t)}\Vert\bb(t)\Vert_{L^2(\Omega,\mathbb{R}^d)}.
\end{align}
Equivalently, 
\begin{align}\label{eq:ineqlitydivided}
\frac{\partial_t W(t)}{\sqrt{W(t)}}\leq\sqrt{\frac{2}{\rho}}\Vert\bb(t)\Vert_{L^2(\Omega,\mathbb{R}^d)}.
\end{align}
Integrating the inequality from $0$ to $t$ and squaring both sides gives 
\begin{align}\label{eq:uniformbd}
\mathcal{K}(t) + \mathcal{E}^\epsilon(t) + \mathcal{D}^\epsilon(t)\leq \left(\sqrt{\frac{2}{\rho}}\int_0^t\Vert\bb(\tau)\Vert_{L^2(\Omega,\mathbb{R}^d)}\,d\tau+\sqrt{W(0)}\right)^2-\eta,
\end{align}
and the desired result follows taking $\eta\rightarrow 0$.

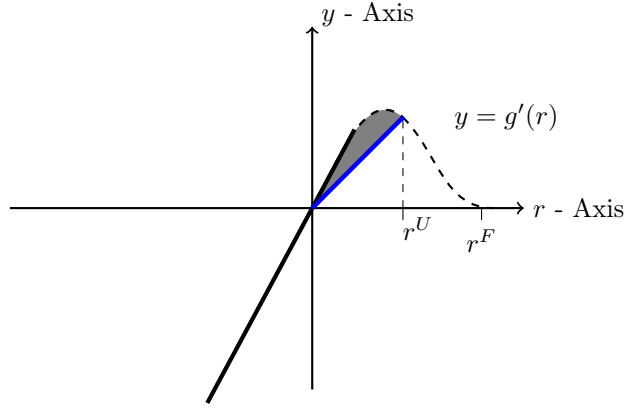
\begin{figure}
    \centering
        \begin{tikzpicture}[xscale=0.8,yscale=0.8]
		    \draw [<-,thick] (0,3) -- (0,-3);
			\draw [->,thick] (-5,0) -- (3.5,0);
            \draw [dashed,thick] (0,0) to [out=60,in=140] (1.5,1.5); 
            \draw [dashed,thick] (1.5,1.5) to [out=-45,in=180] (3,0.0);
			
	    \draw[fill=gray]    [dashed,thick] (0,0) to [out=60,in=140] (1.5,1.5)-- cycle;

			\draw [-,ultra thick] (-1.735,-3.22214) -- (0,0); 
			\draw [-,ultra thick,blue] (0,0) -- (1.5,1.5); 
			\draw [-,ultra thick] (0,0) -- (0.7,1.3); 
			\draw (2.8,-0.2) -- (2.8, 0.0);
			\draw (1.5,-0.2) -- (1.5, 0.0);
			\draw [dashed](1.5,0.0) -- (1.5, 1.5);
			\node [below] at (1.75,-0.0) {${r}^U$};
			
		 \draw [-,ultra thick,blue] (0,0) -- (1.5,1.5); 

			\node [below] at (2.8,-0.2) {${r}^F$};
			\node [right] at (3.5,0) {{$r$}\hbox{ - Axis}};
			\node [right] at (2.2,1.5)  {$y=g'(r)$ };
			\node [right] at (0,3.2)  {$y$ - Axis};
		  \end{tikzpicture}
    \caption{Bonds in damage zone:  
     Bond damage energy/Volume for all points on the blue unloading line connecting $(0,0)$ to  $(r^U,g'(r^U))$ is the area of the dark gray region. 
    }
    \label{Energy}
\end{figure}

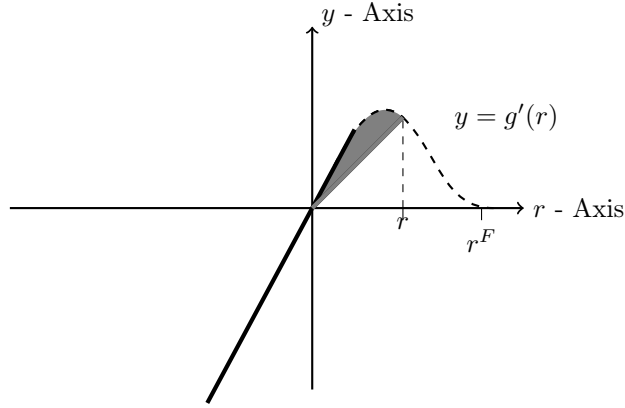
\begin{figure}
    \centering
        \begin{tikzpicture}[xscale=0.8,yscale=0.8]
		    \draw [<-,thick] (0,3) -- (0,-3);
			\draw [->,thick] (-5,0) -- (3.5,0);
            \draw [dashed,thick] (0,0) to [out=60,in=140] (1.5,1.5); 
            \draw [dashed,thick] (1.5,1.5) to [out=-45,in=180] (3,0.0);
			
	    \draw[fill=gray]    [dashed,thick] (0,0) to [out=60,in=140] (1.5,1.5)-- cycle;

			\draw [-,ultra thick] (-1.735,-3.22214) -- (0,0); 
			\draw [-,ultra thick] (0,0) -- (1.5,1.5); 
			\draw [-,ultra thick] (0,0) -- (0.7,1.3); 
			\draw (2.8,-0.2) -- (2.8, 0.0);
			\draw (1.5,-0.2) -- (1.5, 0.0);
			\draw [dashed](1.5,0.0) -- (1.5, 1.5);
		    \node [below] at (1.5,-0.0) {${r}$};

		 \draw [-,ultra thick,gray ] (0,0) -- (1.5,1.5); 

			\node [below] at (2.8,-0.2) {${r}^F$};
			\node [right] at (3.5,0) {${r}$\hbox{ - Axis}};
			\node [right] at (2.2,1.5)  {$y=g'(r)$ };
			\node [right] at (0,3.2)  {$y$ - Axis};
		  \end{tikzpicture}
    \caption{Bonds in process zone: 
    Bond damage energy/Volume for a point on the failure envelope is the area of the dark gray region. 
    }
    \label{processone bond nergy}
\end{figure}

We conclude with an explicit formula for the damage energy at unloading. Recall $r^*$ is given by \eqref{scaledmax} so on the failure envelope 
\begin{align*}
r^*=r(\yy,\xx,\uu(t))=\sqrt{|\yy-\xx|/L}S(\yy,\xx,\uu(t)).
\end{align*}
Replacing variables in \eqref{product4} by the substitution $S=S(\yy,\xx,\uu(\tau))$ with $dS=S(\yy,\xx,\dot\uu(\tau))\,d\tau$
and using the definition of $t^L_{\yy,\xx}$, noting that $t^U_{\yy,\xx}$ corresponds to $S(\yy,\xx,\uu(t^U_{\yy,\xx})):=S^U$, and straight forward but careful time integration  shows  the stress work needed to  degrade the bond up to time $t^U$ is given by
\begin{align}\label{wFfactor8}
 d_{DZ}(t,\yy,\xx)&=\rho^\epsilon(\yy,\xx)L\left( g(r^U)-\frac{g'(r^U)r^U}{2}\right)\nonumber\\
 &=\rho^\epsilon(\yy,\xx)\left( Lg(r^U)-|\yy-\xx|\mu(\gamma(\uu)(\yy,\xx,t^U_{\yy,\xx}))\frac{S^2(\yy,\xx,t^U_{\yy,\xx}))}{2}\right),
\end{align}
where $r^U=\sqrt{|\yy-\xx|/L}S(\yy,\xx,\uu(t^U_{\yy,\xx}))$.
The damage energy of the softened bond associated with  \eqref{wFfactor8} is  depicted in Figure \ref{Energy}.  This assessment holds for all bonds in the damage zone at time $t$. When a bond is in the process zone  $S^*(t,\yy,\xx,\uu)=S(\yy,\xx,\uu(t))>S^L$  then, the bond damage energy is depicted at time $t$ in Figure \ref{processone bond nergy}

\section{Energy balance for displacement-controlled and the initial boundary value problem of fracture evolution}
\label{energybalanceprocesszone2}

In this section, we establish the power and energy balance for displacement-controlled fracture evolutions. Here, the boundary displacement $\UU(t,\xx)$ is prescribed, while the body force is set to zero. Multiplying the linear momentum equation given in \eqref{eq: linearmomentumbal2} with $\dot{\uu}(t,\xx)$ and integrating over the domain $\Omega$ results in 
 \begin{equation}
	\int_{\Omega} \rho \ddot{\uu}(\xx, t) \cdot \dot{\uu}(\xx, t) \,d\xx + 
	\int_{\Omega} \LL^\epsilon [\uu](t,\xx) \cdot \dot{\uu}(\xx, t) \,d\xx = 0.
 \label{eq: linearmomentumbalInt2}
\end{equation}
We apply the identity $\chi_{\Omega^*}=\chi_\Omega+\chi_{\Omega_D^\epsilon}$ integrate by parts and after careful calculation find
\begin{align}\label{eq:prebalance2}
&\dot{\mathcal{K}}(t)+\frac{1}{2}\int_\Omega\int_\Omega\,|\yy-\xx|\rho^\epsilon(\yy,\xx)\mu(\gamma(\uu)(\yy,\xx,t)) S(\yy,\xx,\uu(t)) S(\yy,\xx,\dot\uu(t))\,d\yy\,d\xx\nonumber\\
& + \dot{\mathcal{I}}^\epsilon(t)- {\mathcal{R}}^\epsilon(t)=0,
\end{align}
where
\begin{align}
    \label{interaction}
    \dot{\mathcal{I}}^\epsilon(t)&=\frac{1}{2}\partial_t\int_\Omega\int_{\Omega_D^\epsilon}|\yy-\xx|\frac{J^\epsilon(|\yy-\xx|)}{\epsilon V_d^\epsilon}\bar{\mu}\frac{S^2(\yy,\xx,\uu(t)}{2}d\yy d\xx\nonumber\\
    &+\frac{1}{2}\partial_t\int_{\Omega_D^\epsilon}\int_{\Omega}|\yy-\xx|\frac{J^\epsilon(|\yy-\xx|)}{\epsilon V_d^\epsilon}\bar{\mu}\frac{S^2(\yy,\xx,\uu(t)}{2}d\yy d\xx
\end{align}
and
\begin{align}
    \label{forceflux}
    {\mathcal{R}}^\epsilon(t)=\int_{\Omega_D^\epsilon}\int_{\Omega}\frac{J^\epsilon(|\yy-\xx|)}{\epsilon V_d^\epsilon}\bar{\mu}S(\yy,\xx,\uu(t))\ee_{\xx-\yy}\cdot\UU(t,\xx)d\yy d\xx.
\end{align}

As in Section \ref{energybalanceprocesszone} the dissipation is non-negative  $\dot{\mathcal D}^\epsilon(t)$ and given by \eqref{damagepower}, the kinetic energy is $\dot{\mathcal{K}}(t)$ and given by \eqref{kenitic}. Now recall $\dot{\mathcal{E}}_1^\epsilon$ and $\dot{\mathcal{E}}_2^\epsilon$ given by \eqref{elasticpower} and \eqref{elasticpzpower2} and set $\dot{\mathcal{E}}^\epsilon=\dot{\mathcal{E}}_1^\epsilon+\dot{\mathcal{E}}_2^\epsilon$
then the same considerations used in Section \ref{energybalanceprocesszone} give the power balance

\noindent{\bf Power Balance displacement-controlled fracture evolutions}
\begin{align}\label{eq:rateenergybalance2}
\dot{\mathcal{K}}(t) + \dot{\mathcal{E}}^\epsilon(t) + \dot{\mathcal{D}}^\epsilon(t)-\dot{\mathcal{I}}^\epsilon(t)-{\mathcal{R}}^\epsilon(t)=0.
\end{align}

Noting that $\dot{\mathcal{D}}^\epsilon(t)\geq 0$ we summarize observations in the following Lemma.
\begin{lemma}[{\bf Growth of the Damage Energy and Process Zone}]
\label{damagezonegrowth2}
We have the power balance:
\begin{align}\label{eq:processzonepowerbalance3}
\dot{\mathcal{D}}^\epsilon(t) ={\mathcal{R}}^\epsilon(t)-\dot{\mathcal{K}}(t) - \dot{\mathcal{E}}^\epsilon(t)  -\dot{\mathcal{I}}^\epsilon(t) \geq 0.
\end{align}
\end{lemma}

Now we allow the body force to be nonzero in \eqref{eq: linearmomentumbal2} and collect arguments to obtain

\noindent{\bf Power balance for initial boundary value problem for fracture evolution}
\begin{align}\label{eq:rateenergybalance4}
\dot{\mathcal{K}}(t) + \dot{\mathcal{E}}^\epsilon(t) + \dot{\mathcal{D}}^\epsilon(t)-\dot{\mathcal{I}}^\epsilon(t)={\mathcal{R}}^\epsilon(t)+\int_\Omega\,\bb(t)\cdot\dot\uu(t)\,d\xx.
\end{align}
Since $\dot{\mathcal{D}}^\epsilon(t)\geq 0$, these observations are summarized in the following Lemma.
\begin{lemma}[{\bf Growth of the damage energy and process zone for initial boundary value problem for fracture evolution}]
\label{damagezonegrowth4}
\begin{align}\label{eq:processzonepowerbalance4}
\dot{\mathcal{D}}^\epsilon(t) =\int_\Omega\,\bb(t)\cdot\dot\uu(t)\,d\xx+{\mathcal{R}}^\epsilon(t)-\dot{\mathcal{K}}(t) - \dot{\mathcal{E}}^\epsilon(t)  -\dot{\mathcal{I}}^\epsilon(t) \geq 0.
\end{align}
\end{lemma}

\begin{remark}[{\bf  Condition for damage in the initial boundary value problem for fracture evolution}]
\label{damagecondition2}
{\em If the rate of energy put into the system exceeds the material's capacity to generate kinetic and elastic energy through displacement and velocity, then damage occurs.}
\end{remark}
Note here that \eqref{eq:processzonepowerbalance4} is  consistent with thermodynamics and in this formulation follows directly from the equations of motion \eqref{eq: linearmomentumbal2} and the Dirichlet boundary conditions.

The energy balance follows and is given by

\noindent{\bf Energy balance for the initial boundary value problem for fracture evolution}
\begin{align}\label{eq:processzoneenergybalance2}
{\mathcal{D}}^\epsilon(t) =\int_0^t\left(\int_\Omega\,\bb(\tau)\cdot\dot\uu(\tau)\,d\xx+{\mathcal{R}}^\epsilon(\tau)\, \right)d\tau-({\mathcal{K}}(t) + {\mathcal{E}}^\epsilon(t)+{\mathcal{I}}^\epsilon(t)  - {\mathcal{K}}(0) - {\mathcal{E}}^\epsilon(0) -{\mathcal{I}}^\epsilon(0)),
\end{align}
where $\mathcal{D}^\epsilon(t)$ is given by \eqref{damageenergy}.

We apply a Gronwall inequality  using the power balance \eqref{eq:rateenergybalance4} and argue similar to the proof  of Theorem \ref{energybound} to find that the total energy of the fracture evolution initial boundary value problem is bounded uniformly for $0\leq t \leq T$ and if $u_0=0$ it is also uniformly bounded in horizon size $0<
\epsilon$ with the bound depending only on the initial condition $\vv_0$ loading data $\bb(t,\xx)$ and $\UU(t,\xx)$.
\vspace*{1\baselineskip}

\begin{theorem}[{\bf Energy Bound}]
\label{energybound2}
\begin{equation}\label{boundedenergy2}
\max_{0<t<T}\left\{ \mathcal{K}(t)+\mathcal{E}^\epsilon(t)+\mathcal{I}^\epsilon(t)+\mathcal{D}^\epsilon(t)\right\}< C
\end{equation}
where the constant $ C$ only depends on the initial conditions and the loading data.
\end{theorem}

\section{Calibration, fast cracks and edge cracks} 
\label{sec.Straight Cracks}

In this section, we calibrate the model using quantities obtained directly from the dynamics and equating them to the material properties of the specimen. The elastic moduli and critical energy release rate follow from $\Gamma$-convergence as $\epsilon\rightarrow 0$, see \cite{lipton2016cohesive}, \cite{lipton2024energy}. But in the former case are obtained here by Taylor series and in the latter case are obtained using rigorous geometric measure theory arguments presented in \cite{lipton2024energy}. 
\subsection{Elastic properties}
The horizon length scale $\epsilon$ is chosen small enough to resolve the process zone hence smaller than the length scale of the process zone of the material.
The Lam\'e constants for undamaged material are used to calibrate $\overline{\mu}$. 
Inside undamaged material containing a neighborhood $\mathcal{H}_\epsilon({\bf x})$ the elastic energy density is given by
\begin{align}\label{elastictundamaged}
\mathcal{W}^\epsilon(t,{\bf x})&=\int_{\Omega} F_1(S(\yy,\xx,\uu(t))) \,d\yy=\frac{1}{2}\int_{\Omega} \rho^\epsilon(\yy,\xx)|\yy-\xx|\overline{\mu} \frac{S^2(\yy,\xx,\uu(t))}{2} \,d\yy\nonumber\\
&=\frac{1}{2V_d^\epsilon}\int_{{\mathcal{H}}_{\epsilon}({\bf x})}\frac{|\yy-\xx|}{\epsilon}J\left(\frac{|\yy-\xx|}{\epsilon}\right)\overline{\mu}\frac{S^2(\yy,\xx,\uu(t))}{2} \,d\yy.
\end{align}
 When the displacement is linear across $\mathcal{H}_\epsilon(\xx)$, i.e., $\uu=L{\bf x}$ where $L$  is a constant matrix, then ${S(\yy,\xx,\uu(t))}=F\ee\cdot \ee$, where $F=(L+L^T)/2$, and changing variables $\yy=\xx+\epsilon\zeta$, with $|\zeta|<1$, gives
 \begin{align}
\mathcal{W}^\epsilon(t,{\bf x})&=\frac{1}{4\omega_d}\int_{\mathcal{H}_1(0)}|\zeta| J(|\zeta|)\overline{\mu}(F\ee\cdot\ee)^2\,d\zeta.
\label{LEFMequality}
\end{align}
Observe next that $(F\ee\cdot \ee)^2=\sum_{ijkl}F_{ij}F_{kl}e_ie_je_ke_l$ so
\begin{eqnarray}
\mathcal{W}^\epsilon(t,{\bf x})=\frac{1}{2}\sum_{ijkl}\mathbb{C}_{ijkl}F_{ij}F_{kl}
\label{leading order}
\end{eqnarray}
where
\begin{eqnarray}
\mathbb{C}_{ijkl}=\frac{\overline{\mu}}{2\omega_d}\int_{\mathcal{H}_1(0)}|\xi|J(|\xi|)\,e_i e_j e_k e_l\,d\xi.
\label{elasticpart3}
\end{eqnarray}
On the other hand, the potential elastic energy per unit volume of a material characterized by shear moduli $G$ and Lam\'e constant $\lambda$ is given by
\begin{align}
U&=\frac{1}{2}\sum_{ijkl}\mathbb{\tilde{C}}_{ijkl}F_{ij}F_{kl}
\label{LEFMequality13}
\end{align}
where
\begin{eqnarray}
\mathbb{\tilde{C}}_{ijkl}=G (\delta_{ik} \delta_{jl} +\delta_{il}\delta_{kj})+ \lambda\delta_{ij}\delta_{kl}
\label{elasticpart4}
\end{eqnarray}
 Setting $U=\mathcal{W}^\epsilon(t,{\bf x})$, we get $\mathbb{\tilde{C}}=\mathbb{C}$ and we arrive at the calibration for determining $\overline{\mu}$ given by
\begin{eqnarray}\label{lambdamu}
G=\lambda=\frac{\overline{\mu}}{8}\int_{0}^1r^2J(r)dr, \hbox{ $d=2$}&\hbox{ and }& G=\lambda=\frac{\overline{\mu}}{10} \int_{0}^1r^3J(r)dr, \hbox{ $d=3$}.\label{calibrate1}
\end{eqnarray}
With Possion ratio $\nu=1/3$ for $d=2$ and $\nu=1/4$ for $d=3$. These formulas are consistent with those obtained using $\Gamma$-convergence obtained in \cite{lipton2016cohesive}.

 \subsection{Strain hardening and Strength}
The domain of elastic behavior is calibrated to that of the material.  Here, we use the tensile stress of a material corresponding to the onset of nonlinear behavior $\sigma^L$. The value of $r^L$  is determined by
\begin{align}\label{lossoflinearity}
\sigma^L=g'(r^L).
\end{align}
The critical stain $r^C$ is determined by
\begin{align}\label{strength}
\sigma^C=g'(r^C).
\end{align}

\subsection{Critical energy release rate}
\label{criticalenergyrelease}

The measured value of the critical energy release rate of the material is used to calibrate the  stress work necessary to take undamaged material and fail it irrevocably. 
Define $t^F_{\yy,\xx}$ as the time the bond fails, i.e.,  $\gamma(\uu)(\yy,\xx,t^F_{\yy,\xx})=0$ and the energy associated with the set of all damaged bonds inside the sample that have failed up to the present time $t$ is given by
 \begin{align}\label{product6}
\mathcal{F}^\epsilon(t)=&\frac{1}{2}\int_\Omega\int_\Omega\left(-\chi_{\Gamma^\epsilon(t)}(\yy,\xx)\int_{t^L_{\yy,\xx}}^{t^F_{\yy,\xx}} |\yy-\xx| \rho^\epsilon(\yy,\xx)\partial_{\tau}\mu(\gamma(\uu)(\yy,\xx,\tau))\frac{S^2(\yy,\xx,\uu(\tau))}{2}d\tau\right)\,d\yy d\xx.
\end{align}
The stress work to fail a bond of length $|\yy-\xx|$ in tension is given by $d_F(\yy,\xx)$ 
\begin{align}\label{wFfactor}
d_F(\yy,\xx)=-\int_{t^L_{\yy,\xx}}^{t^F_{\yy,\xx}}|\yy-\xx|\rho^\epsilon(\yy,\xx)\overline{\mu}\,\partial_{\tau}\gamma(\uu)(\yy,\xx,\tau)\frac{S^2(\yy,\xx,\uu(\tau))}{2}\,d\tau.
\end{align}
Thus, for $t>t^F_{\yy,\xx}$ one has the failure energy $\mathcal{F}^\epsilon({t})$ given by
\begin{align}\label{product7}
\mathcal{F}^\epsilon(t)=&\frac{1}{2}\int_\Omega\int_\Omega\,\,d_F\,\chi_{\Gamma^\epsilon(t)}(\yy,\xx)\,d\yy d\xx.
\end{align}
where $\chi_{\Gamma^\epsilon(t)}$ is the indicator function of the failure set \eqref{damageset}.

We now use \eqref{wFfactor} and \eqref{product7} to calibrate  $r^F$ to measured values of $\mathcal{G}_c$.  From  \eqref{wFfactor8}, we see $d_{DZ}=d_F$ when the unloading curve degenerates to the abscissa and $ \mu(\gamma(\uu)(\yy,\xx,\uu(t^F))=0$,  so  the stress work needed to fail the bond in tension is given by the area underneath the failure envelope, and
\begin{align}\label{wFfactor3}
 d_F=\rho^\epsilon(\yy,\xx)L g(r^F).
\end{align}

We now use \eqref{wFfactor3} to first rewrite formula \eqref{product7} representing the energy needed to break all bonds up to time $t$. To do this, we change the coordinates and choose a unit direction $\ee$ on the unit sphere ${\mathbb{S}}^{d-1}$. We write $\xx$ in the body $\Omega$ according to $\xx=\zz+s\ee$ where $\zz$ is any point on the plane perpendicular to $\ee$ and $s$ is the $\ee$ coordinate. We denote the plane $\zz\perp \ee$ by $\Pi^\zz$. The set $\Omega_{\zz}^\ee$ is the intersection of $\Omega$ and the line $\zz+s\ee$. The set $\Omega^\ee$ is the set of all lines $\zz+s\ee$ perpendicular to $\Pi^\zz$ that intersect $\Omega$. 
Proceeding as in \cite{lipton2024energy}, direct calculation shows that the failure energy for general set of failed bonds up to time $t$ is given by
 \vspace*{1\baselineskip}

\noindent{\bf  Geometric integral representation for Failure energy}
\label{geometricintegral}

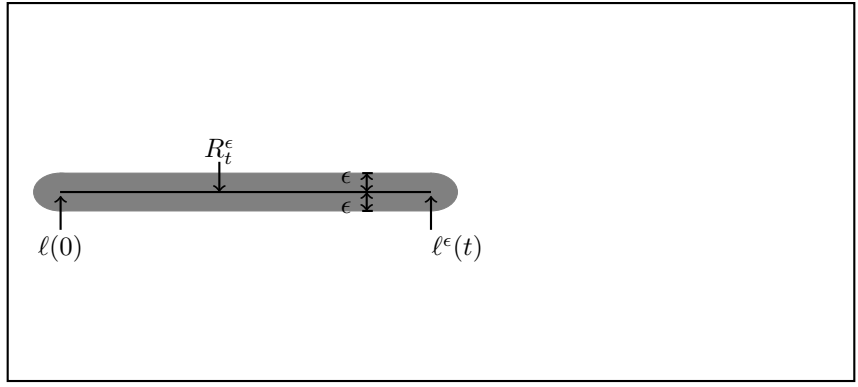
\begin{figure} 
\centering
\begin{tikzpicture}[xscale=0.70,yscale=0.50]

\draw [fill=gray, gray] (-7.01,-0.5) rectangle (0.0,0.5);






\node[above] at (-4.0,0.5) {$R_t^\epsilon$};

\draw [->,thick] (-4.0,0.8) -- (-4.0,0.0);

\draw[fill=gray, gray] (0,-0.5) arc (-90:90:0.5)  -- cycle;


\draw[fill=gray, gray](-7.01,0.0) circle (0.5);












\draw [-,thick] (-7.01,0) -- (-0.0,0);




\draw [thick] (-8,-5) rectangle (8,5);












\node [above] at (-1.6,0) {$\epsilon$};

\node [below] at (-1.6,0) {$\epsilon$};

\draw [thick] (-1.3,.5) -- (-1.1,.5);

\draw [<->,thick] (-1.215,.5) -- (-1.215,0);

\draw [<->,thick] (-1.215,-.5) -- (-1.215,0);

\draw [thick] (-1.3,-.5) -- (-1.1,-.5);

\draw [->,thick] (0.0,-1.0) -- (0.0,-0.1);

\node  at (0.5,-1.5) { $\ell^\epsilon(t)$};

\draw [->,thick] (-7,-1.0) -- (-7,-0.1);

\node  at (-7,-1.5) { $\ell(0)$};

\end{tikzpicture} 
\caption{{  Flat cracks and bond alignment. 
The crack $R_t^\epsilon=\{\ell(0)<x_1<\ell(t)\}$ and all broken bonds are given by the gray shaded region. We call this alignment because the broken bonds are only those that cross the flat crack set $R_t$. The energy of all the broken bonds are integrated up in the calculation of  the failure energy ${\mathcal{F}}^\epsilon(t)$} associated with the crack.}
 \label{P}
\end{figure}

 \vspace*{1\baselineskip}
\begin{align}
\label{3rdtenergy}
\mathcal{F}^\epsilon({t})=\frac{{L  g(r^F)}}{2\omega_d}\int_{\mathbb{S}^{d-1}}\int_{\Omega^\ee}\int_0^1\,J(|\bzz|)|\bzz|^{d} m^{\epsilon}({t},\ee,\zz,|\bzz|)\,d|\bzz|\,d\zz\,d\ee,
\end{align}
with
\begin{align}
\label{1stkernel}
m^{{\epsilon}}({t},\ee,\zz,|\bzz|)=\frac{1}{\epsilon|\bzz|}\int_{\Omega^\ee_\zz}\chi^+(\zz+\ee(s+\epsilon|\bzz|),\zz+\ee s,{t})\,ds,
\end{align}
where
\begin{align}
\label{kernelkernel}
\chi^+(\zz+\ee(s+{\epsilon}|\bzz|),\zz+\ee s,{t})=\left\{\begin{array}{ll}1,&\hbox{if the pair $(\zz+\ee(s+{\epsilon}|\bzz|),\zz+\ee s)$ is in $\Gamma^\epsilon({t})$}\\
0,& \hbox{otherwise}.\end{array}\right.
\end{align}
The function $m^{\epsilon}({t},\ee,\zz,|\bzz|)$ is associated with the intersection of the line $\xx=\zz+\ee s$ with the subset of bonds of length $|\yy-\xx|$ in $\Gamma^\epsilon(t)$ divided by the length of the bond $|\yy-\xx|$. On considering the case of a flat crack (see Figure \ref{P}), we  find below that the failure energy as represented by \eqref{3rdtenergy} trivially factors into two parts, one given by the length/area of the crack and the second given by a critical energy release rate.

Consider the flat crack given by the failure set $\Gamma^\epsilon(t)$ defined by a flat $d-1$ dimensional piece of surface (line segment) $R_t$ a distance $\epsilon$ away from the boundary.  Points above the surface are no longer influenced by forces due to points below the surface and vice versa. This is the case of alignment, i.e., all bonds connecting points $\yy$  above  $R_t$ to points $\xx$ below are broken and vice versa see Figure \ref{P}. 
The calculation shows that the failure energy $\mathcal{F}^\epsilon$  does not depend upon $\epsilon$ and factors into two parts given by
\begin{align}
\label{crofton}
\mathcal{F}^\epsilon({t})=\mathcal{F}({t})=\mathcal{G}_c\times\frac{1}{2\omega_{d-1}}\int_{\mathbb{S}^{d-1}}\int_{\Omega^\ee}N(t,\ee,\zz)\,d\zz\,d\ee,
\end{align}
with each factor independent of $\epsilon$.
Here 
\begin{eqnarray}
\mathcal{G}_c=\frac{L g(r^F)}{2}\frac{2\omega_{d-1}}{\omega_d}\, \int_{0}^1 r^{d}J(r)dr, \hbox{  for $d=2,3$}
\label{epsilonfracttough}
\end{eqnarray}
is the energy release rate for $d=2,3$ and is consistent with the formula for energy release rate obtained from $\Gamma$-convergence \cite{lipton2016cohesive}. The second factor is Crofton's formula
\begin{align}
\label{crofton2}
\frac{1}{2\omega_{d-1}}\int_{\mathbb{S}^{d-1}}\int_{\Omega^\ee}N(t,\ee,\zz)\,d\zz\,d\ee,
\end{align}
where $N(t,\ee,\zz)$ is the multiplicity function of the line with normal $\zz$ along a direction $\ee$ giving the value one if it pierces the crack and zero if it does not intersect the crack. In general, Crofton's formula delivers the arc-length of plane curves $d=2$ and surface area of surfaces $d=3$. It also provides the $d-1$ dimensional Hausdorff measure denoted by $\mathcal{H}^{d-1}$ for countably rectifiable surfaces \cite{Federer}, \cite{Morgan}.
With this observation, the second factor is immediately identified as Crofton's formula, which delivers the length $(d=2)$ or area ($d=3$) for flat cracks. Here, the failure  energy is the fracture energy and is given by the product
\begin{align}
\label{last}
\mathcal{F}({t})=\mathcal{G}_c\times\mathcal{H}^{d-1}(R_t),
\end{align}
where the area or length of $R_t$ is written as  $\mathcal{H}^{d-1}(R_t)$ which is the surface area of the crack for $d=3$ and length of the crack  for $d=2$. 
This is the well-known formula for Griffith fracture energy, but now derived directly from the equations of motion without any external hypothesis.  
The formula for $\mathcal{G}_c$ given by \eqref{epsilonfracttough} is set equal to the critical energy release rate of the material $G_c$  and this determines $L$. Since the failure envelope $g(r)$ is one to one in $r$ one has the natural constraint $r^F\geq r^C$.
Since $g(r)$ is of a general form $g(r^F)$ and $\partial_r g(r^C)$ can be specified independently so strength and fracture toughness can be specified independently.

One can immediately extend the formula \eqref{last} to a system of dispersed flat cracks separated by the distance $\epsilon$ with different orientations. More generally the formula for the damage energy given in this paper delivers the Griffith fracture energy for countably rectifiable crack shapes when $\epsilon\rightarrow 0$. This is proved in Lipton and Bhattacharya (2025). What is distinctive is that the Griffith fracture energy for a straight crack is the failure energy  of the model itself, obtained without sending any parameter such as  $\epsilon$ to zero.

In all simulations we apply the bilinear model and choose influence function $J(r)$ given by $1$ for $0<r<1$ and zero for $1\leq r$ and $L$ is determined by
\begin{eqnarray}
L =\frac{3\pi G_c}{r^F\sigma^C},\hbox{  $r^F\geq r^C$}=\frac{\sigma^C}{9 E}. 
\label{bilinearfracttoughstrength}
\end{eqnarray}

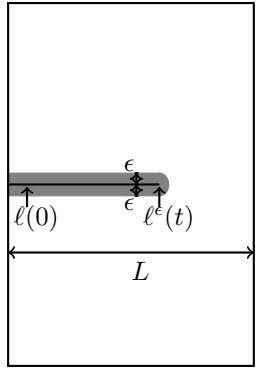
\begin{figure} 
\centering
\begin{tikzpicture}[xscale=0.25,yscale=0.30]
















\draw [fill=gray, gray] (-8,-0.5) rectangle (0.0,0.5);

\draw[fill=gray, gray] (0,-0.5) arc (-90:90:0.5)  -- cycle;

\node [above] at (-1.6,0.2) {$\epsilon$};

\node [below] at (-1.6,-0.2) {$\epsilon$};

\draw [<->,thick] (-1.215,.5) -- (-1.215,0);

\draw [<->,thick] (-1.215,-.5) -- (-1.215,0);






\draw [-,thick] (-8,0) -- (-0.0,0);




\draw [thick] (-8,-8) rectangle (5,8);

\draw [<->,thick] (-8,-3) -- (5,-3);

\node[below] at (-1.0,-3) {$L$};













\draw [thick] (-1.3,.5) -- (-1.1,.5);



\draw [thick] (-1.3,-.5) -- (-1.1,-.5);

\draw [->,thick] (0.0,-1.0) -- (0.0,-0.1);

\node  at (0.5,-1.5) { $\ell^\epsilon(t)$};

\draw [->,thick] (-7,-1.0) -- (-7,-0.1);

\node  at (-6.5,-1.5) { $\ell(0)$};

\end{tikzpicture} 
\caption{ Representative length of sample. Pre-crack goes from edge into specimen to $\ell(0)$. The crack  tip at  time $t$ is $\ell^\epsilon(t)$. The choice of representative length of sample is $L$. The set of broken bonds is of width $2\epsilon$.}
 \label{P2}
\end{figure}

\subsection{Fast crack evolution}

Note that on the softening envelope we have $r^*=r$, $S(\yy,\xx,\uu(t))>S^L$, and
\begin{align}
   \label{softening envelope}
    \overline{\mu}\gamma(\uu)(\yy,\xx,t)S(\yy,\xx,\uu(t))=\frac{\partial_rg(r)}{\sqrt{|\yy-\xx|/L}}.
\end{align}
With this in mind, we can choose boundary forces such that bonds strained beyond the elastic regime monotonically increase in strain and ultimately fail. This is the setting of fast crack evolution and is consistent with the formulation given in \cite{lipton2024energy}. For fast fracture, the failure energy factors for flat cracks and is precisely the Griffith fracture energy. Fast fracture simulations are given in Section \ref{sec.dynamic-crack}.

\subsection{Edge Cracks}
\label{edge}
No special treatment of bond breaking energy is needed for a crack originating from the notched edge (or corner) of a specimen in the blended model. The traction-free boundary conditions enforced by the kernel $\rho^\epsilon(\yy,\xx)$ implies that any bond originating at $\xx$ with the terminus $\yy$ outside the domain does not influence the force on $\xx$.
As calculated earlier in \eqref{wFfactor3} the stress work needed to fail a bond is given by
\begin{align*}
 d_F=\rho^\epsilon(\yy,\xx)L g(r^F),
\end{align*}
and and the energy associated with the set of all damaged bonds inside the sample that have failed up to the present time $t$ is given by
\begin{align}\label{product20}
\mathcal{F}^\epsilon(t)=&\frac{1}{2}\int_\Omega\int_\Omega\,d_F\,\chi_{\Gamma^\epsilon(t)}(\yy,\xx)\,d\yy d\xx.
\end{align}
For this case, the energy of failure will depend on the geometry of the specimen boundary and $\epsilon$. For example, it will depend on the radius of curvature of the notch and $\epsilon$ for an idealized smooth notched specimen. For a pre-crack perpendicular to a traction free edge the choice of characteristic length $L$ for a sample is given in Figure \ref{P2}.

\section{Dimension-Free Formulation}
\label{nondimensionalization}
In this section, we formulate the dynamics in dimension-free form. In doing so, we identify the dimension-free group for the problem. Recall that $L$ is a characteristic length associated with the evolution, e.g., \eqref{bilinearfracttoughstrength}, and the dimension-free domain is denoted by $\bar{\Omega}^\ast$ where ${\Omega}^\ast=L\bar{\Omega}^\ast$. Similarly, we define the dimension-free points inside $\bar{\Omega}^\ast$    by ${\xx}=L\bar\xx$, ${\yy}=L\bar\yy$. The dimension-free horizon horizon $\bar\epsilon$ is given by ${\epsilon}=L\bar{\epsilon}$.   Let $T_0$ be a characteristic time scale and the dimension-free time $\bar t$ is defined by  $t=T_0\bar{t}$.  The dimension-free displacement $\bar\uu$ is defined by ${\uu}=L\bar{\uu}$. The time derivatives with respect to to dimension-free time are given by $\partial_{{t}}=T_0^{-1}\partial_{\bar{t}}$. It is pointed out that $r$ is defined by \eqref{def of r} and is dimension-free. Finally, the material density ${\rho}$ is of dimension mass per unit volume given by $M/L^d$.

The kernel given by \eqref{scale2} written in terms of the dimension-free quantities is
\begin{equation}\label{eq:scalllle}
\rho^{\ast,\bar \epsilon}({\bar\yy},{\bar\xx}):=\frac{\chi_{{\bar\Omega}^\ast}({\bar\yy})J^{{\bar\epsilon}}(|{\bar\yy}-{\bar\xx}|)}{\bar\epsilon V_d^{{\bar\epsilon}}},
\end{equation}
where $V_d^{\bar\epsilon}=\omega_d{\bar\epsilon}^{d}$.
The bond stiffness $\bar\mu$ has dimensions of force per unit area  $ML^{-2}T_0^{-2}$. 
Recall that the damage factor $\gamma$ is dimensionless,  $0\leq \gamma\leq 1$ and depends on the non-dimensional scaled strain $r$ and satisfies $\gamma(\bar{\uu})(\bar{\yy},\bar{\xx},\bar{t})=\gamma({\uu})({\yy},{\xx},{t})$.
On substitution of the non-dimensional variables the dimension-free form of the force $\LL^{\bar\epsilon} [\bar\uu](\bar t,\bar\xx)$
is found through
\begin{align}
       \label{dim free elastic}
     \LL^{\epsilon} [\uu]( t,\xx)=&\frac{\overline{\mu}}{L}\LL^{\bar\epsilon} [\bar\uu](\bar t,\bar\xx) = -\frac{\overline{\mu}}{L}\int_{\bar\Omega} \boldsymbol{f}^{\bar\epsilon}(\bar t,\bar\yy,\bar\xx,\bar\uu) \,d\bar\yy\nonumber\\
     &=-\frac{\overline{\mu}}{L}\int_{\bar\Omega}\rho^{\bar\epsilon}(\bar\yy,\bar\xx)\overline{\mu}^{-1}(\gamma(\bar\uu)(\bar\yy,\bar\xx,\bar t))S(\bar \yy,\bar\xx,\bar\uu(\bar t))\boldsymbol{e} d\bar\yy,
\end{align}
and the dimension-free form of the force balance becomes

\begin{align}\label{eq: linearmmmmmomentumbal}
\bar{\rho}\ddot{\bar{\uu}}(\bar{t},\bar{\xx})+ \LL^{\bar\epsilon} [\bar{\uu}](\bar{t},\bar{\xx}) =\bar{\bb}(\bar{t},\bar{\xx}).
\end{align}
The dimension-free group is given by the dimension-free density 
$\bar\rho=\beta^{-2}L^{2}T^{-2}$,
where $\beta$ is the shear wave speed of the material $\beta=\sqrt{{\overline{\mu}}/{\rho}}$.
The dimension-free group $\bar\rho$ provides the ratio of inertial force to elastic force, with $L$ proportional to the ratio of fracture toughness to strength, e.g., \eqref{bilinearfracttoughstrength}.

\section{Results}
\label{sec.numerics}
This section presents the numerical implementation of the proposed theoretical framework, providing details of how the model is incorporated into computational simulations. A mesh convergence study is then conducted to ensure numerical accuracy. Following this, the numerical results are presented and compared with experimental data across five benchmark problems: (i) monotonic and cyclic Mode I fracture, (ii) cyclic mixed-mode fracture, (iii) mixed-mode fracture induced by a corner singularity, (iv) dynamic crack branching, and (v) the size effect in concrete beams. Throughout this section, the contour plots of the local damage variable are used to present the crack path resulting from the numerical analysis. The damage variable, i.e. local damage, can be any value between zero and one, indicating intact and completely damaged material points, respectively. Local damage at the material point $\xx$ at time \textit{t}, $D(\xx, t)$ can be calculated as

\begin{equation} \label{eq:localDamage}
	D(\xx, t) = 1 - \frac{\int_{\HH_\epsilon(\xx)} \gamma(\uu)(\yy,\xx,t) \,d\yy} {\int_{\HH_\epsilon(\xx)} \,d\yy}
\end{equation}

\noindent
where $\gamma(\uu)(\yy,\xx,t)$ is called the two-point phase field described in Figure \ref{ConvexConcaveC}. A simple bilinear failure envelope is used to describe the constitutive law throughout this section, see \eqref{bilinear} and Figure \ref{Failure envelope}. Eq. \eqref{bilinearphasefield} provides the explicit form of the two-point phase field used within this study. The details of the calibration of the material constants are provided in Section \ref{sec.Straight Cracks}.

The quasi-static results were obtained using the dynamic relaxation method \cite{underwood1983}. To utilize this method, a damping term is added to the equation of motion given in \eqref{eq: linearmomentumbal2}, and the central difference time integration algorithm is used to obtain the displacement field satisfying the static equilibrium at each time step. However, the time step corresponds to the loading step rather than the physical time step used in transient analyses. Since this technique is an explicit iterative algorithm, the critical time step should be calculated accordingly. In this study, the critical time step is calculated as described by \cite{sillingaskari2005}. The numerical algorithm for the utilized dynamic relaxation method is provided in the supplementary materials of this study.

Since this study focuses on the quasi-brittle failure in which tension softening occurs, we choose a bilinear constitutive law given by \eqref{bilinearphasefield} in the numerical simulations. Hence, the bond force densities can be calculated using \eqref{constituitive}. Using \eqref{strains}, and \eqref{lambdamu},  with an influence function of $J(r)=1$ for two-dimensional domains, the calibration of the constant $\overline{\mu}$ results in $\overline{\mu} = 9E$.
Here $E$ is the Young's Modulus of the material. Then, using \eqref{strength}, and \eqref{bilinear} \eqref{strains}, 
the critical bond strain is
\begin{equation} \label{sc}
    S^C = \sqrt{\frac{L}{|\yy-\xx|}} \frac{\sigma^C}{9E},
\end{equation}
\noindent
where $\sigma^C$ is the tensile strength of the material, and $L$ is the characteristic length of the domain. For static quasi-brittle failure, the characteristic length is the smallest dimension of the domain. The failure bond strain is calibrated using \eqref{epsilonfracttough}, \eqref{bilinear}, and is

\begin{equation}\label{sf}
    S^F = \frac{3 \pi \; {G}_c}{\sqrt{L |\yy-\xx|} \; \sigma^C}
\end{equation}

\noindent where ${G}_c$ is the critical energy release rate of the material, and \eqref{sf} is consistent with  \eqref{bilinearfracttoughstrength}. 

For all fracture simulations, we choose $\epsilon \leq L$, where $L$ is determined by \eqref{bilinearfracttoughstrength} with $r^F=r^C$.
Only uniform meshes and 2D domains are considered here. The displacement boundary conditions are applied at least to the horizon size of the layer nodes. The damping coefficient is selected as $2.0\times 10^7$ kg$/$m$^3$ unless otherwise stated. Finally, the horizon size is selected as 3.015 times of the uniform grid size. As \cite{trageser2020bond} stated earlier, Poisson's ratio is limited inherently due to the two-parameter formulation of the bond-based peridynamics. Therefore, Poisson's ratio is used as 0.33 throughout this study.

For comparison purposes, experimental and numerical data from the literature are extracted using WebPlotDigitizer by \cite{WebPlotDigitizer}. The same software is utilized when comparing the crack trajectories obtained by the proposed blended model with the experimental trajectories. In addition to damage contour plots, displacement contour plots prove valuable for determining resultant crack trajectories, as illustrated in the figures that can be found in the supplementary materials of this study. 
\subsection{Mesh Convergence Study}
\label{sec.meshConvergence}

A three-point bending test (TBP) is selected to perform the numerical convergence analysis of the proposed model. The beam has a depth of 150 mm, a thickness of 50 mm, and a span of 600 mm, with a 45 mm pre-notch. The material properties of the concrete are provided in Table \ref{table:chenLiu-material}. A uniform grid spacing is used throughout Section \ref{sec.numerics}. Therefore, the convergence analysis of non-uniform grid spacing, mesh orientation, and related factors will be addressed in a separate study.

\begin{figure}
    \centering
    \includegraphics[width=1.0\linewidth]{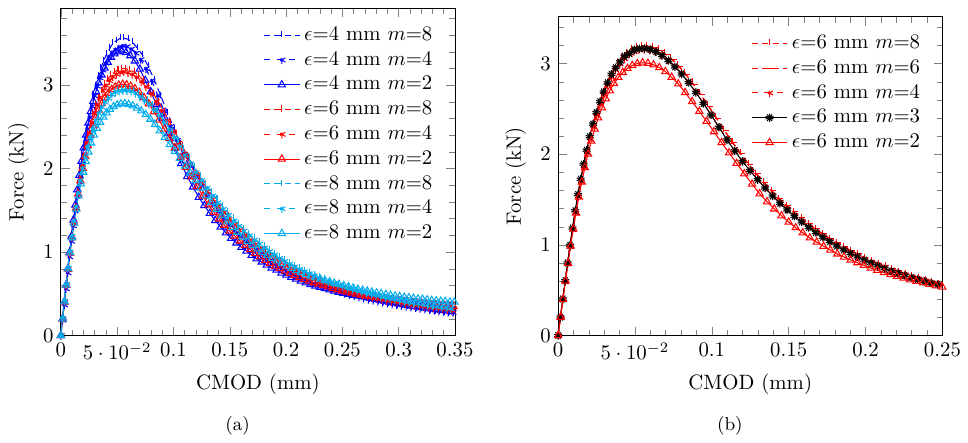}
    \caption{The convergence study for decreasing uniform grid spacing for fixed horizon sizes for (a) 4 mm, 6 mm, and 8 mm, (b) 6 mm with detailed analysis. The $m$ value is the ratio of the horizon to grid spacing used in the numerical simulation.}
    \label{fig:convergence-1}
\end{figure}

Two main variables are considered in the mesh convergence analysis: horizon size and  grid spacing. The characteristic length is not treated as a mesh parameter, as it is tied to the smallest dimension of the 2D problem domain (e.g., the beam depth in a TPB test) in a static quasi-brittle failure modeling. Figure \ref{fig:convergence-1}a presents the force vs. crack mouth opening displacement (CMOD) results for fixed horizon sizes ranging from 4 mm to 8 mm with decreasing grid sizes. In the peridynamics literature, this type of convergence study is referred to as $m$-convergence, as described by \cite{bobaru2009convergence}. The horizon size can be expressed as $\epsilon = m h$, where $h$ is the size of the uniform grid. As the value of $m$ increases, the numerical results approximate the exact solution of the nonlocal initial value problem in \eqref{eq: linearmomentumbal} and \eqref{initialconditions} for a given $\epsilon$. Figure \ref{fig:convergence-1} shows that numerical results for $m$ values greater than 2 exhibit rapid convergence for fixed horizon sizes. Figure \ref{fig:convergence-1}b provides a more detailed $m$-convergence analysis for a fixed horizon size of 6 mm. It is clearly seen that the differences between the numerical results for $m$ values greater than 4 are negligible. Therefore, $m = 3$ can be safely chosen to reduce computational cost. It is important to note that as the value of $m$ increases, both the number of neighbors per node and the total number of nodes in the domain increase, resulting in a significant increase in computational cost.

\begin{figure}
	\centering
	\includegraphics[width=0.6\linewidth]{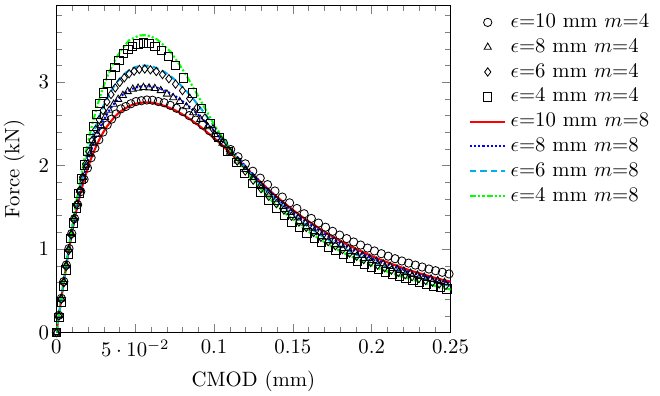}
	\caption{The convergence study for decreasing horizon size, $\epsilon$, for fixed $m$ values.}
	\label{fig:convergence-3}
\end{figure}

Figure \ref{fig:convergence-3} presents the so-called $\delta$-convergence study of the proposed blended model. As shown in Figure \ref{fig:convergence-3}, two different $m$-values 4 and 8 are considered, while the horizon size decreases from 10 mm to 4 mm. As described by \cite{bobaru2009convergence}, it is expected that the numerical solution of the nonlocal model approaches the local solution as the horizon size becomes smaller. However, since the problem lacks an exact local solution, a direct comparison between local and nonlocal results is not possible. Nevertheless, the numerical results can be compared with the experimental data presented in Figure \ref{figure:chenLiuResults-a}. The experimentally measured load-carrying capacity was approximately 3.4 kN. The numerical results show improved agreement with the experimental data as the horizon size decreases, yielding a peak load of 2.7 kN for $\epsilon = 10$ mm and 3.6 kN for $\epsilon = 4$ mm.

This section demonstrates that the proposed blended approach converges to a nonlocal solution for a given $\epsilon$ as the number of neighbors inside the horizon increases. 
However, the horizon size $\epsilon$ must be small enough to match experimental results while still being large enough to keep computational costs reasonable.

\subsection{Mode-I fracture test on high-strength concrete beam under monotonic and cyclic loading}
\label{sec.mode-I-mono-cyclic}
\begin{figure} 
	\centering
	\includegraphics[width=0.65\textwidth]{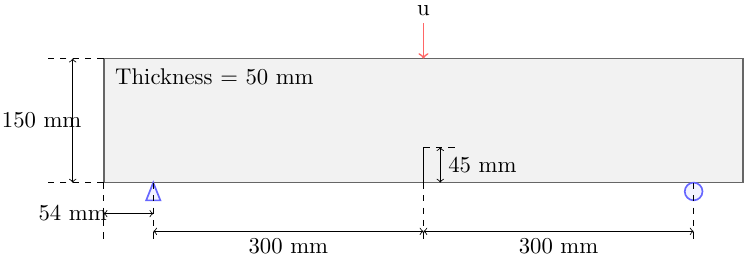}
	\caption{Dimensions and boundary conditions of the mode-I fracture test with a high strength concrete beam.}
	\label{figure:chenLiu-modeI}
\end{figure}

As the first numerical example, the proposed model is used to simulate the mode-I fracture test on a concrete beam under both monotonic and cyclic loading cases. The reference experiments were conducted by \cite{chen2023fracture} in a displacement-controlled manner during loading and a force-controlled manner during unloading. However, in the numerical simulations, both the loading and unloading regimes are modeled in a displacement-controlled manner. The applied loading is provided in the supplementary materials of this study. The dimensions and boundary conditions are presented in Figure \ref{figure:chenLiu-modeI}. Material properties and loading-unloading information are also provided in the reference study and are directly used in our numerical model. The material properties are listed in Table \ref{table:chenLiu-material}.

\begin{table} 
	\small
	\centering
	\caption{Material constants used in the mode-I fracture test of the high-strength concrete beam.}
	\begin{tabular}{cccc}
		\hline \hline
		&Symbol & Values &Units \\
		\hline
		&$E$ & $33.89$ & GPa\\
		&$\sigma^C$ &  $3.39$ &  MPa\\
		&$\mathcal{G}_c$ &  $116.15$ &  N/m\\
		\hline\hline
	\end{tabular}
	\label{table:chenLiu-material}
\end{table}

The uniform grid spacing is 2 mm, and the horizon is set to 6.03 mm. The stable time step is calculated as $4.64\times 10^{-7}$ seconds, and the local damping coefficient is set to $2.0\times 10^{7}$ kg/m$^3 \cdot $s. The characteristic dimension, L, is selected as the beam depth, which is 150 mm.

\begin{figure} 
	\centering
	\begin{subfigure}[b]{0.8\textwidth}
		\includegraphics[width=1.0\textwidth]{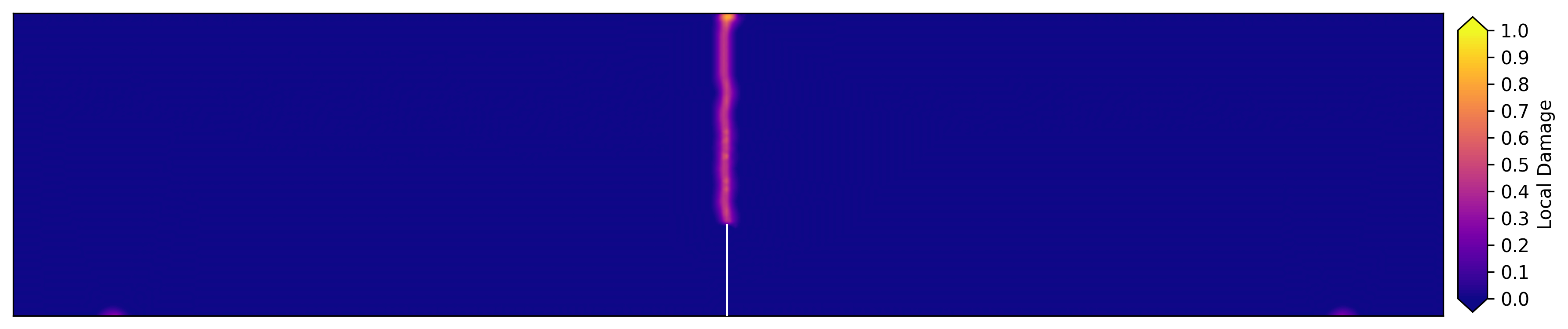}
		\caption[]{}
		\label{fig:chen-mono-cracks}
	\end{subfigure}
	\vfill
	\begin{subfigure}[b]{0.8\textwidth}
		\includegraphics[width=1.0\textwidth]{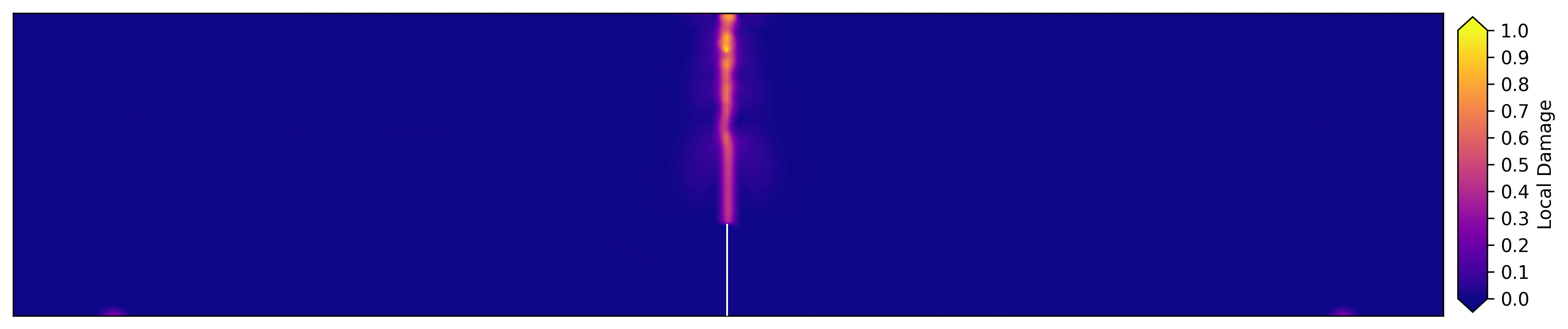}
		\caption[]{}
		\label{fig:chen-cyclic-cracks}
	\end{subfigure}
	\caption%
	{Local damage contour plots showing the crack trajectories obtained under (a) monolithic loading, and (b) cyclic loading.}
	\label{figure:chen-cracks}
\end{figure}

The local damage contour plots, shown in Figure \ref{figure:chen-cracks}, illustrate the numerically predicted cracks for both monotonic and cyclic loading cases. As seen in Figure \ref{figure:chen-cracks}, beams under both loading conditions fail due to straight crack propagation from the tip of the pre-notch to the loading zone at the top middle part of the beam, consistent with mode-I failure. The damage localizes in a narrow band, forming a well-defined crack path that propagates vertically from the notch tip toward the loading point. In addition, the similarity in crack patterns between the monotonic and cyclic loading cases indicates that the fundamental fracture mechanism remains unchanged regardless of the loading history.

\begin{figure} 
	\centering
	\begin{subfigure}{0.48\textwidth}
		\label{fig:chenLiuResults-a}
		\includegraphics[width=1.0\textwidth]{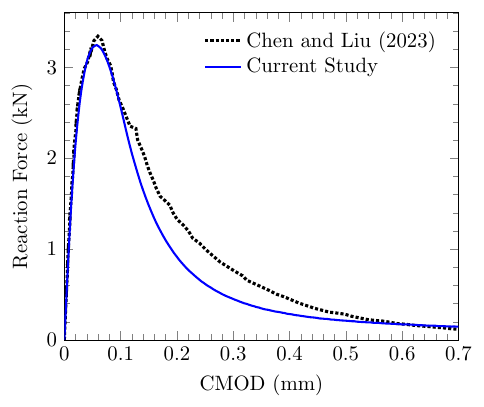}
		\caption{}
        \label{figure:chenLiuResults-a}
	\end{subfigure}
	\hfill
	\begin{subfigure}{0.48\textwidth}
		\centering
		\label{fig:chenLiuResults-b}
		\includegraphics[width=1.0\textwidth]{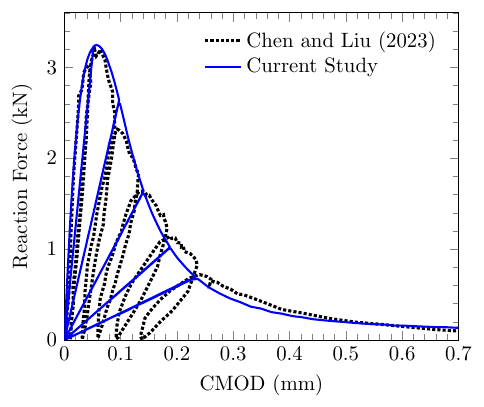}
		\caption{}
        \label{figure:chenLiuResults-b}
	\end{subfigure}
	\caption{Comparison of experimental and numerical force vs CMOD curves for (a) monolithic, and (b) cyclic loading. Experimental data: black dotted Line; numerical results: blue solid line.}
	\label{figure:chenLiuResults}
\end{figure}

The comparison of numerical and experimental load vs. Crack Mouth Opening Displacement (CMOD) curves is presented in Figure \ref{figure:chenLiuResults} for both monotonic and cyclic loading cases. In both cases, an excellent agreement is observed between the numerical and experimental results. For the monotonic loading case, provided in Figure \ref{figure:chenLiuResults-a}, the numerical simulation accurately captures the initial linear elastic response, the peak load, and the general softening behavior. The peak load of approximately 3.3 kN closely matches the experimental value. For the cyclic loading case, presented in Figure \ref{figure:chenLiuResults-b}, the numerical model successfully reproduces the characteristic hysteresis loops observed in the experiment. The loading paths of each cycle and the peak loads at various CMOD values align well with the experimental data. It is worth noting that the numerical curves return to the origin at the end of each unloading cycle, whereas the experimental curves do not due to accumulated plastic deformation. This occurs because the current model only considers fully elastic unloading and reloading. However, since our focus is primarily on brittle and quasi-brittle materials and failure mechanisms, plastic deformations are not as significant as in ductile failure. Therefore, this difference is considered acceptable for this study.

The results demonstrate that the proposed blended formulation can effectively model the mode-I fracture behavior of high-strength concrete under both monotonic and cyclic loading conditions. Although the current model does not account for permanent deformations during unloading cycles, it provides a reasonable approximation for the overall mechanical response of quasi-brittle materials like high-strength concrete.
\subsection{Mixed-mode fracture tests on concrete beams under cyclic loading}
\label{sec.mixed-mode-tpb-cyclic}
\begin{figure} 
	\centering
	\includegraphics[width=0.65\textwidth]{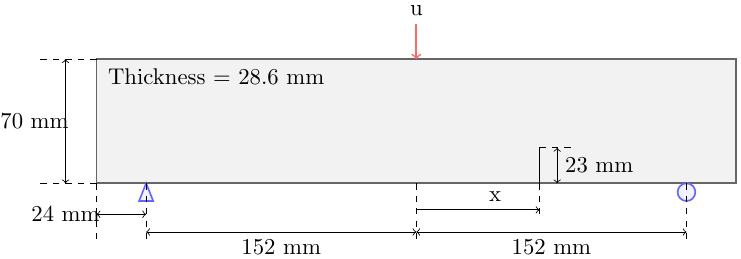}
	\caption{Dimensions and boundary conditions of the mixed-mode fracture tests on concrete beams.}
	\label{fig:geometry-mixedMode}
\end{figure}

Mixed-mode failure of concrete, where a combination of tensile and shear modes exists, is commonly observed in concrete structures. Meanwhile, as \cite{jenq1988mixed} stated earlier, crack trajectories resulting from mixed-mode failure are far more complex than those occurring due to mode-I failure. \cite{jenq1988mixed} studied crack initiation theories using linear elastic fracture mechanics (LEFM) and proposed a mixed-mode crack stability criterion. To verify their proposed model, they conducted a series of experiments, and we use their experimental findings to evaluate the performance of our numerical model. The dimensions and boundary conditions of the selected beams are presented in Figure \ref{fig:geometry-mixedMode}. The compressive strength and Young’s modulus are provided in the reference study as 34.3 MPa and 32.8 GPa, respectively. However, the tensile strength and critical fracture energy release rate are not specified. Therefore, the values given in Table \ref{table:jenqShah-material} are selected and kept constant for all tests conducted in this section.

In the reference study, the offset ratio, $\gamma$, is defined as the ratio of distance \textit{x} (see Figure \ref{fig:geometry-mixedMode}) to half the span length (half the distance between the two supports). Four sets of experiments were performed for each test, corresponding to $\gamma = 0, \; 1/6, \; 1/3, \; \text{and } 1/2$. It is important to note that the force vs. CMOD curves obtained from the study by \cite{jenq1988mixed} are presented as "typical force-CMOD" curves. Unfortunately, the study does not provide all four experimental curves for each test. Therefore, we cannot assess the variation in the force-displacement responses of the beams. In addition, the exact loading and unloading protocol is not provided in the experimental study; therefore, the applied displacements (please see the supplementary materials) are selected such that the unloading points match with the experimental ones as closely as possible. However, the final crack paths for each test are available in the reference study by \cite{jenq1988mixed} and are used to compare the predicted crack trajectories with the experimental ones.

\begin{table} 
	\small
	\centering
	\caption{Material constants used in the mixed-mode fracture tests of the concrete beams.}
	\begin{tabular}{cccc}
		\hline \hline
		&Symbol & Values &Units \\
		\hline
		&$E$ & $32.8$ & GPa\\
		&$\sigma^C$* &  $3.25$ &  MPa\\
		&$\mathcal{G}_c$* &  $115$ &  N/m\\
		\hline\hline
	\end{tabular}
	\vfill
	*Not provided in the experimental study by \cite{jenq1988mixed}.
	\label{table:jenqShah-material}
\end{table}

In the numerical models, the uniform grid spacing is 1 mm and the horizon is selected as 3.015 mm. The local damping coefficient is used as $4.0\times 10^{7}$ kg/m$^3 \cdot $s, and the time step is selected as $2.36\times 10^{-7}$ seconds. The characteristic dimension, L, is chosen as the beam depth, which is 70 mm.

\begin{figure} 
	\centering
	\small
	\begin{subfigure}[b]{0.49\textwidth} 
		\includegraphics[width=1.0\textwidth]{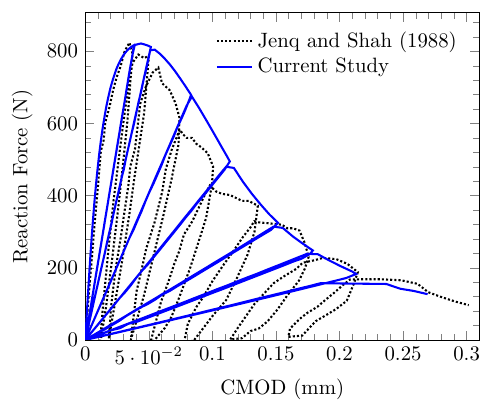}
		\caption[]%
		{{\small x = 0 mm ($\gamma = 0$)}}
		\label{fig:Jenq-Shah-test-forceDisp-1}
	\end{subfigure}
	\hfill
	\begin{subfigure}[b]{0.49\textwidth}
		\includegraphics[width=1.0\textwidth]{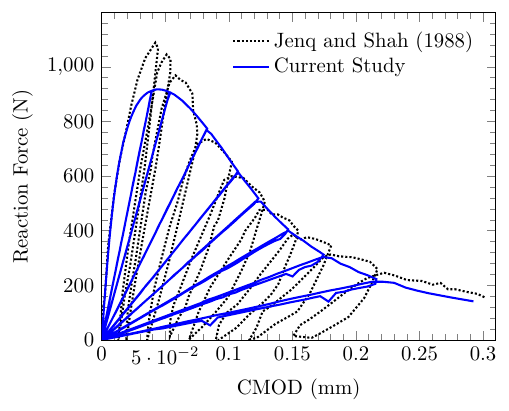}
		\caption[]%
		{{\small x = 25 mm ($\gamma = 1/6$)}}
		\label{fig:Jenq-Shah-test-forceDisp-2}
	\end{subfigure}
	\vfill
	\begin{subfigure}[b]{0.49\textwidth}
		\includegraphics[width=1.0\textwidth]{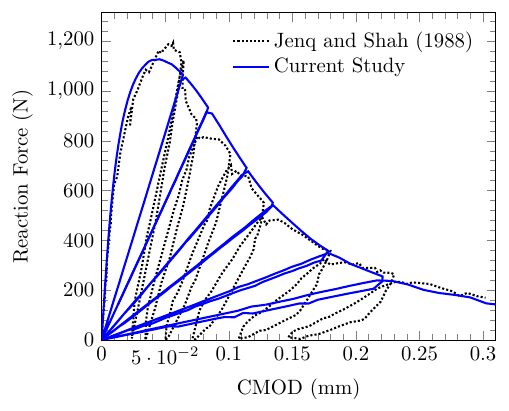}
		\caption[]%
		{{\small x = 50 mm ($\gamma = 1/3$)}}
		\label{fig:Jenq-Shah-test-forceDisp-3}
	\end{subfigure}
	\hfill
	\begin{subfigure}[b]{0.49\textwidth}
		\includegraphics[width=1.0\textwidth]{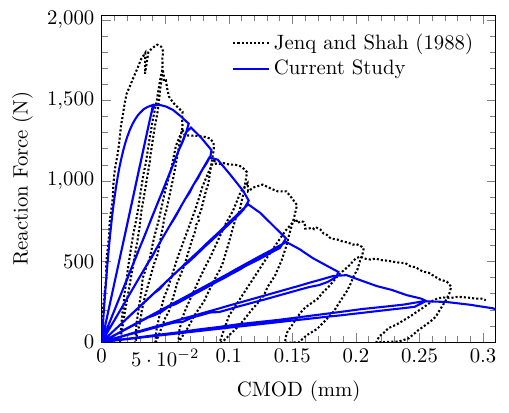}
		\caption[]%
		{{\small x = 76 mm ($\gamma = 1/2$)}}
		\label{fig:Jenq-Shah-test-forceDisp-4}
	\end{subfigure}
	\caption%
	{Comparison of experimental and numerical force vs CMOD curves for the offset values: (a) x=0, (b) x= 25 mm, (c) x=50 mm, and (d) x=76 mm. Experimental data: black dotted Line; numerical results: blue solid lines.}
	\label{fig:Jenq-Shah-test-forceDisp}
\end{figure}

It is important to note that since the values for $\sigma^C$ and $\mathcal{G}_c$ are not provided in the reference study conducted by \cite{jenq1988mixed}, we calibrated these values by comparing the force vs CMOD curves for the mode-I test shown in Figure \ref{fig:Jenq-Shah-test-forceDisp-1}. However, these values remain unchanged for the remaining tests, i.e., no additional calibration was performed for the mixed-mode failure cases.

By comparing the typical force vs. CMOD response from the experimental study, we observe that the force-carrying capacity of the beams is underestimated for the mixed-mode tests, as shown in Figures \ref{fig:Jenq-Shah-test-forceDisp-2} - \ref{fig:Jenq-Shah-test-forceDisp-4}. 

\begin{figure}  
	\centering
	\begin{subfigure}[b]{0.24\textwidth}
		\centering
		\includegraphics[width=1.0\textwidth]{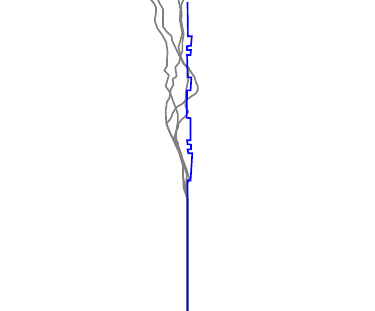}
		\caption[]%
		{{\small x = 0 mm}}
	\end{subfigure}
	\hfill
	\begin{subfigure}[b]{0.24\textwidth}
		\centering
		\includegraphics[width=1.0\textwidth]{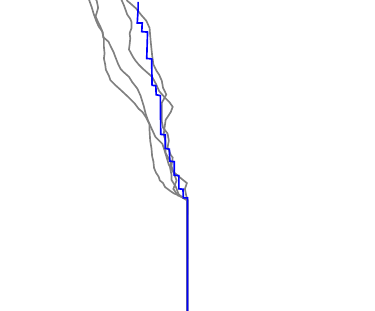}
		\caption[]%
		{{\small x = 25 mm}}
	\end{subfigure}
	\hfill
	\begin{subfigure}[b]{0.24\textwidth}
		\centering
		\includegraphics[width=1.0\textwidth]{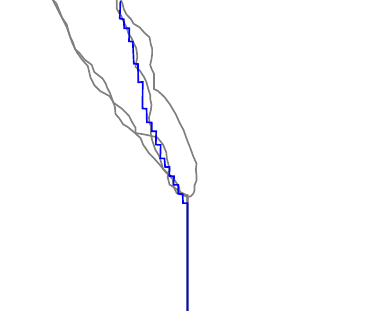}
		\caption[]%
		{{\small x = 50 mm}}
	\end{subfigure}
	\hfill
	\begin{subfigure}[b]{0.24\textwidth}
		\centering
		\includegraphics[width=1.0\textwidth]{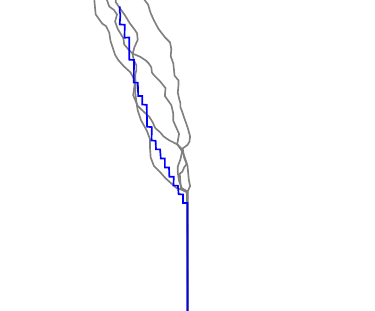}
		\caption[]%
		{{\small x = 76 mm}}
	\end{subfigure}
	\caption%
	{Comparison of the crack trajectories for the offset distances (a) x=0, (b) x= 25 mm, (c) x=50 mm, and (d) x=76 mm. While gray solid lines represents the experimental crack paths borrowed from \cite{jenq1988mixed}, blue solid lines are obtained from the local damage plots.}
	\label{fig:Jenq-Shah-test-cracks}
\end{figure}

The final crack paths obtained from the numerical simulations are compared with the experimental ones in Figure \ref{fig:Jenq-Shah-test-cracks}. Here, four experimental tests for each pre-notch location are shown as solid gray lines, while the blue solid lines represent the numerically predicted crack trajectories, drawn from the damage variable contour plots. In order to obtain crack trajectories from the numerical results, horizontal displacement contour plots are used. The horizontal displacement contour plots as well as the damage contour plots can be found in the supplementary materials of this study.

We believe Figure \ref{fig:Jenq-Shah-test-cracks} highlights two important conclusions. Firstly, the crack paths obtained from the numerical simulations align well with the experimentally observed ones. Secondly, noticeable variations exist within the experimental results, even though the material properties and boundary conditions were intended to be the same. Concrete is a composite material where micro-cracks and voids are inevitable, even in undamaged specimens. In addition to the interaction between micro-cracks, the stiffness variations among its constituents (e.g., the mortar matrix and different sizes/types of aggregates) contribute to these discrepancies in the experimental results.

Furthermore, our numerical model treats concrete as a homogeneous material and does not account for micro-scale effects. Nevertheless, our numerical results exhibit strong agreement with the experimental findings, providing an accurate modeling approach for fracture simulations in quasi-brittle materials. 

\subsection{Mixed-mode fracture of concrete L-shaped panel}
\label{sec.mixed-mode-Lshaped}
\begin{figure} 
	\centering
	\includegraphics[width=0.45\textwidth]{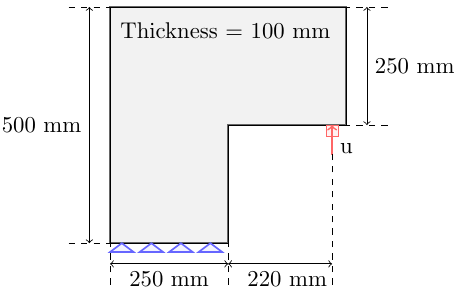}
	\caption{Dimensions and boundary conditions of the mixed mode fracture test with l-shaped concrete panel.}
	\label{figure:lShapedDimensions}
\end{figure}

As the third numerical example, we study the mixed-mode fracture test of an L-shaped concrete panel without a pre-notch. The experimental test was performed in a displacement-controlled manner by \cite{winkler2001traglastuntersuchungen}. The dimensions and boundary conditions are presented in Figure \ref{figure:lShapedDimensions}.

The L-shaped specimen has external dimensions of 500 mm $\times$ 500 mm with a thickness of 100 mm. The internal corner forms a 90$^\circ$ angle with dimensions of 250 mm $\times$ 250 mm. The specimen geometry creates a stress concentration at the internal corner, which initiates a mixed-mode fracture.

Since the numerical model does not incorporate rotational degrees of freedom, the clamped boundary conditions are applied by constraining the displacements in both directions through a layer with a height equal to the horizon size, as shown in Figure \ref{figure:lShapedDimensions}. The vertical displacements are applied to a square region as a linearly increasing function, reaching a final value of 0.65 mm over 125 time steps. The stable time is calculated as $8.77\times 10^{-8}$ seconds and the local damping coefficient is set to $2.0\times 10^{7}$ kg/m$^3 \cdot $s. The uniform grid size is 2.5 mm and the horizon size is selected as 7.875 mm. The characteristic dimension, L, is selected as the length of the domain where the crack propagates, which is 250 mm.

The material properties listed in Table \ref{table:lShaped-material} are taken as it is given in the experimental work conducted by \cite{winkler2001traglastuntersuchungen} except the Young's modulus. The Young's modulus is selected as 18 GPa, which better represents the initial slope of the experimental force-displacement curves.

\begin{table} 
	\small
	\centering
	\caption{Material constants used in the mixed-mode fracture test of the concrete panel.}
	\begin{tabular}{cccc}
		\hline \hline
		&Symbol & Values &Units \\
		\hline
		&$E$* & $18$ & GPa\\
		&$\sigma^C$ &  $2.7$ &  MPa\\
		&$\mathcal{G}_c$ &  $90.0$ &  N/m\\
		\hline\hline
	\end{tabular}
	\vfill
	*Different than the experimental values provided by \cite{winkler2001traglastuntersuchungen}: E=25.85 GPa.
	\label{table:lShaped-material}
\end{table}

\begin{figure}
	\centering
	\includegraphics[width=0.65\textwidth]{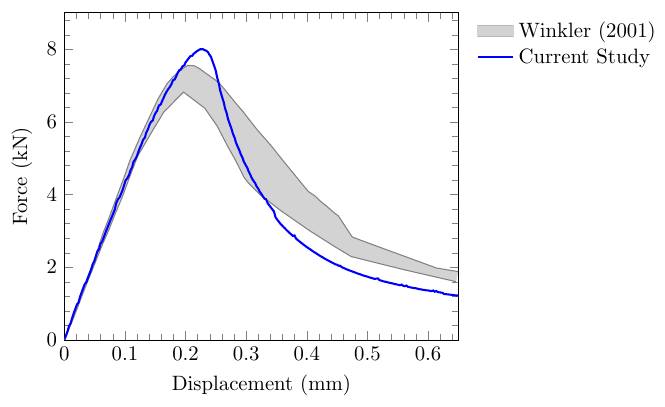}
	\caption{Comparison of experimental and numerical force vs displacement curves. Experiment: gray shaded region, numerical: solid blue line.}
	\label{figure:lShaped-results}
\end{figure}

Figure \ref{figure:lShaped-results} presents the comparison of the experimental and numerical load vs displacement curves for the mixed-mode fracture test of the concrete l-shaped panel. As can be seen from this figure, the load-carrying capacity as well as the softening behavior are predicted by the proposed model within an acceptable accuracy. The numerical model predicts a peak load of approximately 8 kN, which is slightly higher than the experimental average of around 7 kN. The initial elastic response up to approximately 0.15 mm displacement closely matches the experimental observations, confirming the appropriateness of the adjusted Young's modulus. The post-peak softening behavior follows the general trend of the experimental data, though our model predicts a slightly more rapid decrease in load capacity between 0.3 mm and 0.45 mm displacement compared to the experimental range.

\begin{figure} 
	\centering
	\begin{subfigure}{0.48\textwidth}
		\centering
		\includegraphics[width=0.9\linewidth]{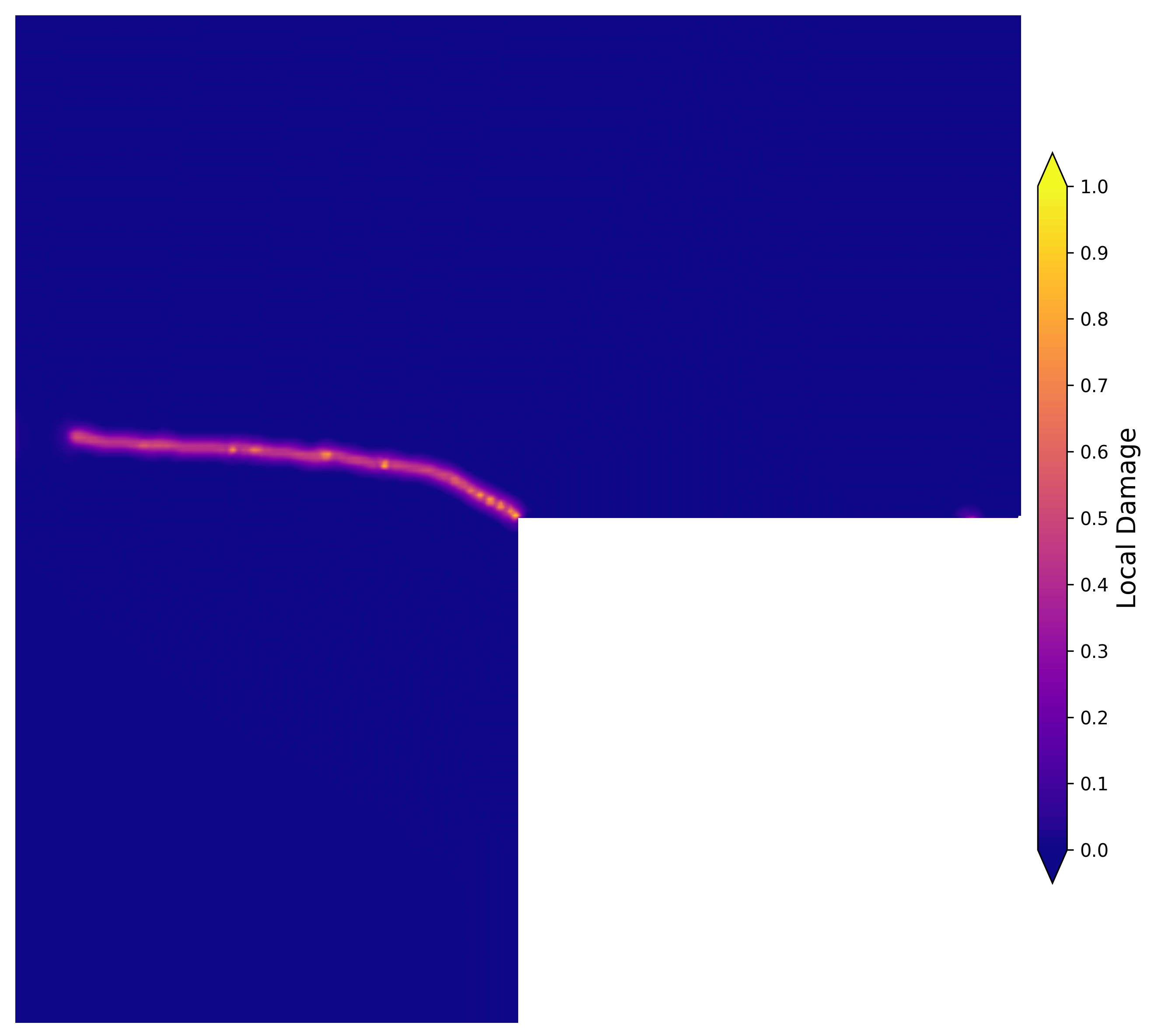}
		\caption{}
        \label{figure:lShaped-cracks-a}
	\end{subfigure}
	\hfill
	\begin{subfigure}{0.48\textwidth}
		\centering
		\includegraphics[width=0.9\textwidth]{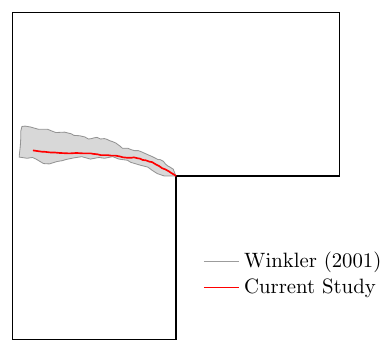}
		\caption{}
        \label{figure:lShaped-cracks-b}
	\end{subfigure}
	\caption{(a) Local damage contour plot and (b) comparison of the crack trajectories.}
	\label{figure:lShaped-cracks}
\end{figure}

Figure \ref{figure:lShaped-cracks-a} illustrates the local damage contour plot at the final displacement step. The damage localizes along a curved path starting from the inner corner of the L-shaped panel, which is the region of the highest stress concentration. The crack initiates in a mixed-mode fashion due to the combined tension and shear stresses at the inner corner, and it propagates toward the upper edge of the specimen.

Figure \ref{figure:lShaped-cracks-b} provides a comparison between the numerically predicted crack trajectory and the experimental observations. The gray shaded region represents the experimental crack paths observed across multiple specimens, showing some natural variability in the experiments. The red solid line indicates our numerical prediction, which falls well within the experimental region. The crack path exhibits a characteristic curved trajectory that starts perpendicular to the maximum principal stress direction at the inner corner and gradually curves upward as it propagates.

The agreement between the numerical and experimental crack trajectories demonstrates the capability of the proposed model to accurately capture the mixed-mode fracture behavior in concrete structures without the need for pre-defined crack paths or special interface elements.

\subsection{Dynamic crack propagation in a glass sheet}
\label{sec.dynamic-crack}
Dynamic crack propagation in brittle materials has been extensively studied (\cite{freund1990dynamic}, \cite{fineberg1992instability}, \cite{ravi2004dynamic}, \cite{zhou2005rate}, \cite{BobaruZhang}, \cite{bleyer2017dynamic}, \cite{rakici2023discrete}) due to its importance in understanding failure mechanisms. The phase field method of \cite{miehe2010phase} developed for quasi-static and  quasi-brittle materials was extended to dynamic quasi-brittle fracture in the work of \cite{borden2012phase}. In this section, we apply the blended model and present numerical simulations of dynamic crack propagation in a glass sheet under suddenly applied tensile stress and compare our results with those of  \cite{borden2012phase}. 

\begin{figure} 
	\centering
	\includegraphics[width=0.6\textwidth]{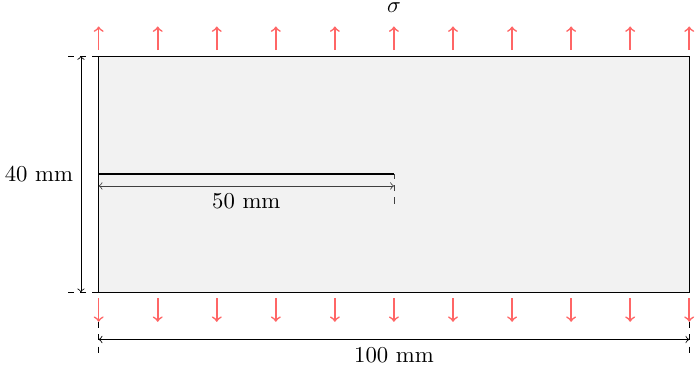}
	\caption{Dimensions and boundary conditions of the dynamic problem.}
	\label{figure:glass-plate}
\end{figure}

The computational domain consists of a rectangular specimen with dimensions 100 mm $\times$ 40 mm containing a pre-notch of 50 mm length positioned at the mid-height of the left edge, as illustrated in Figure \ref{figure:glass-plate}. The specimen is subjected to a 1 MPa tensile stress, $\sigma$ = 1 MPa, applied uniformly along the top and bottom edges. The damage evolution and crack propagation characteristics were monitored throughout the simulation time frame. Except for the tensile strength, the material properties are borrowed from \cite{song2008comparative} and \cite{borden2012phase}, are summarized in Table \ref{table:glass-material}. An average tensile strength value for a glass material is chosen to be used in the numerical simulations since it was not provided in the reference study.

\begin{table} 
	\small
	\centering
	\caption{Material constants for the glass sheet.}
	\begin{tabular}{cccc}
		\hline \hline
		&Symbol & Values &Units \\
		\hline
        &$\rho$ & $2450$ & kg/$\text{m}^3$\\
		&$E$ & $32$ & GPa\\
		&$\sigma^C$ &  $50$ &  MPa\\
		&$\mathcal{G}_c$ &  $3$ &  N/m\\
		\hline\hline
	\end{tabular}
	\label{table:glass-material}
\end{table}

\begin{figure} 
	\centering
	\begin{subfigure}{0.65\textwidth}
		\includegraphics[width=1.0\textwidth]{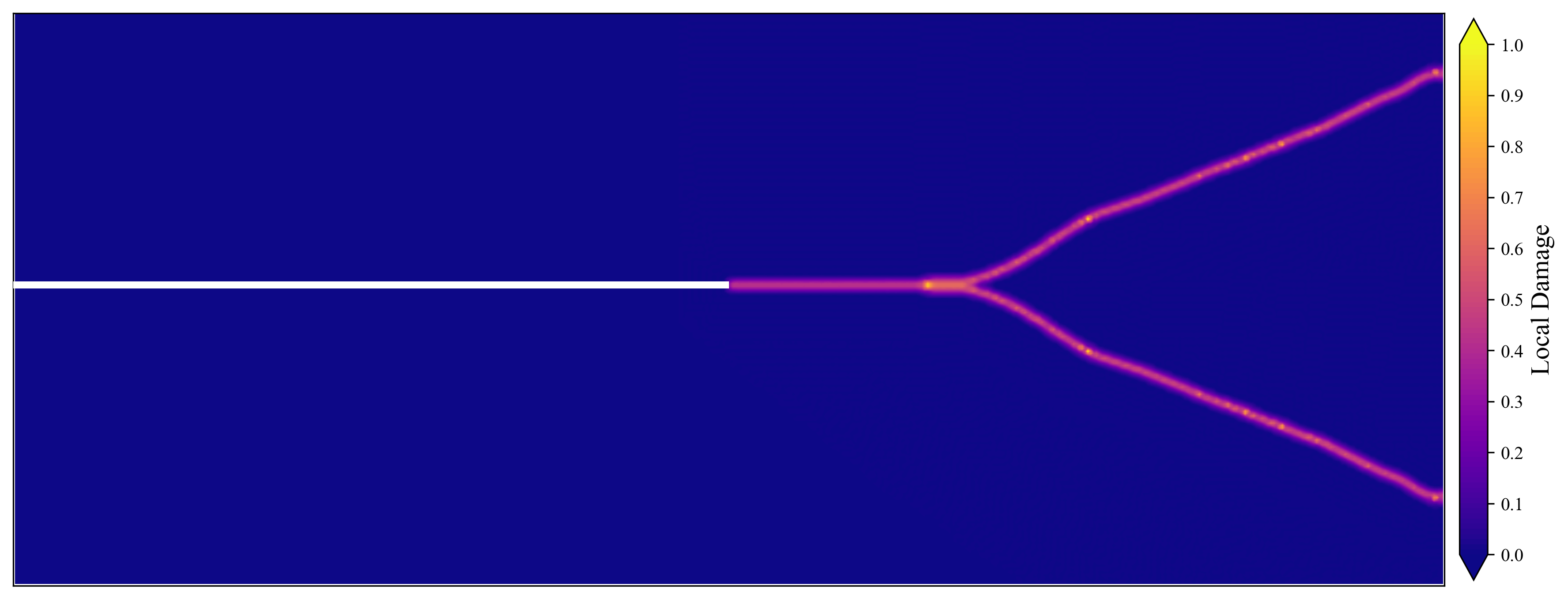}
		\caption{}
        \label{figure:dynamicCrack-a}
	\end{subfigure}
	\vfill
	\begin{subfigure}{0.65\textwidth}
		\includegraphics[width=1.0\textwidth]{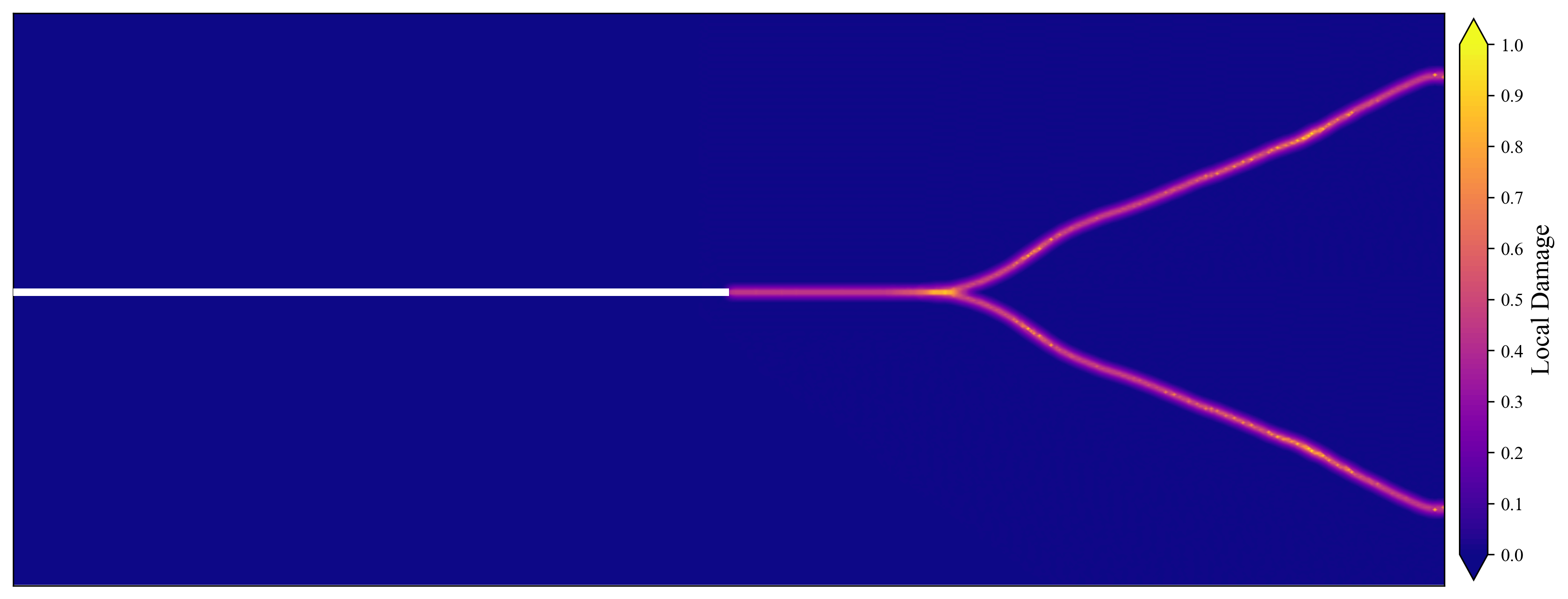}
		\caption{}
        \label{figure:dynamicCrack-b}
	\end{subfigure}
    \vfill
	\begin{subfigure}{0.65\textwidth}
		\includegraphics[width=1.0\textwidth]{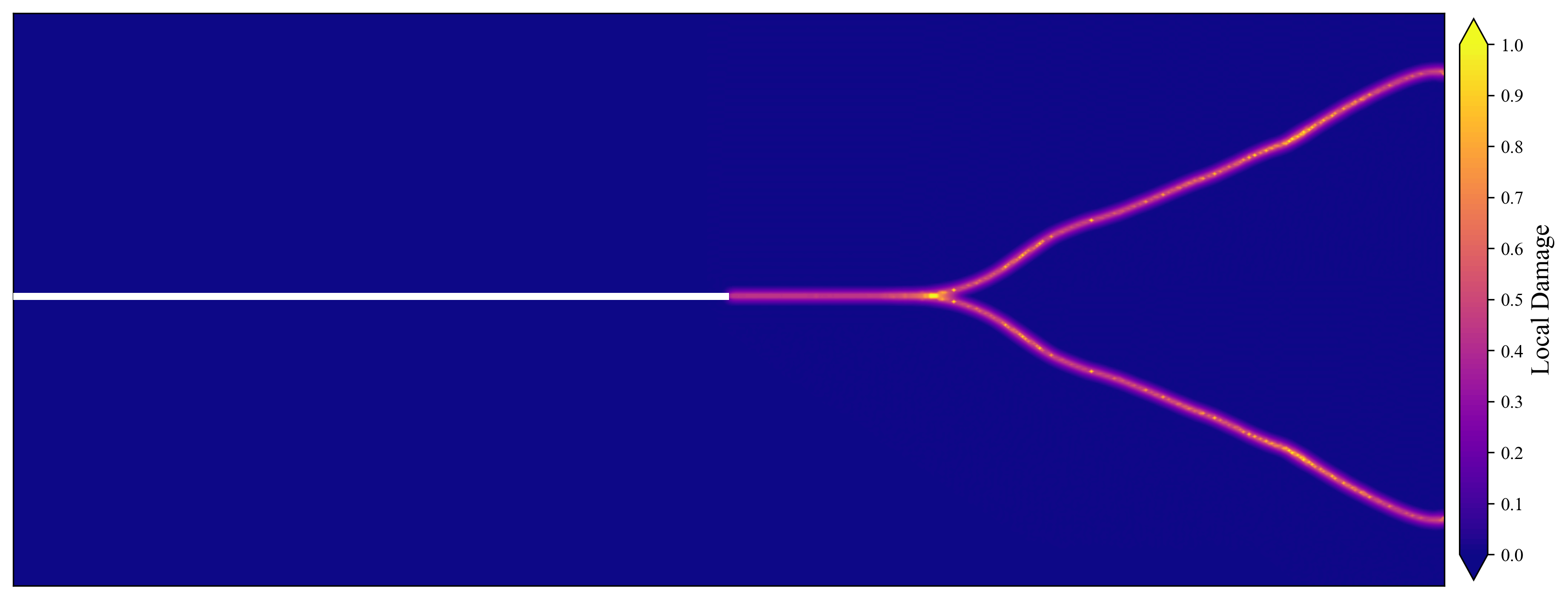}
		\caption{}
        \label{figure:dynamicCrack-c}
	\end{subfigure}
	\caption{Damage variable contour plots at 80 $\mu$s for (a) Mesh I, and (b) Mesh II, and Mesh III. White solid line represents the pre-notch.}
	\label{figure:dynamicCrack}
\end{figure}

Similar to \cite{borden2012phase}, three different uniform grid spacings, $h$, are used: 0.25, 0.125, and 0.0625 mm. The horizon value is fixed as 0.75 mm for each grid spacing. Since the horizon directly affects the thickness of the cracks emerging in the numerical simulations, its value is carefully selected by taking into account this fact as well as the computational cost of the numerical analysis.  In the reference study, $h = $ 0.25, 0.125, and 0.0625 mm correspond to Mesh I, Mesh II, and Mesh III, respectively. Therefore, we adopt the Mesh I, Mesh II, and Mesh III naming convention while comparing the results of the current study with those of \cite{borden2012phase}.  Since this numerical problem is transient, the local damping coefficient in the dynamic relaxation technique used to obtain the static result is set to zero. The stable time steps calculated by \cite{borden2012phase} are also utilized in this study as 100, 50, and 25 nanoseconds for the Mesh I, Mesh II, and Mesh III, respectively. These spatial and temporal parameters remain constant throughout each simulation. The characteristic dimension, L is calculated and is 0.75 mm, and $\epsilon$  was chosen to be $L$.
Figure \ref{figure:dynamicCrack} presents the local damage variable contour plots  at 80 $\mu$s for the uniform grid spacing $h = $ 0.25, 0.125, and 0.0625 mm, i.e. Mesh I, Mesh II, and Mesh III, respectively.  Initially, the crack propagates horizontally, but at approximately 35 $\mu s$, it bifurcates into two distinct crack paths forming a Y-shaped pattern. This branching behavior is a well-documented phenomenon in dynamic fracture mechanics of brittle materials under high-intensity loading conditions, where the energy release rate exceeds a critical threshold that makes a single crack path energetically unfavorable. Unlike the results of the reference study, each crack branch successfully reaches the right boundary at 80 $\mu$s. Therefore, we did not observe the effect of grid spacing limiting the crack tip velocity, as reported for the Mesh I case in the reference study.

Since there is no need for a crack tracking algorithm in the proposed model, a post-processing algorithm was used to determine the crack tip location and velocity to present these results. The crack region was determined using a damage threshold criterion which was selected as 0.38. Nodes with damage values exceeding this threshold limit were identified as a part of the crack region. It is important to note that the local damage, calculated in accordance with \eqref{eq:localDamage}, is not a primary variable obtained directly by solving the governing equations. Instead, it is computed based on the two-point phase-field variable, $\gamma(\uu)(\yy,\xx,t)$, and the neighborhood of the material point to quantify the percentage of degradation experienced by that point relative to its original, intact condition. For a newly formed straight crack, the material point closest to the crack loses half of its neighbors. As a result, the local damage value for that point is 0.50. Since we aim to capture crack initiation, a local damage threshold of 0.38 is chosen, which is lower than 0.50. Since the coordinates and the corresponding damage values were available as output of the solver, the crack tip was identified as the rightmost x-coordinate where damage exceeded this threshold. In order to detect the crack branching, the distribution of the y-coordinates at this rightmost position were analyzed to determine the distance exceeding 2.0 mm, which indicates a crack branching. The crack tip velocity was calculated using a linear regression approach with a window size of 20 data points in order to minimize the noise in the measurement. Hence, this regression provided the instantaneous velocity which will be then compared with the results presented by \cite{borden2012phase}.

\begin{figure} 
	\centering
	\includegraphics[width=0.8\textwidth]{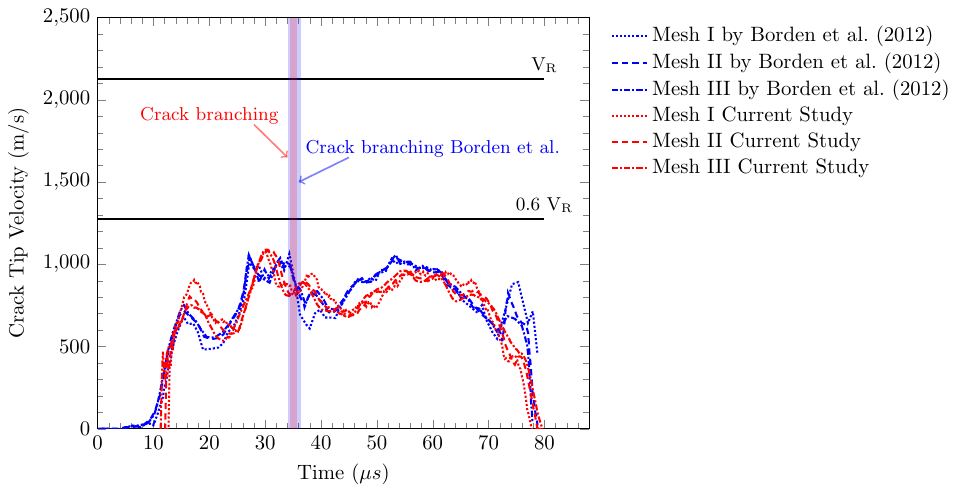}
	\caption{Comparison of the crack tip velocity and crack branching times of current study and with those of \cite{borden2012phase}.}
	\label{figure:dynamicCrack-comparison}
\end{figure}

Figure \ref{figure:dynamicCrack-comparison} presents a comparison of the crack tip velocities and crack branching times between the results of the current study and those of \cite{borden2012phase} for three different mesh configurations. The results from both numerical studies remain well below the Rayleigh wave speed, calculated to be 2125 m/s. In fact, none of the numerical results exceed the 60\% of the Rayleigh wave speed, $c_R$, which is consistent with experimental findings \cite{ChandarKnauss1984,guerraamaro2009}. Although the initial acceleration phase between 10-16 $\mu$s shows a notable difference only for the coarse mesh (Mesh I), the remaining portions of the crack tip velocity profiles exhibit good agreement. Similarly, the crack initiation time ranges in our study and in the phase field study by \cite{borden2012phase} exhibit a good agreement.

\begin{figure} 
	\centering
	\includegraphics[width=0.45\textwidth]{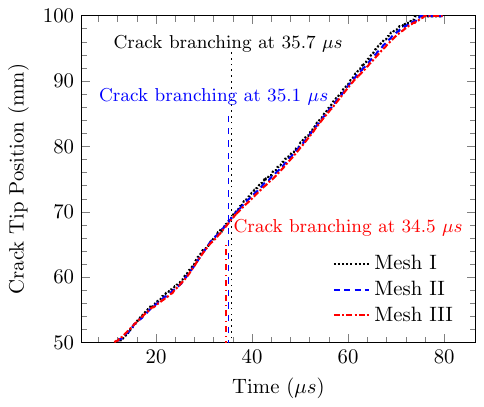}
	\caption{Crack tip position with respect to time.}
	\label{figure:dynamicCrack-position}
\end{figure}

The positions of the crack tip with respect to time are presented for Mesh I, II, and II and associated crack branching times are indicated in Figure \ref{figure:dynamicCrack-position}. It is seen that the Mesh I and Mesh II results are almost identical, indicating a reasonable convergence has been reached. Based on the data presented in Figures \ref{figure:dynamicCrack-comparison} and \ref{figure:dynamicCrack-position}, it is shown that the crack tip first reaches a velocity around 750 m/s ($\approx 0.35c_R$) and then accelerate even faster than the previous regime and reaches just below the $0.6c_R$. After the second  acceleration phase, velocity profile exhibit fluctuations, indicating unstable crack growth. Once the crack branches at $\approx 35\ \mu$s, the fluctuations in the velocity profile of the crack tip are decreasing, which may indicate a more stable crack propagation than before the branching. After 60 $\mu$s, a deceleration phase follows, where the crack gradually slows down as it approaches the right boundary.

The simulation demonstrates that the blended model captures the dynamic crack branching phenomenon in a glass sheet. The velocity profiles of the crack tip and the crack branching times agree well with the results of the phase field model reported by \cite{borden2012phase}. In addition, the maximum crack velocity observed in our simulations remains well below $0.6c_R$, consistent with the theoretical limit proposed by \cite{freund1990dynamic}. 
This limitation is attributed to energy dissipation mechanisms and the increasing instability of the crack path at higher velocities. On the other hand, dynamic fracture experiments on glass can exhibit certain chaotic behaviors, such as curved crack growth from the prenotch and micro-branching. In contrast, our numerical simulations show a perfectly straight crack trajectory from the prenotch and symmetric crack branching. This behavior is expected due to the perfectly symmetric geometry, loading, and initial boundary conditions, as well as the absence of material impurities in the model.

\subsection{Structural size effect in concrete beams}
\label{sec.size-effect-problem}
\cite{BazantPlanas1998} defines the structural size-effect as the deviation of the actual load-carrying capacity of the structure, due to the change of its size, from the one predicted by any deterministic theory where the failure of the material is expressed in terms of stress and/or strain. However, classical failure theories such as plastic limit analysis or maximum allowable stress or strain criterion do not depend on the size of the specimen under consideration, which is problematic since experiments show a strong size-effect in the failure of brittle and quasi-brittle materials.

To study the size-effect, we choose to work on the mode-I failure tests for different sized  beams made from the same concrete mix. The displacement-controlled experiments were performed by \cite{garcia2012analysis}. The dimensions of the beams are summarized in Table \ref{table:sizeEffect-beamDim} and the material properties of the concrete are given in Table \ref{table:sizeEffect-material}.

\begin{table} 
	\small
	\centering
	\caption{Dimensions of the beams tested for the mode-I failure test.}
	\begin{tabular}{cccccc}
		\hline \hline
		&Specimen &Depth (mm) &Span (mm) &Thickness (mm) &Prenotch Length (mm)\\
		\hline
		&Beam 1 &80  &200  &50 &20\\
		&Beam 2 &160 &400  &50 &40\\
		&Beam 3 &320 &800  &50 &80\\
		\hline\hline
	\end{tabular}
	\label{table:sizeEffect-beamDim}
\end{table}

\begin{table} 
	\small
	\centering
	\caption{Material constants used in the mode-I fracture test of the concrete beams.}
	\begin{tabular}{cccc}
		\hline \hline
		&Symbol & Values &Units \\
		\hline
		&$E$ & $33.8$ & GPa\\
		&$\sigma^C$ &  $3.5$ &  MPa\\
		&$\mathcal{G}_c$ &  $80.0$ &  N/m\\
		\hline\hline
	\end{tabular}
	\label{table:sizeEffect-material}
\end{table}

\begin{figure} 
    \small
	\centering
	\includegraphics[width=1.0\textwidth]{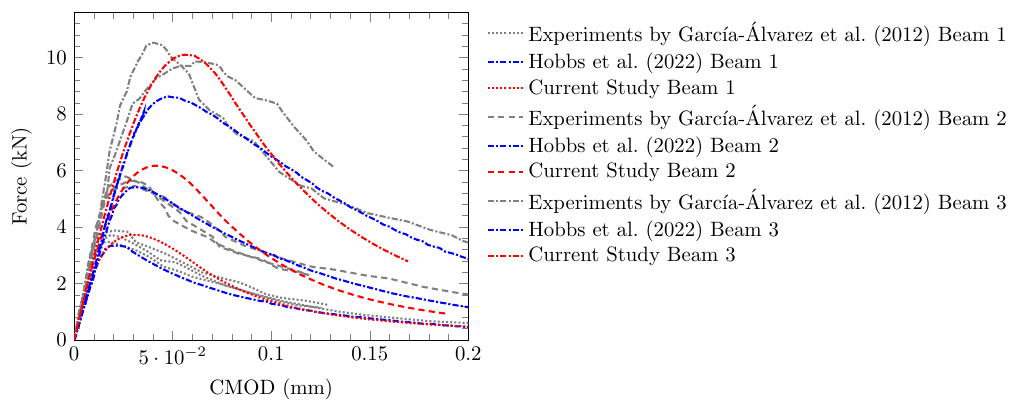}
	\caption{Comparison of the force  vs CMOD curves for the mode-I fracture test.}
    \label{figure:sizeEffect-loadDisp}
\end{figure}

Since the size-effect represents the dependence of the strength of the structure on its size, one first needs to define a measure of the strength of the structure and then compare the strength values obtained for different sizes with similar geometry of the specimen. The strength of the structure is conveniently characterized by the nominal stress at the maximum load. For this purpose, we calculate the ligament stress values at the ultimate load and use this value as the nominal strength of the beam. Since the beams are loaded under the three-point-bending set-up, the nominal strength of the beams can be calculated as

\begin{equation} \label{eq:nominalStrength}
	\sigma_{\text{lig}} = 1.5 \frac{F_{ult} \; s}{b \; h^2}
\end{equation}

\noindent
where $F_{ult}$ is the ultimate (maximum) load that the beam carries, $s$ is the length of the span, $b$ is the thickness of the beam, and $h$ is the length of the ligament (difference between the depth of the beam and the length of the pre-notch). Here we take $F_{ult}$ to be the maximum of our computed red dashed curves in Figure \ref{figure:sizeEffect-loadDisp}. The nominal strength of the structure $\sigma_{\text{lig}}$ is defined to be the state of stress where the load is applied necessary to fail the beam. Formula \eqref{eq:nominalStrength} is recovered from beam theory and given by the formula $\sigma_{\text{lig}}=My/I$ where the bending moment is $M=F_{ult} s/4$, the moment of inertia is $I=bh^3/12$, and $y=h/2$. 

\begin{figure} 
	\centering
	\includegraphics[width=1.0\textwidth]{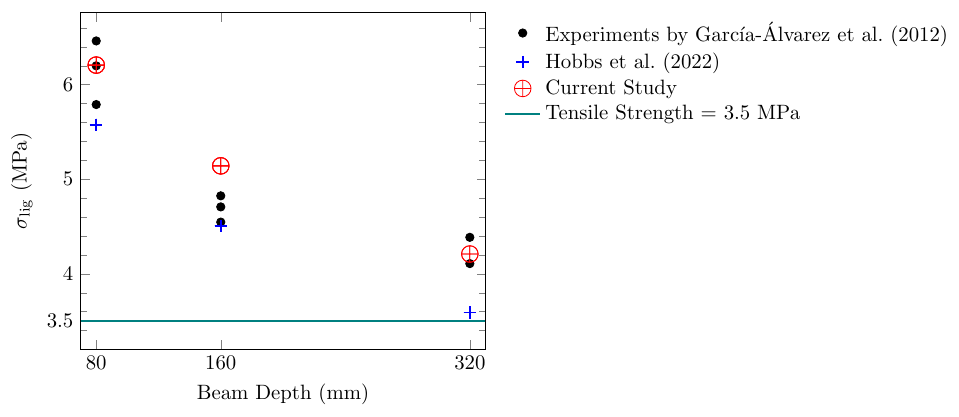}
	\caption{Nominal strength values with respect to beam depth.}
    \label{figure:sizeEffect-strengthSize}
\end{figure}

In the numerical simulations, the uniform grid spacing is 2 mm and the horizon is selected as 6.03 mm. The characteristic dimension, L is consistent with \eqref{bilinearfracttoughstrength}  and  delivers the beam depth of the medium size beam, which is 160 mm.
Here, our results are compared with the experimental findings of \cite{garcia2012analysis} and the peridynamic bond based model presented by \cite{hobbs}.  Once the maximum load carrying capacities are obtained from Figure \ref{figure:sizeEffect-loadDisp} for each experimental and numerical tests, the strength values can be calculated using \eqref{eq:nominalStrength} and plotted with respect to beam size as shown in Figure \ref{figure:sizeEffect-strengthSize}. Hence, the relation between the size and strength of the structure obtained from the experimental findings can be clearly seen, and the performance of two different peridynamic models to capturing the size-effect can be discussed.

In this study, we implemented a simple bilinear model to capture the softening effect observed in the quasi-brittle materials. Whereas, \cite{hobbs} uses a constitutive relation where there is initial exponential decay with a linear tail. Both models use the elastic modulus and the tensile strength of the material given in Table \ref{table:sizeEffect-material}. On the other hand, \cite{hobbs} use an estimated value for the fracture energy of 125.2 N/m which is obtained from an equation given in a design code. Here, we stay with the value provided in the original experimental work conducted by \cite{garcia2012analysis}.

Figure \ref{figure:sizeEffect-loadDisp} presents the load-CMOD results for beam 1, beam 2, and beam 3 whose dimensions are given in Table \ref{table:sizeEffect-beamDim}. The experimental curves are plotted by gray, \cite{hobbs}'s results by blue, and current results by red colored lines. The results of beams 1, 2, and 3 are plotted with dotted, dashed, and dash-dotted lines, respectively. As can be seen from Table \ref{table:sizeEffect-beamDim}, as beam 1 represents the smallest, beam 3 represents the largest among the three beams. The comparison of the peridynamic and blended models shows a subtle difference in the post-peak regime mainly because of the difference in the implemented constitutive models. In addition, the peak loads, or load-carrying capacities, show variation for the two numerical models, which may be a result of the selected fracture energy values as well as the constitutive models.

The nominal strength values, corresponding to the ligament stress at the peak load and calculated using \eqref{eq:nominalStrength}, are plotted with respect to the depth of the beam in Figure \ref{figure:sizeEffect-strengthSize}. While the tensile strength of the material is plotted by the teal solid line, the nominal strength values corresponding to each experimental test are shown with black-filled circles. As can be seen from the experimental values, there exists a strong dependence of the strength of the concrete beam on its size such that the average nominal strength values are obtained as 6.15 MPa, 4.7 MPa, and 4.25 MPa for the small, medium, and large beams, respectively. If the maximum allowable stress criterion is selected to be used as the ultimate tensile strength, the strength of beams 1 and 2 are significantly overestimated (43\% and 26\%, respectively). On the other hand, both blended and peridynamic bond based models exhibit a great performance capturing the size dependency on the strength values regardless of the differences in their constitutive models. 
It is important to note that none of the blended and peridynamic models employ an explicit rule to determine material strength based on sample size.


\section{Conclusions}
\label{sec.conclusion}

The blended model is formulated as a mathematically well-posed initial value boundary value problem for the displacement inside a quasi-brittle material. The solution is given by the displacement, damage field, and crack set history. The model inherently satisfies energy balance principles, with a positive energy dissipation rate that is consistent with thermodynamics. 
These properties are not artificially imposed, but emerge naturally from the material's constitutive law and the evolution equation.

In the numerical examples, the simplest possible constitutive rule is intentionally chosen to minimize the number of material and numerical constants. By utilizing Young's modulus, tensile strength, and fracture energy, the numerical constants can be determined, yielding both qualitative and quantitative results that align closely with the experimental data. The numerical simulations demonstrate excellent agreement with the experimental results for several benchmark problems, including mode-I fracture in concrete beams, mixed-mode fracture in notched specimens, fracture in L-shaped panels, and dynamic crack propagation. In particular, the same set of material constants and constitutive model successfully captures the size-effect phenomenon across three different beam sizes without incorporating any explicit rules for determining strength based on sample size. These results confirm the model's ability to replicate real-world material behavior with minimal computational complexity.

\section{Acknowledgment}
This material is based upon work supported by the U.S. Army Research Laboratory and the U.S. Army Research Office under Contract/Grant Number W911NF-19-1-0245 and W911NF-24- 2-0184.

\bigskip

\noindent {\bf Data availability} The code used to generate the results in this study is available at \\ https://github.com/SemsiCoskun/PDBlend.



\bibliographystyle{elsarticle-harv} 
\bibliography{references}






\end{document}